\documentclass[11pt]{article}
\usepackage{amsfonts,amsmath,amssymb}
\usepackage{enumerate}
\usepackage{hyperref}
\usepackage{bbm}
\usepackage{nicefrac}
\usepackage[all]{xy}
\usepackage{graphicx}
\usepackage{bm}
\usepackage{makecell}
\usepackage[table]{xcolor}

\usepackage{upgreek}
\usepackage{cancel}			% NEW

\usepackage{framed}

\usepackage{booktabs}
\newcommand{\ra}[1]{\renewcommand{\arraystretch}{#1}}

\newcommand{\be}{\begin{equation}}
\newcommand{\ee}{\end{equation}}

\newcommand{\bea}{\begin{eqnarray}}
\newcommand{\eea}{\end{eqnarray}}

\newcommand{\bes}{\begin{subequations}}
\newcommand{\ees}{\end{subequations}}

\newcommand{\cN}{{\cal N}}

\def\sst#1{{\scriptscriptstyle #1}}

\def\0{{\sst{(0)}}}
\def\1{{\sst{(1)}}}
\def\2{{\sst{(2)}}}
\def\3{{\sst{(3)}}}
\def\4{{\sst{(4)}}}
\def\5{{\sst{(5)}}}
\def\6{{\sst{(6)}}}
\def\7{{\sst{(7)}}}
\def\8{{\sst{(8)}}}

\def\cM{{{\cal M}}}

\newcommand{\vol}{\textrm{vol}}

\allowdisplaybreaks  %Introduces page breaks inside \eqnarray%

\usepackage{multirow}
\usepackage{rotating}

\newcommand{\ba}{\begin{align}}
\newcommand{\ea}{\end{align}}

\newcommand{\bse}{\begin{subequations}}
\newcommand{\ese}{\end{subequations}}

\allowdisplaybreaks

\global\long\def\ca{\mathcal{A}}
\global\long\def\ca{\mathcal{A}}
\global\long\def\cb{\mathcal{B}}
\global\long\def\cc{\mathcal{C}}

\global\long\def\cj{\mathcal{J}}

\global\long\def\cptwo{\mathbb{CP}^{2}}
\global\long\def\cpthree{\mathbb{CP}^{3}}

\global\long\def\tilzeta{\tilde{\zeta}}

\global\long\def\rp{\text{Re}}
\global\long\def\ip{\text{Im}}
\global\long\def\metriceleven{d\hat{s}_{11}^{2}}

\begin{document}

\makeatletter
\renewcommand{\theequation}{\thesection.\arabic{equation}}
\@addtoreset{equation}{section}
\makeatother

\begin{titlepage}

\begin{flushright}
IFT-UAM/CSIC-19-088 \\
%
%\today
\end{flushright}

\vspace{5pt}

   \begin{center}
   \baselineskip=16pt

   \begin{Large}\textbf{
\mbox{Embedding the SU(3) sector of SO(8) supergravity in $D=11$}
}
   \end{Large}

\vspace{25pt}

{\large  Gabriel Larios$^{1}$ ,\, Praxitelis Ntokos$^{2}$ \,and \,  Oscar Varela$^{1,2}$}
		
\vspace{25pt}

	\begin{small}

	  {\it $^{1}$ Departamento de F\'\i sica Te\'orica and Instituto de F\'\i sica Te\'orica UAM/CSIC , \\
   Universidad Aut\'onoma de Madrid, Cantoblanco, 28049 Madrid, Spain}   \\

	\vspace{15pt}
	
	  {\it $^{2}$ Department of Physics, Utah State University, Logan, UT 84322, USA}  \\

	\end{small}

\vskip 50pt

\end{center}

\begin{center}
\textbf{Abstract}
\end{center}

\begin{quote}

The SU(3)--invariant sector of maximal supergravity in four dimensions with an SO(8) gauging is uplifted to $D=11$ supergravity. In order to do this, the SU(3)--neutral sector of the tensor and duality hierarchies of the $D=4$ ${\cal N}=8$ supergravity is first worked out. The consistent $D=11$ embedding of the full, dynamical SU(3) sector is then expressed at the level of the $D=11$ metric and three-form gauge field in terms of these $D=4$ tensors. The redundancies introduced by this approach are eliminated at the level of the $D=11$ four-form field strength by making use of the $D=4$ duality hierarchy. Our results encompass previously known truncations of $D=11$ supergravity down to sectors of SO(8) supergravity with symmetry larger than SU(3), and include new ones. In particular, we obtain a new consistent truncation of $D=11$ supergravity to minimal $D=4$ ${\cal N}=2$ gauged supergravity.

\end{quote}

\vfill

\end{titlepage}

\tableofcontents

%%%%%%%%%%%%%%%%%%%%%%%%%%%%%%%%%%%%%%%%%%%%%%%%%%%%%%%%%%%%%%%%%%%%%%%%%%

%%%%%%%%%%%%%%%
%%%%%%%%%%%%%%%

\section{Introduction}

%%%%%%%%%%%%%%%
%%%%%%%%%%%%%%%

Being complicated theories with large field contents, it proves useful for applications to truncate maximal gauged supergravities to smaller subsectors that are invariant under some symmetry group. In this paper, we will be interested in $D=4$ $\cN=8$ supergravity with an electric SO(8) gauging \cite{deWit:1982ig} and one of its most fruitful sectors: the one invariant under the SU(3) subgroup of SO(8). This sector preserves $\cN=2$ supersymmetry and retains, along with the $\cN=2$ gravity multiplet, a vector multiplet and a hypermultiplet with an Abelian gauging. The (AdS) vacuum structure in this sector has been completely charted \cite{Warner:1983vz} and the corresponding mass spectra within the full $\cN=8$ theory determined \cite{Nicolai:1985hs,Bobev:2010ib}. Holographic duals have been established for some of these vacua as distinct superconformal phases \cite{Aharony:2008ug,Benna:2008zy} of the M2-brane field theory. Other interesting solutions of, for example, domain wall \cite{Ahn:2000mf,Bobev:2009ms}, defect \cite{Bobev:2013yra}, black hole \cite{Bobev:2018uxk} or Euclidean \cite{Bobev:2018wbt} type have been constructed in this sector that enjoy precise holographic interpretations \cite{Benna:2008zy,Freedman:2013ryh}. 

The relevance for holography of $D=4$ $\cN=8$ SO(8)--gauged supergravity \cite{deWit:1982ig} is intimately linked to the fact that it can be obtained as a consistent truncation of $D=11$ supergravity \cite{Cremmer:1978km} on the seven-sphere, $S^7$ \cite{deWit:1986iy,deWit:1984nz}. Further results on the consistency of the truncation have been given more recently in \cite{Nicolai:2011cy,deWit:2013ija,Godazgar:2013nma,Godazgar:2013dma,Godazgar:2013pfa,Lee:2014mla,Godazgar:2014nqa,Hohm:2014qga,Godazgar:2015qia,Varela:2015ywx,Kruger:2016agp}. The goal of this paper is to provide the consistent uplift of the SU(3) sector of SO(8) gauged supergravity into $D=11$ by using the uplifting formulae of \cite{Varela:2015ywx}, thus putting them to the test. We extend previous results on the consistent $D=11$ embedding of further subsectors contained in the SU(3) sector \cite{Corrado:2001nv,Azizi:2016noi,Bobev:2010ib}, and provide a unified treatment. We make contact with those previously known consistent truncations and establish new ones. In particular, we construct a new consistent embedding of $D=4$ $\cN=2$ pure gauged supergravity into $D=11$, where the internal geometry on $S^7$ corresponds to the $\cN=2$ $\textrm{SU}(3) \times \textrm{U}(1)$--invariant solution obtained by Corrado-Pilch-Warner (CPW) \cite{Corrado:2001nv}.

A systematic approach to the consistent uplift of $D=4$ $\cN =8$ SO(8) supergravity to $D=11$ was proposed in \cite{Varela:2015ywx}, similar to the method employed in \cite{Guarino:2015vca,Guarino:2015jca} to uplift $D=4$ $\cN=8$ ISO(7) supergravity \cite{Guarino:2015qaa} into type IIA. This approach relies on the tensor hierachy \cite{deWit:2008ta,deWit:2008gc} of maximal four-dimensional supergravity --the extension of its field content to include the magnetic gauge fields along with higher rank potentials in representations of E$_{7(7)}$. The full $D=11$ embedding of the bosonic sector of SO(8) supergravity can be expressed at the level of the $D=11$ metric and three-form potential in terms of a subset, dubbed {\it restricted} in \cite{Varela:2015ywx}, of the $D=4$ tensor hierarchy that is still $\cN=8$ but only covariant under $\textrm{SL}(8) \subset \textrm{E}_{7(7)}$. The $D=4$ tensor hierarchy carries redundant degrees of freedom beyond those contained in the conventional $\cN=8$ Lagrangian, and these are carried over to the $D=11$ embedding. These redundancies can be eliminated in $D=4$ by imposing suitable duality relations among the field strengths of the tensor hierarchy \cite{Bergshoeff:2009ph}. Expressing the $D=11$ embedding at the level of the four-form field strength and employing these $D=4$ dualisations, redundancy-free uplifting formulae are obtained that contain only the dynamically-independent fields (that is, the metric, the scalars and the electric vectors) that feature in the conventional $D=4$ $\cN=8$ Lagrangian. 

Some aspects of the SU(3)--invariant sector of SO(8)--gauged supergravity are summarised in section \ref{sec:SU3ofSO8}, and the SU(3)--invariant restricted tensor and duality hierarchies are constructed. Section \ref{sec:SU3ofSO8in11D} discusses the consistent uplift of the SU(3)--invariant sector into $D=11$ supergravity following the tensor and duality hierarchy approach. Contact with the consistent uplift of previously known subsectors is made and a new $D=11$ embedding of $D=4$ $\cN=2$ pure gauged supergravity is established. Section \ref{sec:11Dsolutions} further tests our formalism by recovering known AdS$_4$ solutions in $D=11$ from uplift of critical points, and section \ref{sec:Discussion} concludes. Some technical details are contained in the appendices. Our conventions for $D=11$ and $D=4$ $\cN=8$ supergravity are those of \cite{Varela:2015ywx}.

%%%%%%%%%%%%%%%
%%%%%%%%%%%%%%%

\section{The SU(3)--invariant sector of SO(8) supergravity} \label{sec:SU3ofSO8}

%%%%%%%%%%%%%%%
%%%%%%%%%%%%%%%

Let us start by reviewing some aspects of the SU(3) sector of SO(8)--gauged supergravity. We choose a triangular, or Iwasawa, parametrisation for the (SU(3)--invariant truncation of the) E$_{7(7)}/\textrm{SU}$(8) coset representative. Since previous literature often chooses the unitary gauge for the coset, we believe that our presentation has some intrinsic value even if the material that is covered (the Lagrangian in section \ref{sec:Lagrangian}, the further subsectors in \ref{sec:D=4subsectors}, and the vacuum structure in \ref{sec:vacua}) is mostly review. The SU(3)--invariant, restricted tensor and duality hierarchies worked out in section \ref{sec:RestTDH} are new.

\subsection{Field content and Lagrangian} \label{sec:Lagrangian}

The SU(3)--invariant sector of SO(8)--gauged maximal four-dimensional supergravity \cite{deWit:1982ig} corresponds to an $\cN=2$ supergravity coupled to a vector and a hypermultiplet. In addition to the fields entering these $\cN=2$ multiplets, we wish to consider the SU(3)--singlets in the (restricted, in the sense of \cite{Varela:2015ywx}) $\cN=8$ tensor hierarchy \cite{deWit:2008ta,deWit:2008gc}. The relevant bosonic matter content thus includes
\begin{eqnarray} \label{fieldContentHierarchy}
\textrm{the metric} & : & \quad ds_4^2 \; ,  \nonumber \\
\textrm{6 scalars} & : & \quad \varphi \; , \;  \chi \; , \; \phi \; , \;  a\; , \;  \zeta \; , \;  \tilde \zeta \; , \nonumber \\
\textrm{2 electric vectors and their magnetic duals} & : & \quad A^0 \; , \;  A^1 \; , \; \tilde{A}_0 \; , \;  \tilde{A}_1 \; ,  \\
\textrm{5 two-form potentials} & : & \quad B^0 \; ,  \; B^2 \; , \; B^{ab} = B^{(ab)} \; , 
\nonumber \\
\textrm{4 three-form potentials} & : & \quad C^1 \; , \;  C^{ab} = C^{(ab)} \; , \nonumber
\end{eqnarray}
all of them real. The superscripts on $B^0$, $B^2$ and $C^1$ are just labels without further meaning. The electric and magnetic vectors can be collectively denoted $A^\Lambda$ and $\tilde{A}_\Lambda$, with the index $\Lambda = 0 ,1$ formally labelling ``half" the fundamental representation of Sp$(4,\mathbb{R})$. The indices on $B^{ab}$ and $C^{ab}$ take on two values which, for convenience, are labelled $a=7,8$. The index $a$ formally labels a doublet of SL(2), but we do not attach any significance to its position as it can be raised and lowered with $\delta_{ab}$. See appendix \ref{sec:Construction} for the embedding of the SU(3)--invariant fields (\ref{fieldContentHierarchy}) into their parent $\cN=8$ counterparts.

Only the metric, the scalars and the vector fields enter the conventional Lagrangian. The fields $\varphi$, $\phi$ and $a$ are proper scalars, while $\chi$, $\zeta$ and $\tilde{\zeta}$ are pseudoscalars. All of these parametrise a submanifold
\begin{eqnarray} 
\label{ScalManN=2}
\frac{\textrm{SU}(1,1)}{\textrm{U}(1)} \times  \frac{\textrm{SU}(2,1)}{\textrm{SU}(2) \times \textrm{U}(1)} \,   
\end{eqnarray}
of E$_{7(7)}/$SU(8), where each factor respectively contains the vector-, $(\varphi, \chi)$, and the hypermultiplet, $q^u \equiv (\phi, a , \zeta , \tilde{\zeta})$, $u = 1, \ldots, 4$, (pseudo)scalars\footnote{We will rarely need indices to label the scalars but, when needed, the local indices will be denoted $m=1, \ldots, 6$, on the entire manifold (\ref{ScalManN=2}), $\alpha =1,2$ on the first factor, and $u= 1 , \ldots , 4$ on the second.}. The vectors gauge (electrically, in the usual symplectic frame), the $\textrm{U}(1)^2$, compact Cartan subgroup of the hypermultiplet isotropy group. In the Iwasawa parametrisation of the scalar manifold (\ref{ScalManN=2}), the bosonic Lagrangian reads
\begin{eqnarray}	\label{eq:Lagrangian}
{\cal L} &=&  R \, \textrm{vol}_4 + \tfrac{3}{2} ( d\varphi )^2  +  \tfrac{3}{2} e^{2 \varphi} \, ( d\chi )^2 +  2(D\phi )^2 + \tfrac{1}{2} \, e^{4 \phi} \,   \big( Da +  \tfrac{1}{2}  ( \zeta D \tilde{\zeta} - \tilde{\zeta} D \zeta  ) \big)^2    \\[5pt]
&&  +  \tfrac{1}{2} \, e^{2 \phi} \,  ( D\zeta )^2    +  \tfrac{1}{2} \, e^{2 \phi} \,   (D\tilde{\zeta} )^2 +  \tfrac{1}{2} \, \mathcal{I}_{\Lambda\Sigma} \, H_{\2}^{\Lambda} \wedge  * H_{\2}^{\Sigma}  + \tfrac{1}{2} \, \mathcal{R}_{\Lambda\Sigma} \, H_{\2}^{\Lambda} \wedge H_{\2}^{\Sigma}  - V \, \textrm{vol}_4  \nonumber  \ ,
\end{eqnarray}
with $ ( d\varphi )^ 2 \equiv d\varphi  \wedge * d\varphi $, etc. The covariant derivatives of the hyperscalars take on the form
\begin{eqnarray}	\label{eq:covder}
& D\phi = d\phi - g  A^0 \, a \; , \qquad 
Da = da + g A^0 \big( 1 +e^{-4\phi} (Z^2 -Y^2) \big) \; ,  \\[5pt]
 & D \zeta = d\zeta + g A^0  \,  e^{-2\phi} \big(  \zeta \, Z - \tilde{\zeta} \, Y \big)  -3g A^1 \,  \tilde{\zeta} \; , \;
 D \tilde{\zeta} = d\tilde{\zeta} + g A^0  \,  e^{-2\phi} \big(  \tilde{\zeta} \, Z + \zeta \, Y \big)  +3g A^1 \,  \zeta \; , \nonumber
\end{eqnarray}
where $g$ is the gauge coupling constant. Following \cite{Guarino:2015qaa}, here and throughout we have employed the shorthand definitions
\begin{equation} \label{scalDefs}
X \equiv 1+ e^{2\varphi} \chi^2
\hspace{5mm} , \hspace{5mm}
Y \equiv 1+\tfrac{1}{4} \, e^{2\phi} \, (\zeta^2+\tilde{\zeta}^2)
\hspace{5mm} , \hspace{5mm}
Z \equiv  e^{2\phi} \, a  \ .
\end{equation}
The covariant derivatives (\ref{eq:covder}) correspond to an electric gauging of the U$(1)^2$ Cartan subgroup of $\textrm{SU}(2) \times \textrm{U}(1) \subset \textrm{SU}(2,1)$ generated by
\begin{eqnarray} \label{GaugedIsometries}
k_0 = \tfrac{1}{\sqrt{2}} \,  (k[E_2] - k[F_2] ) \; , \qquad 
k_1 = -k[H_2]  \; , 
\end{eqnarray}
where $k[E_2]$, etc., are $\textrm{SU}(2,1)$ Killing vectors: see (\ref{eq:KillingSL2}) and (\ref{eq:KillingSU21}) for the explicit expressions for the Killing vectors of the scalar manifold (\ref{ScalManN=2}) in our parametrisation.

The scalar potential $V$ in (\ref{eq:Lagrangian}) reads
{\setlength\arraycolsep{1pt}
\begin{eqnarray}	\label{eq:scalarpot}
g^{-2} V & = & -12 e^\varphi -6 e^{-2\phi -\varphi} XY \big( e^{4\phi} +Y^2+Z^2 \big) -12 e^\varphi (Y-1) \big( 1+Y -\tfrac32 XY \big)   \\[5pt]
&& +6 e^{-2\phi -\varphi} (Y-1) \big( e^{4\phi} +Y^2 +Z^2 \big) X^2  
 + e^{-3\varphi} \Big[ \tfrac12 e^{-4\phi} +a^2 -1 + \tfrac12 e^{4\phi} (1+a^2)^2 \nonumber \\
&& \qquad \qquad \quad  +\tfrac12 e^{-4\phi} (Y-1) \big( 1+2 Z^2 - 2 e^{4\phi} +Y(1+2 e^{4\phi} +2Z^2)+ Y^2 + Y^3  \big)  \Big] X^3 \nonumber \; ,
\end{eqnarray}
}and derives from the following real superpotential (squared) 
{\setlength\arraycolsep{1pt}
\begin{eqnarray} \label{eq:superpot}
		W^2&=& \tfrac{1}{32} \, g^2 \, X\left[12 e^{-\varphi -2 \phi } (X-2) (Y-2) \left(Y^2+Z^2+e^{4 \phi }\right)+ 36 e^{\varphi } Y^2	
				\vphantom{\sqrt{(X-1) (Y-1) \left[\left(e^{4 \phi }-Y^2+Z^2\right)^2+4Y^2Z^2\right]}}	\right. \nonumber \\
			& &\qquad \qquad +e^{-3 \varphi -4 \phi} X^2\left(Y^2+Z^2+e^{4 \phi }\right)^2-16 e^{-3 \varphi } X^2 (Y-1) \nonumber 	\\
			&& \qquad \qquad  \left.-48 e^{-\varphi -2 \phi } \sqrt{(X-1) (Y-1) \left[\left(e^{4 \phi }-Y^2+Z^2\right)^2+4Y^2Z^2\right]}
		\right]\,,	
\end{eqnarray}
}through the usual formula
\begin{equation}
	\tfrac14V=2G^{mn}\partial_m W\partial_n W-3W^2\, .
\end{equation}
Here, $G_{mn}$, $m=1, \ldots , 6$, denotes the nonlinear sigma model metric on (\ref{ScalManN=2}), and $G^{mn}$ its inverse, which can be read off from the scalar kinetic terms in the Lagrangian (\ref{eq:Lagrangian}).

Finally, the gauge kinetic matrix is
\begin{equation}
\label{NMatrix}
\mathcal{N}_{\Lambda \Sigma} = \mathcal{R}_{\Lambda \Sigma} + i \, \mathcal{I}_{\Lambda \Sigma} = 
\frac{1}{(2\, e^{\varphi } \, \chi +i )}
\left(
\begin{array}{cc}
 -\dfrac{e^{3 \varphi }}{(e^{\varphi } \, \chi -i )^2} & \dfrac{3 \, e^{2 \varphi } \, \chi }{(e^{\varphi} \, \chi -i )} \\[5mm]
 \dfrac{3 \, e^{2 \varphi } \, \chi }{(e^{\varphi } \, \chi -i)} & 3 \,  ( e^{\varphi } \, \chi^2+e^{-\varphi})
\end{array}
\right) \; ,
\end{equation}
and the (electric) gauge two-form field strengths that appear in (\ref{eq:Lagrangian}) are simply
\begin{equation} \label{electricH2}
H_\2^\Lambda = dA^\Lambda \; , \qquad \Lambda = 0 ,1.
\end{equation}

We have computed the SU(3)--invariant Lagrangian (\ref{eq:Lagrangian}) and the quantities that define it using the $D=4$ $\cN=8$ embedding tensor formalism \cite{deWit:2007mt} (see \cite{Trigiante:2016mnt} for a recent review) with the conventions of \cite{Varela:2015ywx} for the SO(8) gauging \cite{deWit:1982ig}. The superpotential (\ref{eq:superpot}) corresponds to one of the eigenvalues of the $\cN=8$ gravitino mass matrix restricted to the SU(3)--singlet space. See \cite{Bobev:2010ib} for the $\cN=2$ special geometry of the model, in unitary gauge for the scalar coset. Superpotentials have previously appeared, also in unitary gauge, in \cite{Bobev:2009ms,Ahn:2009as}.

%%%%%%%%%%%%%%%
\subsection{Restricted tensor and duality hierarchies} \label{sec:RestTDH}
%%%%%%%%%%%%%%%

Besides the electric gauge fields that enter the conventional supergravity Lagrangian, one may consider a set of other gauge potentials in the so-called tensor hierarchy. The full $\cN=8$ tensor hierarchy includes all vectors, both electric and magnetic, along with higher-rank (two-, three-, and four-form) gauge potentials, in representations of the duality group of the ungauged theory, E$_{7(7)}$ \cite{deWit:2008ta,deWit:2008gc}. The full tensor hierarchy corresponding to the $\cN=2$ subsector at hand is obtained by retaining the singlets under the decomposition of those E$_{7(7)}$ representations under SU(3). Here, we are only interested in a subset of the $\cN=8$ tensor hierarchy. The reason is that not all E$_{7(7)}$--covariant fields in the hierarchy are necessary to describe the full $D=11$ embedding of $\cN=8$ SO(8)--gauged supergravity, as argued in \cite{Varela:2015ywx}. Only the vectors and some two- and three-form potentials in representations of the maximal SL$(8, \mathbb{R})$ subgroup of E$_{7(7)}$ are relevant for this purpose. This subset was dubbed the {\it restricted} tensor hierarchy in \cite{Varela:2015ywx}. Thus, the tensor fields that we want to consider are the singlets under $\textrm{SU}(3) \subset \textrm{SL}(8 , \mathbb{R})$ of the $\cN=8$ restricted tensor hierarchy. The complete list is given in (\ref{fieldContentHierarchy}). See appendix \ref{sec:Construction} for further details. 

The field strengths of the SU(3)--invariant, restricted tensor hierarchy fields can be obtained by particularising the $\cN=8$ expressions given in \cite{Varela:2015ywx}, with the help of the expressions contained in appendix \ref{sec:Construction} for their embedding into their $\cN=8$ counterparts. The electric vector field strengths have already been given in (\ref{electricH2}), while the magnetic field strengths are 
\begin{equation} \label{magnH2}
\tilde{H}_{\2 0} = d\tilde{A}_{0}+g B^{0} \; , \qquad  
\tilde{H}_{\2 1} = d\tilde{A}_{1}-2g B^{2} \; .
\end{equation}
The three-form field strengths read, in turn,
{\setlength\arraycolsep{0pt}
\begin{eqnarray} \label{H3def}
&& H_{\3}^0 = dB^{0} \; , \qquad \quad 
H_{\3}^{ 2} = dB^{2} \; ,  \\
&& H_{\3}^{ab} = DB^{ab} + \tfrac14 \big(  3 A^{0}\wedge d\tilde{A}_{0} + 3 \tilde{A}_{0}\wedge dA^{0} - A^{1}\wedge d\tilde{A}_{1} - \tilde{A}_{1}\wedge dA^{1} \big) \, \delta^{ab} \nonumber \\
&& \qquad \qquad + 3g\, C^{1} \, \delta^{ab}-4g \, C^{ab} + \tfrac12 g \,  C^c{}_c \, \delta^{ab} \; , \nonumber
\end{eqnarray}
}where $DB^{ab} = dB^{ab}+ 2g \epsilon^{c(a} A^{0}\wedge B^{b)}{}_{c}$. Finally, the four-form field strengths are
\begin{equation} \label{H4def}
H_{\4}^{1} = dC^{1}-\tfrac13 \, H_{\2}^{1}\wedge B^{2} \; , \qquad 
H_{\4}^{ab} = DC^{ab}+ \tfrac12 \,  H_{\2}^{0}\wedge \big( \epsilon^{(a}{}_{c} \, B^{b)c}+ B^{0}\, \delta^{ab} \big)  \; ,
\end{equation}
with $DC^{ab} = dC^{ab}+ 2g \epsilon^{c(a} A^{0}\wedge C^{b)}{}_{c}$. 

The field strengths (\ref{electricH2})--(\ref{H4def}) are subject to the Bianchi identities
\begin{eqnarray} \label{eq:Bianchis}
& dH_{\2}^{0}= 0 \; , \qquad 
dH_{\2}^{1}= 0 \; , \qquad 
d\tilde{H}_{\2 0}=g H_{\3 0} \; , \qquad 
d\tilde{H}_{\2 1}=-2g H_{\3 2} \; , \nonumber \\
&  DH_{\3}^{ab} = \Big(\tfrac{3}{2} H_{\2}^{0}\wedge\tilde{H}_{\2 0}-\tfrac{1}{2} H_{\2}^{1}\wedge\tilde{H}_{\2 1} 
 +3g H_{\4}^{1}+\tfrac{1}{2}g H_{\4 c}{}^c \Big)\delta^{ab}-4g H_{\4 }^{ab} \; , \nonumber \\
&  dH_{\3}^{0}= 0 \; , \qquad 
dH_{\3}^{2}=0 \; , \qquad 
dH_{\4}^{1} \equiv 0 \; , \qquad 
dH_{\4}^{ab} \equiv 0 \; ,
\end{eqnarray}
where we have defined  $DH_{\3}^{ab} = dH_{\3}^{ab}-2g \epsilon^{(a}{}_{c} \, A^{0} \wedge H_{\3}^{b)c}$. These expressions particularise the Bianchi identities (14) of \cite{Varela:2015ywx} to the present case.

All of the fields in the restricted tensor hierarchy carry degrees of freedom, although not independent ones. They are instead subject to a duality hierarchy \cite{Bergshoeff:2009ph}. The magnetic two-form field strengths can be written as scalar-dependent combinations of the electric gauge field strengths and their Hodge duals:
{\setlength\arraycolsep{1pt}
\begin{eqnarray} \label{duality:vector/vector}
\tilde{H}_{\2 0}	&=& \dfrac{1}{X^2(4X-3)}\left[-e^{3 \varphi } (3X-2)*H_{\2}^{0} + 3 e^{\varphi}X(X-1) *H_{\2}^{1} \right. \nonumber \\
&& \qquad \qquad\qquad \; \left. - 2e^{6\varphi} \chi^3 \, H_{\2}^{0} +3 \chi\, e^{2\varphi}X(2X-1) \, H_{\2}^{1}\right] \; , \nonumber \\[6pt]
\tilde{H}_{\2 1}	&=&  \dfrac{1}{X(4X-3)}\left[3 e^{\varphi}(X-1)*H_{\2}^{0} - 3e^{-\varphi}X^2*H_{\2}^{1} 	\right.	\nonumber
\\
&& \qquad \qquad\qquad \; \left. +3\chi e^{2\varphi}(2X-1)  \, H_{\2}^{0} +6\chi X^2\, H_{\2}^{1}\right] \; .
\end{eqnarray}
}The three-form field strengths are dual to scalar-dependent combinations of derivatives of scalars:
{\setlength\arraycolsep{1pt}
\begin{eqnarray}	\label{duality:3forms}
H_{\3}^{ 0}	&= & -* \Big[ \big(Y^2-2Y+Z^2 + e^{4 \phi }\big) \big(Da+\tfrac{1}{2} (\zeta  D\tilde{\zeta}-\tilde{\zeta}D\zeta)\big)+ Y \big(\zeta  D\tilde{\zeta}-\tilde{\zeta}D\zeta\big)   \nonumber \\
&& \qquad +2a DY-4a Y D\phi \Big] \; , \nonumber \\[6pt]
H_{\3}^{ 2}  &= & 3 \, e^{2\phi} * \Big[(Y-1) \big( Da+\tfrac{1}{2} (\zeta D\tilde{\zeta}-\tilde{\zeta} D\zeta )\big)+\tfrac{1}{2} \big(\zeta D\tilde{\zeta}-\tilde{\zeta} D\zeta \big)  \Big] \; , \nonumber \\[6pt]
H_{\3}^{ 77}	&=& * \Big[ 2Z e^{2\phi} \big (Da+\tfrac{1}{2} (\zeta  D\tilde{\zeta}-\tilde{\zeta} D\zeta )\big )+ 2DY-4Y D\phi +  3\big(d\varphi - e^{2\varphi}\chi d\chi  \big) \Big]		\; , 	\\[6pt]
H_{\3}^{ 78}	&= & * \Big[ \big(Y^2-2Y+Z^2-e^{4 \phi }\big) \big(Da+\tfrac{1}{2} (\zeta  D\tilde{\zeta}-\tilde{\zeta}D\zeta)\big)+ Y \big(\zeta  D\tilde{\zeta}-\tilde{\zeta}D\zeta\big)  \nonumber \\
&& \qquad +2a DY-4a Y D\phi \Big] \; , \nonumber \\[6pt]
H_{\3}^{ 88}	&=& -* \Big[ 2Z e^{2\phi} \big (Da+\tfrac{1}{2} (\zeta  D\tilde{\zeta}-\tilde{\zeta} D\zeta )\big )+ 2DY-4Y D\phi -  3\big(d\varphi - e^{2\varphi}\chi d\chi  \big) \Big]	 \; . \nonumber  \; 
\end{eqnarray}
}Finally, the four-form field strengths correspond to the following scalar-dependent top forms on four-dimensional spacetime:
{\setlength\arraycolsep{1pt}
\begin{eqnarray}	\label{duality:4forms}
H_{\4}^{1}	&=& g\left[2e^{\varphi} Y \big(3X+2 Y-3XY \big)+e^{-\varphi-2 \phi} X\big(X+Y-XY\big)\big(Y^2+Z^2+e^{4 \phi }\big)\right]\textrm{vol}_4\; ,	\nonumber	\\[6pt]
H_{\4}^{77} &=& -g X \left[ e^{-3 \varphi}  X^2 \big(Y^2-2 Y+Z^2+e^{4\phi}\big)+6 e^{-\varphi+2\phi} \big(X Y-X-Y \big)\right]\textrm{vol}_4		\; ,		\nonumber	\\[6pt]
H_{\4}^{78} &=&  -g X Z \left[ e^{-3 \varphi -2 \phi } X^2  \big(Y^2+Z^2+e^{4 \phi }\big)+6 e^{-\varphi } (X Y-X-Y)\right]\textrm{vol}_4	\; ,				\\[6pt]
H_{\4}^{88} &=&  -g X  \left[e^{-3 \varphi } X^2 \left(Y^2-2 Y+Z^2\right) +6 e^{-\varphi -2 \phi }  (XY-X-Y) \left(Y^2+Z^2\right) \right.						\nonumber	\\
						&& \qquad    \quad \left. + e^{-3 \varphi -4 \phi } X^2 \big(Y^2+Z^2\big)^2 \right]\textrm{vol}_4 \;  . \nonumber
\end{eqnarray}
}The dualisations (\ref{duality:vector/vector})--(\ref{duality:4forms}) particularise (16) of \cite{Varela:2015ywx} to the SU(3)--invariant case. 

It can be checked that the scalar potential (\ref{eq:scalarpot}) can be recovered from the dualised four-forms (\ref{duality:4forms}) via
\begin{equation} \label{eq:Potfrom4forms}
g \, \big( 6H_\4^1+ H_\4^{77} +H_\4^{88} \big) = -2 \, V \, \textrm{vol}_4 \; .
\end{equation}
Likewise, the Bianchi identities (\ref{eq:Bianchis}) combined with the dualisation conditions (\ref{duality:vector/vector})--(\ref{duality:4forms}) partially reproduce the equations of motion that derive from the Lagrangian (\ref{eq:Lagrangian}). The list of identities needed to verify this includes the action of the $\textrm{SL}(2, \mathbb{R})$ Killing vector $k[H_0]$ in (\ref{eq:KillingSL2}) on the gauge kinetic matrix (\ref{NMatrix}),
\begin{equation} \label{idsEomsFromBianchisNmatrix}
 \partial_\varphi \cN_{00} - \chi \, \partial_\chi \cN_{00} = 3 \,  {\cal N}_{00} \ , \quad
\partial_\varphi \cN_{11} - \chi \, \partial_\chi \cN_{11} = -  {\cal N}_{11} \ , \quad
\partial_\varphi \cN_{01} - \chi \, \partial_\chi \cN_{01} =  {\cal N}_{01} \; ,
\end{equation}
and the following identities that can be checked to hold for the dualised three-form field strengths (\ref{duality:3forms}),
\begin{eqnarray} \label{idsEomsFromBianchisThreeForms}
& H_\3^{77}-H_\3^{88}  = -4 \, h_{uv} \, k^u[H_1] \, *Dq^v    \; , 
\qquad
H_\3^{78}   = -\sqrt{2}  \, h_{uv} \, \big( k^u[E_2] + k^u[F_2] \big)  \, *Dq^v   \; , \nonumber  \\[4pt]
& H_\3^0  = - 2 \, h_{uv} \, k_0^u \, *Dq^v  \; ,  \qquad
H_\3^2  =  h_{uv} \, k_1^u \, *Dq^v  \; ,
\end{eqnarray}
and four-form field strengths (\ref{duality:4forms}) and the potential (\ref{eq:scalarpot}), 
\begin{eqnarray} \label{idsEomsFromBianchisFourForms}
& 3 g  \big( 2H_\4^1 -  H_\4^{77}-  H_\4^{88}  \big) = -  k^\alpha [H_{0}] \, \partial_\alpha V \textrm{vol}_4  \; ,  \nonumber \\[4pt]
& 2g \big( H_{\4}^{77}-H_\4^{88} \big)  =- \,k^u [H_{1}] \, \partial_u V\,\text{vol}_{4}   \; , 
\qquad
4 \sqrt{2} \, g H_{\4}^{78} = - \big(  k^u[E_{2}]+ k^u[F_{2}] \big) \, \partial_u V \,\text{vol}_{4}  \; ,
 \nonumber \\[5pt]
& k_0^u  \, \partial_u V  =0  \; , \qquad 
k_1^u \, \partial_u V  =0 \; .
\end{eqnarray}
In (\ref{idsEomsFromBianchisThreeForms}) and (\ref{idsEomsFromBianchisFourForms}), $Dq^u$, $u = 1, \ldots, 4$, collectively denote the hypermultiplet covariant derivatives (\ref{eq:covder}); $k_0$ and $k_1$ are the hypermultiplet Killing vectors (\ref{GaugedIsometries}) along which the gauging is turned on; $k[H_0]$ and $k[H_1]$ are other Killing vectors (see (\ref{eq:KillingSL2}), (\ref{eq:KillingSU21})) on each factor of the scalar manifold (\ref{ScalManN=2}); and $h_{uv}$ is the metric that can be read off from the hypermultiplet kinetic terms in the Lagrangian (\ref{eq:Lagrangian}).

The last two identities in (\ref{idsEomsFromBianchisFourForms}) reflect the invariance of the potential (\ref{eq:scalarpot}) under the gauged hypermultiplet isometries (\ref{GaugedIsometries}). These are the only symmetries of the SU(3)--invariant potential (\ref{eq:scalarpot}). The symmetry is enhanced in the subsectors that we now turn to  discuss.

% Subsectors
\subsection{Some further subsectors} \label{sec:D=4subsectors}

It is interesting to consider further subsectors contained in the SU(3)--invariant sector in the notation that we are using. A natural way to obtain those is to impose invariance under a subgroup $G$ of SO(8) that contains SU(3).  The relevant tensor hierarchy field strengths and their dualisation conditions are obtained by bringing the $G$--invariant restrictions specified on a case-by-case basis below to (\ref{electricH2})--(\ref{H4def}) and (\ref{duality:vector/vector})--(\ref{duality:4forms}). The field content in each of these subsectors is summarised for convenience in table~\ref{table: subsectors}.

\begin{table}[t]
	\centering\scriptsize
	\ra{1.65}
	\setlength{\tabcolsep}{15pt}
	\begin{tabular}{c|ccccc}
		\Xhline{1pt}
		sector				&	scalars				& 	pseudoscalars 	& 	E\&M vectors 		& 	2-forms 		& 3-forms 				\\%[-3pt]
		\Xhline{1pt}
		SU(3)				&		3				&		3			&		4		&		5		&		4			\\%[-3pt]
		\hline
		SU(3)$\times$U(1)${}^2$	&		1				&		1			&		4		&		1		&		2			\\%[-3pt]
		\hline
		SU(3)$\times$U(1)${}_v$	&		3				&		1			&		4		&		4		&		4			\\[-3pt]
		SU(3)$\times$U(1)${}_c$	&		1				&		3			&		4		&		2		&		2			\\[-3pt]
		SU(3)$\times$U(1)${}_s$	&		1				&		1			&		4		&		1		&		2			\\%[-3pt]
		\hline
		SO(6)${}_v$			&		3				&		0			&		2		&		4		&		4			\\[-3pt]
		SU(4)${}_c$			&		0				&		3			&		2		&		1		&		1			\\[-3pt]
		SU(4)${}_s$			&		0				&		0			&		2		&		1		&		1			\\%[-3pt]
		\hline
		SO(7)${}_v$			&		1				&		0			&		0		&		1		&		2			\\[-3pt]
		SO(7)${}_c$			&		0				&		1			&		0		&		0		&		1			\\[-3pt]
		SO(7)${}_s$			&		0				&		0			&		0		&		0		&		1			\\%[-3pt]
		\hline
		G${}_2$				&		1				&		1			&		0		&		1		&		2			\\%[-3pt]
		\Xhline{1pt}
	\end{tabular}
	\caption{\small{Number of bosonic tensor hierarchy fields in each subsector.}\normalsize}
	\label{table: subsectors}
\end{table}

An obvious yet still interesting sector is attained by requiring an additional invariance under the U$(1)^2$ with which SU(3) commutes inside SO(8). The resulting $\textrm{SU}(3) \times \textrm{U}(1)^2$--invariant sector throws out the hypermultiplet and sets identifications on the restricted tensor hierarchy\footnote{Curiously, $B^0$ and $B^2$ are allowed by group theory to be non-vanishing, but are set to $B^0 =B^2 = 0$ by the duality relations (\ref{duality:3forms}) evaluated with the scalar restrictions (\ref{VMTruncation}). Similar comments apply to the condition $B^2=0$ in (\ref{SU3U1vsector}) and $B^0 = -\tfrac23 B^2$ in (\ref{SU3U1csector}).},
\begin{eqnarray} \label{VMTruncation}
\textrm{SU}(3) \times \textrm{U}(1)^2 \; &:&  \phi = a = \zeta = \tilde{\zeta} = 0 \; ,  \\
&& B^0 = B^2 = B^{78} = 0 \; , \quad B^{77} = B^{88} \; , \quad 
C^{78} = 0 \; , \quad C^{77} = C^{88} \; . \nonumber
\end{eqnarray}
This sector thus reduces to $\cN=2$ supergravity coupled to a vector multiplet with a Fayet-Iliopoulos gauging, namely, to the U$(1)^4$--invariant sector ({\it i.e.}, the gauged STU model)  with all three vector multiplets identified, along with the relevant tensor hierarchy fields. Inserting (\ref{VMTruncation}) in (\ref{eq:Lagrangian}), the Lagrangian indeed reduces to {\it e.g.} (6.28), (6.29) of \cite{Azizi:2016noi} with the fields and coupling constants here and there identified as
\begin{eqnarray} \label{TranslationVMTrunc}
e^{\varphi_\textrm{there}} = e^{-\varphi_\textrm{here}} (1+ e^{2\varphi_\textrm{here}} \chi^2_\textrm{here}) \; , \qquad
\chi_\textrm{there} \, e^{\varphi_\textrm{there}} = \chi_\textrm{here} \, e^{\varphi_\textrm{here}} \; , \nonumber \\[3pt]
\tilde{A}_{\1 \textrm{there}} = -A^0_\textrm{here} \; , \qquad
A_{\1 \textrm{there}} = A^1_\textrm{here} \; , \qquad
g_\textrm{there} = - g_\textrm{here} \; .
\end{eqnarray}
The potential of the $\textrm{SU}(3) \times \textrm{U}(1)^2$--invariant sector, (\ref{eq:scalarpot}) with (\ref{VMTruncation}), acquires a symmetry under the compact generator, $k[E_0] - k[F_0]$  in the notation of (\ref{eq:KillingSL2}), of the vector multiplet scalar manifold. The field redefinition in the first line of (\ref{TranslationVMTrunc}) is a $\textrm{U}(1) \subset \textrm{SL}(2, \mathbb{R})$ transformation generated by this Killing vector, followed by a change of sign of $\chi$. 

One may also consider $\textrm{SU}(3) \times \textrm{U}(1)$--invariant sectors, with U(1) chosen to be one of the three triality--inequivalent\footnote{Under triality, the representations $\bm{8}_v$, $\bm{8}_s$, $\bm{8}_c$ of SO(8) split under the subgroups SO$(7)_v$, SO$(7)_s$, SO$(7)_c$ as in {\it e.g.}~(C.1) of \cite{Pang:2017omp}, with labels $(v,+,-)$ there denoted $(v,s,c)$ here. We follow the spectrum conventions of {\it e.g.}~\cite{Klebanov:2008vq} whereby, at the SO(8) vacuum, the (graviton, gravitini, vectors, spinors, scalars, pseudoscalars) of $\cN=8$ supergravity lie in the $(\bm{1} , \bm{8}_s , \bm{28}  , \bm{56}_s , \bm{35}_v , \bm{35}_c )$ of SO(8).}  U$(1)_v$, U$(1)_s$ or U$(1)_c$, factors with which SU(3) commutes inside SO(8). These invariant sectors are attained by setting
\begin{eqnarray} \label{SU3U1vsector}
\textrm{SU}(3) \times \textrm{U}(1)_v & : & \zeta = \tilde{\zeta} = 0 \; ,  \qquad B^2 = 0 \; , \\[5pt]
\label{SU3U1csector}
\textrm{SU}(3) \times \textrm{U}(1)_c & : & e^{-2\phi} = 1-\tfrac14 ( \zeta^2 + \tilde{\zeta}^2 ) \; ,  \quad a =0 \; , \\
&& B^0 = -\tfrac23 B^2 \; , \;  B^{78} = 0 \; , \quad B^{77} = B^{88} \; , \quad 
C^{78} = 0 \; , \quad C^{77} = C^{88} \; , \nonumber
 \\[5pt] 
\label{SU3U1ssector}
\textrm{SU}(3) \times \textrm{U}(1)_s & : & \phi = a = \zeta = \tilde{\zeta} = 0 \; ,  \\
&& B^0 = B^2 = B^{78} = 0 \; , \quad B^{77} = B^{88} \; , \quad 
C^{78} = 0 \; , \quad C^{77} = C^{88} \; , \nonumber
\end{eqnarray}
while retaining both vectors and their magnetic duals. Only the $\textrm{SU}(3) \times \textrm{U}(1)_s$--invariant subtruncation is supersymmetric, and coincides with the $\textrm{SU}(3) \times \textrm{U}(1)^2$ sector discussed above --in other words, invariance under U$(1)_s$ cannot be enforced on top of SU(3) without also imposing U$(1)_c$ invariance, but not the other way around. The other two subtruncations retain the would-be vector multiplet and `half' a hypermultiplet: either the scalars $\phi$, $a$ in the $\textrm{SU}(3) \times \textrm{U}(1)_v$ sector, or the pseudoscalars $\zeta$, $\tilde{\zeta}$ in the $\textrm{SU}(3) \times \textrm{U}(1)_c$ sector, with $\phi$ a function of the pseudoscalars in the latter case. The covariant derivatives (\ref{eq:covder}) simplify accordingly. In the $\textrm{SU}(3) \times \textrm{U}(1)_v$ sector, $\phi$, $a$ remain charged under $A^0$ and no field is charged under $A^1$. In the $\textrm{SU}(3) \times \textrm{U}(1)_c$ sector the covariant derivatives reduce to
\begin{equation} \label{covDersSU3U1s}
D\zeta = d\zeta - g (A^0 + 3A^1) \,  \tilde{\zeta} \; , \qquad
D\tilde{\zeta} = d\tilde{\zeta} + g (A^0 + 3A^1) \, \zeta \; , 
\end{equation}
showing that $\zeta$, $\tilde{\zeta}$ become a doublet charged only under the combined gauge field $A^0 + 3A^1$. 

It is possible to further truncate the $\textrm{SU}(3) \times \textrm{U}(1)_c$ sector to a two-scalar model retaining $(\varphi , \zeta)$ along with $B^{77} = B^{88}$ and $C^1$, $C^{77} = C^{88}$ by imposing (\ref{SU3U1csector}) together with $\chi = 0$, $\tilde{\zeta} = \zeta$, $A^0 = A^1 = 0$ and $B^0 = -\tfrac23 B^2 = 0$. The Lagrangian is (\ref{eq:Lagrangian}) with these identifications and the superpotential reduces, from (\ref{eq:superpot}), to
\begin{equation} \label{superpotCPW}
	W=\tfrac{1}{2\sqrt{2}} \, g  \, e^{-\frac32 \varphi } \big(e^{2 \phi } - 3 e^{2 \phi  + 2 \varphi } - 2\big)\;,
\end{equation}
where $e^{2\phi}$ is shorthand for the expression in terms of $\zeta = \tilde{\zeta}$ that appears in (\ref{SU3U1csector}). This is the model considered in \cite{Corrado:2001nv}. The identifications
\begin{equation} \label{eq:CompCPW}
e^{-\varphi_{\textrm{here}}} = \rho^4_\textrm{there} \; , \qquad 
\zeta^2_\textrm{here} =\tilde{\zeta}^2_\textrm{here} = 2 \tanh^2 \chi_\textrm{there} \; 
\end{equation}
(the second equation implies $ e^{2\phi_{\textrm{here}}} = \cosh^2 \chi_\textrm{there}$ on (\ref{SU3U1csector})) indeed bring the superpotential (\ref{superpotCPW}) to (3.9) of \cite{Corrado:2001nv}, up to normalisation. 

The $\textrm{SU}(3) \times \textrm{U}(1)$--invariant sectors can be further reduced by imposing a larger $\textrm{SO} (6) \sim \textrm{SU}(4)$ symmetry. The corresponding sectors are obtained by letting
\begin{eqnarray} \label{SU4vsector}
\textrm{SO}(6)_v & : & \zeta = \tilde{\zeta} = \chi =0 \; ,  \qquad A^1 = \tilde{A}_1 = 0 \; , \qquad B^2 = 0  \; ,  
 \\[5pt] 
\label{SU4csector}
\textrm{SU}(4)_c & : &  e^{-2\phi} = 1-\tfrac14 ( \zeta^2 + \tilde{\zeta}^2 ) \; ,  \quad a =0 \; ,  \quad 
e^{-2\varphi} = 1- \chi^2 \; ,  \qquad A^1 = A^0 \equiv A \; , \\
&& \tilde{A}_1 = 3 \tilde{A}_0 \; , \;  B^0 = -\tfrac23 B^2 \; , \;  B^{ab} = 0 \; , \;
C^{1} =C^{77} = C^{88}  \; , \; C^{78} = 0 \; , \nonumber 
\\[5pt]
\label{SU4ssector}
\textrm{SU}(4)_s & : & \phi = a = \zeta = \tilde{\zeta} = \varphi = \chi = 0 \; , \\
&& A^1 = -A^0 \; , \;  \tilde{A}_1 = -3 \tilde{A}_0 \; , \;  B^0 = \tfrac23 B^2 \; , \;  B^{ab} = 0 \; , \;
C^{1} =C^{77} = C^{88}  \; , \; C^{78} = 0 \; . \nonumber 
\end{eqnarray}
Again, only the $\textrm{SU}(4)_s$--invariant sector is supersymmetric: it truncates out the vector multiplet of the $\textrm{SU}(3) \times \textrm{U}(1)_s$ sector, leading to minimal $\cN=2$ gauged supergravity. Setting all scalars to zero as in (\ref{SU4ssector}), further setting consistently $B^0 = \tfrac23 B^2 = 0$, and rescaling for convenience the metric and the graviphoton as 
\begin{equation} \label{IdsMinSugraSO8}
g_{\mu\nu} \equiv \tfrac14 \, \bar{g}_{\mu\nu} \; , \qquad A^1 = - A^0 \equiv  \tfrac14 \bar{A} \; , 
\end{equation}
equation (\ref{eq:Lagrangian}) reduces to the bosonic Lagrangian of pure $\cN=2$ gauged supergravity,
\begin{equation} \label{minimalN=2}
{\cal L}  =  \bar{R} \, \overline{\textrm{vol}}_4 -\tfrac12 \, \bar{F} \wedge \bar{*} \bar{F} + 6 g^2 \, \overline{\textrm{vol}}_4 \; ,
\end{equation}
with $\bar{F} \equiv d\bar{A}$. For later reference, we note that the only tensor hierarchy field strengths that are active in the SU$(4)_s$ sector are
\begin{equation} \label{eq:SU4sFS}
H^1_\2 = - H^0_\2 \equiv \tfrac14 \bar{F} \; , \quad \
\tilde{H}_{\2 0} = -\tfrac13 \tilde{H}_{\2 1} = \tfrac14 \bar{*} \bar{F} \; , \quad \
H_{\4}^{1}=H_{\4}^{77}=H_{\4}^{88}= \tfrac{3}{8} g \overline{\text{vol}}_{4}  \; ,
\end{equation}
where the bars refer to the rescaled quantities (\ref{IdsMinSugraSO8}). The other two truncations (\ref{SU4vsector}), (\ref{SU4csector}) are manifestly non-supersymmetric.   Imposing invariance under $\textrm{SO}(6)_v$ selects the proper scalars $\varphi$, $\phi$, $a$ along with the gauge field $A^0$, while invariance under $\textrm{SU}(4)_c$ retains the pseudoscalars $\chi$, $\zeta$, $\tilde{\zeta}$ along with $A^0 + A^1$. In the latter case, the scalars become functions of the pseudoscalars as indicated in (\ref{SU4csector}). 

It was noted in \cite{Bobev:2010ib} that the $\textrm{SU}(4)_c$--invariant sector coincides with a subtruncation, considered in  \cite{Gauntlett:2009bh}, of the $D=4$ $\cN=2$ gauged supergravity obtained upon consistent truncation of M-theory on any (skew-whiffed) Sasaki-Einstein seven-manifold \cite{Gauntlett:2009zw}. Indeed, using (\ref{SU4csector}) and further identifying the pseudoscalars and vectors here and in \cite{Gauntlett:2009bh} as 
\begin{eqnarray}  \label{TranslationSU4sTrunc}
\chi_\textrm{here} =  h_\textrm{there} \; , \qquad
\zeta_\textrm{here} = -\sqrt{3} \, \textrm{Im} \, \chi_\textrm{there} \; ,&& \qquad
\tilde{\zeta}_\textrm{here} = -\sqrt{3} \, \textrm{Re} \, \chi_\textrm{there} \; , \nonumber\\
A^0_\textrm{here} = A^1_\textrm{here} = -A_{1 \, \textrm{there}} \;,	\quad\quad
&&g_{\text{here}}=-\left(2L\right)^{-1}_{\text{there}}\; 
\end{eqnarray}
(which further imply $\varphi_\textrm{here} = -2U_\textrm{there}- V_\textrm{there}$ and $\phi_\textrm{here} = -3U_\textrm{there}$, with $\varphi$, $\phi$ here subject to (\ref{SU4csector}) and $U$, $V$ there subject to their (4.1)), the Lagrangian (\ref{eq:Lagrangian}) here reproduces (4.3) of \cite{Gauntlett:2009bh}. Neither the SO$(6)_v$ nor the SU$(4)_c$ sectors admit a further truncation to the Einstein-Maxwell, bosonic Lagrangian (\ref{minimalN=2}) of minimal $\cN=2$ supergravity. 

It is possible to enlarge the symmetry to the three different SO(7) subgroups of SO(8) by further imposing
\begin{eqnarray} \label{SO7vsector}
\textrm{SO}(7)_v & : & \zeta = \tilde{\zeta} = \chi =0 \; , \quad \varphi = \phi \; , \quad a= 0 \; , \qquad A^0 =A^1 = \tilde{A}_0 =\tilde{A}_1 = 0  \; , \\
&& B^0 = B^2 = B^{78} = 0 \; , \quad B^{88} = -7 B^{77} \; , \qquad C^1 = C^{77} \; , \quad C^{78} = 0 \; , \nonumber 
 \\[5pt] 
\label{SO7csector}
\textrm{SO}(7)_c & : &  e^{-2\phi} = 1-\tfrac14 ( \zeta^2 + \tilde{\zeta}^2 )  = 1- \chi^2 = e^{-2\varphi}  \; ,  \quad a =0 \; , \quad A^0 =A^1 = \tilde{A}_0 =\tilde{A}_1 = 0  \; , \nonumber \\
&& B^0 = B^2 = 0 \; , \quad B^{ab} = 0 \; , \qquad C^1 = C^{77} = C^{88} \; , \quad C^{78} = 0 \; , 
\\[5pt]
\label{SO7ssector}
\textrm{SO}(7)_s & : & \phi = a = \zeta = \tilde{\zeta} = \varphi = \chi = 0 \; , \quad A^0 = -A^1 =0  \; , \\
&& B^0 = B^2 = 0 \; , \quad B^{ab} = 0 \; , \qquad C^1 = C^{77} = C^{88} \; , \quad C^{78} = 0 \; . \nonumber
\end{eqnarray}
The $\textrm{SO}(7)_s$ truncation gives minimal $\cN=1$ gauged supergravity while the $\textrm{SO}(7)_v$ and the $\textrm{SO}(7)_c$ sectors are non-supersymmetric. They respectively retain one dilaton ($\varphi = \phi$) and one axion $(\chi$, together with the identifications (\ref{SO7csector})), along with the relevant tensors in the hierarchy. 

All three SO(7) sectors are contained within the G$_2$--invariant sector. This corresponds to $\cN=1$ supergravity coupled to a chiral multiplet with a scalar manifold $\textrm{SL}(2)/\textrm{SO}(2)$ which is diagonally embedded in (\ref{ScalManN=2}) via
\begin{eqnarray} \label{eq:G2sector}
\textrm{G}_2 \; & : & %
\phi = \varphi \; , \qquad 
\tilde \zeta = -2\chi \; , \qquad 
a= \zeta = 0 \; , \qquad A^0 =A^1 = \tilde{A}_0 =\tilde{A}_1 = 0  \; , \\
&& B^0 = B^2 = B^{78} = 0 \; , \quad B^{88} = -7 B^{77} \; , \qquad C^1 = C^{77} \; , \quad C^{78} = 0 \; .\nonumber 
\end{eqnarray}
The Lagrangian in this sector is (\ref{eq:Lagrangian}) with the identifications (\ref{eq:G2sector}). It can be cast in canonical $\cN=1$ form, in the conventions of {\it e.g.}~section 4.2 of \cite{Guarino:2015qaa}, in terms of the following K\"ahler potential and holomorphic superpotential
\begin{eqnarray} \label{eq:G2N=1}
K  = -7 \log ( -i (t - \bar{t} ) ) \; , \qquad 
{\cal W}  = 2g \,  ( 7 t^3 + t^7) \; ,
\end{eqnarray}
with $t = -\chi + i e^{-\varphi}$. On the identifications (\ref{eq:G2sector}) that define the G$_2$--invariant sector, the real superpotential (\ref{eq:superpot}) becomes related to (\ref{eq:G2N=1}) via $W^2 = e^{K} \, \overline{ {\cal W} } \,  {\cal W} $. 

All of the above further truncations arise from symmetry principles, by retaining the fields that are neutral under the relevant invariance groups. For this reason, the above truncations can be directly implemented at the level of the Lagrangian (\ref{eq:Lagrangian}). In particular, a consistent truncation to minimal $\cN=2$ supergravity is obtained by retaining singlets under $\textrm{SU}(4)_s$, as noted above. We conclude this section by noting an alternate truncation of the SU(3) sector to minimal $\cN=2$ supergravity that is inequivalent to the $\textrm{SU}(4)_s$--invariant truncation. In fact, this alternative minimal truncation is not driven by symmetry principles in any obvious way, so we have verified its consistency at the level of the field equations. Firstly, freeze the scalars to their vacuum expectation values (vevs) at the $\textrm{SU}(3) \times \textrm{U}(1)_c$--invariant vacuum (see section \ref{sec:vacua}),
\begin{equation} \label{eq:N=2vevs}
e^{-2\varphi} = 3 \; , \qquad \chi = 0 \; , \qquad e^{-2\phi} = 1-\tfrac14 ( \zeta^2 + \tilde{\zeta}^2 ) = \tfrac23 \; ,  \qquad a =0 \; .
\end{equation}
Secondly, identify the electric and magnetic vectors as
\begin{equation} \label{eq:N=2CPWVectors}
A^0 = - 3A^1 \equiv \tfrac12 \, \bar{A} \; , \qquad
\tilde{A}_0 = - \tfrac19 \tilde{A}_1 \equiv  \tfrac1{6\sqrt3} \, \tilde{\bar{A}} \; ,
\end{equation}
turn off the two-form potentials, and retain an auxiliary three-form potential as
\begin{equation} \label{eq:N=2CPW2nad3forms}
B^0=-\tfrac{2}{3}B^2=B^{ab}=0 \; , \qquad C^{78}=0 \; , \quad C^1=C^{77}=C^{88}\; .
\end{equation}
Finally, rescale the metric for convenience:
\begin{equation} \label{eq:N=2CPWMetric}
g_{\mu\nu} \equiv \tfrac{1}{3\sqrt{3}} \, \bar{g}_{\mu\nu} \; .
\end{equation}
We have verified at the level of the bosonic field equations, including Einstein, that these identifications define a consistent truncation of the theory (\ref{eq:Lagrangian}) to minimal $\cN=2$ gauged supergravity (\ref{minimalN=2}). 

The identification of the electric vectors in (\ref{eq:N=2CPWVectors}) retains the $\textrm{SU}(3) \times \textrm{U}(1)_c$--invariant vector (see (\ref{eq:VecEmbedN=8}) with (\ref{U1cgenerator})) that remains massless (see (\ref{covDersSU3U1s})) at the $\cN=2$ vacuum (\ref{eq:N=2vevs}). For future reference, it is also interesting to keep track of the field strengths for this truncation. On (\ref{eq:N=2CPWVectors}), (\ref{eq:N=2CPW2nad3forms}), the two-form potential contributions to the magnetic vector two-form field strengths (\ref{magnH2}) drop out, and the vector field strengths become
\begin{equation} \label{eq:N=2CPWVectorsFS}
H^0 = - 3H^1 \equiv \tfrac12 \, \bar{F} \; , \qquad
\tilde{H}_0 = - \tfrac19 \tilde{H}_1 \equiv  \tfrac1{6\sqrt3} \, \tilde{\bar{F}}  = -\tfrac1{6\sqrt3} \, \bar{*} \bar{F} \; ,
\end{equation}
with $\bar{F} \equiv d\bar{A}$. The relations here for the magnetic field strengths are compatible with the vector duality relations (\ref{duality:vector/vector}) evaluated on the scalar vevs (\ref{eq:N=2vevs}), and the last equality for the magnetic graviphoton field strength $\tilde{\bar{F}}$ is fixed by $\tilde{\bar{F}}=\partial\mathcal{L}/\partial \bar{F}$, with $\mathcal{L}$ as in (\ref{minimalN=2}). Moving on to the three-form field strengths, we find that all of them are zero by bringing (\ref{eq:N=2CPWVectors}), (\ref{eq:N=2CPW2nad3forms}) to their definitions (\ref{H3def}) in terms of potentials. This was expected, as the three-form field strengths are dual to combinations (\ref{duality:3forms}) of (Hodge duals of) derivatives of scalars, and these have been frozen to their vevs (\ref{eq:N=2vevs}). Finally, for the four-form field strengths we obtain, from (\ref{H4def}) with (\ref{eq:N=2CPW2nad3forms}), $H_\4^{78} = 0$, $H_\4^{1}=H_\4^{77}=H_\4^{88} = dC^1$, expressions which are again compatible with the dualisation conditions (\ref{duality:4forms}). Rescaling the volume form using (\ref{eq:N=2CPWMetric}), we find 
\begin{equation} \label{eq:N=2CPWFourFormFS}
H_\4^{1}=H_\4^{77}=H_\4^{88} = \tfrac{1}{2\sqrt{3} } \, g \, \overline{\vol}_4 \; .
\end{equation}
%

% Critical points
\subsection{Vacuum structure } \label{sec:vacua}

\begin{table}[t]
\centering\scriptsize
\ra{2}
\begin{tabular}{cc|cccccc|cc}
\Xhline{1pt}
	$\cN$	&	$G_0$		&	$\chi$	&	$e^{-\varphi}$	&	$e^{-\phi}$	&	$a$	&	$\zeta$	&	$\tilde{\zeta}$	&	$g^{-2}V_0$	&	$L^{2}M^2$	\\
\Xhline{1pt}
	8		&	SO(8)			&	0		&		1		&		1		&	0	&	0		&	0			&	$-24$			
																	&	$\left(-2,\,-2,\,-2,\,-2,\,-2,\,-2\right)	$									\\%[4mm]
	2		&	U$(3)_c$				&	0		&	$\sqrt{3}$		&	$\sqrt{\frac23}$		&	0	&	$\sqrt{\frac23}$		&	$\sqrt{\frac23}$		&	$-18\sqrt{3}$	
																	&	$\left(3\pm\sqrt{17},\,2,\,2,\,2,\,0\right)	$								\\%[4mm]
	1		&	$\text{G}_2$		&	$12^{-\nicefrac{1}{4}}$	&$\left(\frac{25}{12}\right)^{\nicefrac{1}{4}}$	&	$\left(\frac{25}{12}\right)^{\nicefrac{1}{4}}$	&	0	&	0	&	$-2\cdot12^{-\nicefrac{1}{4}}$
																	&	$-\frac{2^{\nicefrac{11}{2}}\;3^{\nicefrac{13}{4}}}{5^{\nicefrac{5}{2}}}$			
														&	$\left(4\pm\sqrt{6},\,\frac{-11\pm\sqrt{6}}{6},\,0,\,0\right)	$								\\[1mm]
\hline
	0		&	$\text{SO}(7)_v$	&	0		&	$5^{\nicefrac{1}{4}}$		&	$5^{\nicefrac{1}{4}}$		&	0	&	0		&	0			&	$-8\times5^{\nicefrac{3}{4}}$	
																	&	$\left(6,\, -\frac{12}5\, -\frac65,\, -\frac65,\, -\frac65,\, 0\right)$					\\%[4mm]
	0		&	$\text{SO}(7)_c$	&	$\frac1{\sqrt{5}}$	&	$\frac2{\sqrt{5}}$	&	$\frac2{\sqrt{5}}$	&	0	&	0	&	$-\frac2{\sqrt{5}}$	&	$-\frac{25\sqrt{5}}2$	
																&	$\left(6,\, -\frac{12}5\, -\frac65,\, -\frac65,\, -\frac65,\, 0\right)$						\\%[4mm]
	0		&	$\text{SU}(4)_c$	&	$0$		&	$1$	&	$\frac1{\sqrt{2}}$	&	0	&	1	&	$1$	&	$-32$	
																&	$\left(6,\, 6\, -\frac34,\, -\frac34,\, 0,\, 0\right)$									\\%[4mm]
\Xhline{1pt}
\end{tabular}
\caption{\small{All critical points of $D=4$ $\cN=8$ supergravity with an electric SO(8) gauging with at least SU(3) invariance, reproducing the results of \cite{Warner:1983vz} in our parametrisation. For each point we give the residual supersymmetry $\cN$ and bosonic symmetry $G_0$ within the full $\cN=8$ theory, their location in the parametrisation that we are using, the cosmological constant $V_0$ and the scalar mass spectrum within the SU(3)--invariant sector. The masses are given in units of the AdS radius, $L^2 =-6/V_0$. We have abbreviated $\textrm{U}(3)_c \equiv \textrm{SU}(3) \times \textrm{U}(1)_c$.}}
\label{table: critical loci}
\end{table}

The list of vacua of $D=4$ $\cN=8$ supergravity with an electric SO(8) gauging \cite{deWit:1982ig} that preserve at least a subgroup SU(3) of SO(8) was elucidated in \cite{Warner:1983vz}. All of them are AdS. These vacua arise as extrema of the scalar potential (\ref{eq:scalarpot}), in our conventions, and for convenience we have summarised them in table \ref{table: critical loci}. The table includes the residual supersymmetry $\cN$ and bosonic symmetry $G_0$ for each vacuum, as well as its location in the scalar space (\ref{ScalManN=2}) in the parametrisation that we are using. The corresponding cosmological constant, given by (\ref{eq:scalarpot}), and the scalar mass spectrum within the SU(3)--invariant sector is also given. See \cite{Bobev:2010ib} for the bosonic spectra within the full $\cN=8$ supergravity. All three supersymmetric points are also extrema of the superpotential (\ref{eq:superpot}). On the SO(8)  and the G$_2$ points, the F-terms that derive from the holomorphic superpotential (\ref{eq:G2N=1}) also vanish. 

It was argued in \cite{Varela:2015ywx} that some combinations of the four-form field strengths of the duality hierarchy ought to vanish at critical points of the scalar potential, thus yielding necessary conditions for critical points. In our SU(3)--invariant case, these conditions read
\begin{equation} \label{NecessaryCondsCritPoints}
8 H_{\4}^1- \big( 6H_{\4}^{1} + \delta_{cd} H_{\4}^{cd} \big) = 0 \; , \qquad
8 H_{\4}^{ab}- \big( 6H_{\4}^{1} + \delta_{cd} H_{\4}^{cd} \big) \, \delta^{ab} = 0 \; .
\end{equation}
Using the dualisation conditions (\ref{duality:4forms}), it can be checked that the relations (\ref{NecessaryCondsCritPoints}) do indeed hold at the critical points summarised in table \ref{table: critical loci}.

%%%%%%%%%%%%%%%
%%%%%%%%%%%%%%%

\section{$D=11$ uplift } \label{sec:SU3ofSO8in11D}

\addtocontents{toc}{\setcounter{tocdepth}{2}}

We now switch gears and present the $D=11$ embedding of the SU(3)--invariant sector considered in the previous section. We will use the consistent $S^7$ uplifting formulae given in \cite{Varela:2015ywx}. It is a tedious, but otherwise mechanical, exercise to particularise the general $\cN=8$ uplifting formulae in that reference to the SU(3)--invariant sector at hand. Section \ref{sec:SU3sectorinD=11} contains the $D=11$ uplift of the entire SU(3)--invariant sector while section \ref{sec:SU3subsectorsinD=11} particularises to some relevant subsectors and makes contact with previous literature. Section \ref{sec:MinN=2fromD=11} contains a new consistent truncation of $D=11$ supergravity to minimal $D=4$ $\cN=2$ gauged supergravity.

\subsection{Uplift of the SU(3) sector} \label{sec:SU3sectorinD=11}

We first find it useful to present the result in terms of $\mathbb{R}^8$ ``embedding coordinates" $\mu^A,\;  A= 1, \ldots, 8$, in the $\bm{8}_v$ of SO(8), that define the $S^7$ as the locus
\begin{equation} \label{eq:S7inR8}
\delta_{AB} \,  \mu^A \mu^B = 1 
\end{equation}
in $\mathbb{R}^8$. Under SU(3), the $\bm{8}_v$ of SO(8) breaks down as $\bm{8}_v \rightarrow \bm{3} + \bm{\bar{3}} + \bm{1} +\bm{1}$. In maintaining an explicitly real notation, it is thus convenient to split $\mathbb{R}^8 = \mathbb{R}^6 \times \mathbb{R}^2$, and the indices as $A = (i,a)$, with $i= 1, \ldots, 6$ and $a=7,8$ respectively labelling the first and second factors. The $D=11$ uplift of the SU(3)--invariant sector utilises the tensors  $\delta_{ij}$, $J_{ij}^\6$ (real) and $\Omega_{ijk}^\6$ (complex) that define the natural Calabi-Yau structure of $\mathbb{R}^6$. See (\ref{eq:SU3invforms}) for our conventions. Inside $\mathbb{R}^8$, these tensors are respectively invariant under SO$(6)_v  \times \textrm{SO}(2)$, $\textrm{SU}(3) \times \textrm{U}(1)^2$ and $\textrm{SU}(3) \times \textrm{U}(1)_c$, where SO(2) rotates the $\mathbb{R}^2$ factor in $\mathbb{R}^8 = \mathbb{R}^6 \times \mathbb{R}^2$. Indices on $\mathbb{R}^6$ and $\mathbb{R}^2$ are raised and lowered with $\delta_{ij}$ and $\delta_{ab}$, respectively.

Only the $D=4$ metric, the scalars, and the electric gauge fields in the SU(3)--invariant restricted duality hierarchy (\ref{fieldContentHierarchy}) enter the $D=11$ metric $d\hat{s}^2_{11}$. In order to express the result, it is useful to introduce a symmetric matrix $h_{ab}$ of $D=4$ scalars and its inverse as\footnote{This matrix $h_{ab}$ should not be confused with the metric $h_{uv}$ on the hypermultiplet scalar manifold.} 
\begin{equation}
		h = 	\begin{pmatrix}	e^{2\phi} 	&	Z 	\\
									Z 		&	e^{-2 \phi } \left(Y^2+Z^2\right) \\
						\end{pmatrix} \; ,
		\qquad
		h^{-1}=
		Y^{-2} \begin{pmatrix}		e^{-2 \phi } \left(Y^2+Z^2\right)		&	-Z 	\\
						-Z	& 	e^{2\phi} 			\\
		\end{pmatrix} \; , 
\end{equation}
and the following combination of $D=4$ scalars and constrained coordinates $\mu^i$, $\mu^a$,
\begin{equation} \label{Deltatilde}
\Delta_1  = e^{2\varphi} \, Y \mu_i \mu^i + X \, h_{ab} \, \mu^a \mu^b \; .
\end{equation}
With these definitions, the embedding into the $D=11$ metric reads
{\setlength\arraycolsep{1pt}
\begin{eqnarray} \label{11DmetricEmbcoords}
d\hat{s}^2_{11}  &=& e^{-\varphi} \, X^{1/3} \, \Delta_1^{2/3} \Big[ ds^2_{4} + g^{-2} \,  e^\varphi \, \Delta_1^{-1} \Big( D\mu_i D\mu^i  + e^{2\varphi} X^{-1} Y (h^{-1})_{ab} D\mu^a D\mu^b \Big)  \\[4pt]
&&  \qquad \qquad \qquad  \quad \;   +g^{-2} e^{3\varphi} X^{-1} Y^{-1} (Y-X) \, \Delta_1^{-2}  \Big( \, Y J^\6_{ij} \mu^i D\mu^j +h_{ab}  \epsilon^{bc} \mu^a D\mu_c \Big)^2 \, \Big] \; , \nonumber
\end{eqnarray}
}where $\epsilon^{ab}$ is the totally antisymmetric symbol with two indices, and the covariant derivatives are defined as
\begin{equation}	\label{CovariantDerivatives}
D\mu^i = d\mu^i - g \, A^1 J^{\6  ij} \mu_j \; , \qquad 
D\mu^a = d\mu^a - g \, A^0 \epsilon^{ab} \mu_b \; .
\end{equation}
For generic values of the $D=4$ scalars, the metric (\ref{11DmetricEmbcoords}) enjoys an $\textrm{SU}(3) \times \textrm{U}(1)_v$ isometry. 

Moving on to the $D=11$ three-form $\hat{A}_\3$, all the $D=4$ fields in the tensor hierarchy (\ref{fieldContentHierarchy}), except for the metric, enter its expression. A long calculation yields 
{\setlength\arraycolsep{1pt}
\begin{eqnarray} \label{11D3formEmbcoords}
\hat{A}_{\3} & = &   C^{1} \mu_i \mu^i + C_{ab} \, \mu^{a}\mu^{b} 
-\tfrac{1}{12}  g^{-1} \big[ \big(  B_a{}^a +2  \, A^{1}\wedge\tilde{A}_{1} \big) \, \delta_{ij} +4  B^{2}\, J^\6_{ij} \big] \wedge\mu^{i}D\mu^{j}\nonumber \\[4pt]
 & & + \tfrac12 g^{-1}\big[  B_{ab}-  A^{0}\wedge\tilde{A}_{0} \, \delta_{ab} + B^{0} \, \epsilon_{ab} \big] \wedge\mu^{a}D\mu^{b} \nonumber \\[4pt]
 && + \tfrac{1}{6} \, g^{-2} \, \tilde{A}_{1} \wedge  J^\6_{ij}  D\mu^{i}\wedge D\mu^{j} +\tfrac{1}{2} \, g^{-2}  \tilde{A}_{0} \wedge \epsilon_{ab}  D\mu^{a}\wedge D\mu^{b} + A \; ,
\end{eqnarray}
}where $A$ is a three-form on the internal $S^7$ that depends on the $D=4$ scalars:
{\setlength\arraycolsep{1pt}
\begin{eqnarray} \label{11D3formEmbcoordsInternal}
A &=&  -  g^{-3} \, \Delta_1^{-1} 
		\Big[ \tfrac12 \, e^{4\varphi} \, \chi  \, X^{-1} Y \, J^\6_{ij}\mu^i D\mu^j \wedge \epsilon_{ab} \, D\mu^a \wedge D\mu^b  
\nonumber \\[4pt]
&& \qquad \qquad \;\; +\tfrac12 \, \chi \, e^{2\varphi}  \big( \, Y J^\6_{ij} \mu^i D\mu^j +h_{ab}  \epsilon^{bc} \mu^a D\mu_c \big)	\wedge J^\6_{kl} D\mu^k\wedge D\mu^l 	 \nonumber	\\[4pt]
&& \qquad \qquad \;\; - \tfrac14 \,  e^{2\varphi} \, \big( V_1 \,  \textrm{Re} \, \Omega^\6_{ijk}
		+V_2 \,   \textrm{Im} \, \Omega^\6_{ijk} \big) \wedge \mu^i D\mu^j\wedge D\mu^k   \nonumber	\\[4pt]
&& \qquad \qquad \;\;
+ \tfrac{1}{12} \,  e^{2\phi} \, X \, \big( v_1  \,  \textrm{Re} \, \Omega^\6_{ijk}
		 + v_2 \,  \textrm{Im} \, \Omega^\6_{ijk} \big) D\mu^i\wedge D\mu^j\wedge D\mu^k \Big] \; .
\end{eqnarray}
}Here, we have defined the shorthand functions
\begin{equation} \label{defSHf}
		v_1 = \mu_7\zeta+\mu_8e^{-2\phi}(\zeta Z+\tilde{\zeta}Y) \; , \qquad 
		v_2	=\mu_7\tilde{\zeta}-\mu_8e^{-2\phi}(\zeta Y-\tilde{\zeta}Z) \; ,
\end{equation}
and one-forms
\begin{equation} \label{defSHV}
		V_1 = (\zeta Y-\tilde{\zeta}Z)D\mu^7+e^{2\phi}\tilde{\zeta}\,D\mu^8 \; , \qquad
		%\
		V_2=(\zeta Z+\tilde{\zeta}Y)D\mu^7-e^{2\phi}\zeta\,D\mu^8 \; .
\end{equation}
The field strength four-form $\hat{F}_{\4}=d\hat{A}_{\3}$ is computed to be 
{\setlength\arraycolsep{1pt}
\begin{eqnarray} \label{11D4formEmbcoords}
\hat{F}_{\4} & = & H_{\4}^{1} \,  \mu_i \mu^{i} +  H_{\4}^{ab} \, \mu_{a} \mu_{b}
-\tfrac{1}{12} g^{-1} \big[  H_{\3 a}{}^a \, \delta_{ij}+4 H_{\3}^2 \, J^\6_{ij} \big] \wedge\mu^{i}D\mu^{j} 
\nonumber \\[4pt]
 &&  + \tfrac{1}{2}  g^{-1} \big[  H_{\3}^{ab}+  H_{\3}^{0} \, \epsilon^{ab} \big] \wedge\mu_{a}D\mu_{b} 
  + \tfrac{1}{6}g^{-2}  \tilde{H}_{\2 1} \wedge J^\6_{ij} D\mu^{i}\wedge D\mu^{j} \nonumber \\[4pt]
 &&+\tfrac{1}{2}g^{-2} \tilde{H}_{\2 0} \wedge \epsilon_{ab}  D\mu^{a} \wedge D\mu^{b}  \nonumber \\[4pt]
&&  +\tfrac14 \, g^{-2} \, e^{2 \varphi} \Delta_1^{-1} 
		\Big[ 4\chi e^{2 \varphi} \, X^{-1}   Y \,   J^\6_{ij}\mu^i D\mu^j   \wedge \mu_k D\mu^k 
\nonumber \\[4pt]
&& \qquad \qquad \qquad \quad   +e^{2\phi}\left(v_2\,  \textrm{Re} \,  \Omega^\6_{ijk} -v_1\,  \textrm{Im} \,  \Omega^\6_{ijk} \right) \mu^i D\mu^j\wedge D\mu^k \Big]	\wedge H_\2^0 	\\[4pt]
& & -  \tfrac{1}{4} g^{-2} \, \Delta_1^{-1}   \Big[ 2 \chi e^{2\varphi} \, X^{-1} Y \mu_k\mu^k \big( X J^\6_{ij} D\mu^i\wedge D\mu^j  +e^{2\varphi}  \epsilon_{ab} \, D\mu^a \wedge D\mu^b \big) \nonumber \\[4pt]
&& \qquad \qquad \quad \;  -4 \chi \, e^{2\varphi} \,  \mu_k D\mu^k \wedge \big( Y J^\6_{ij}\mu^i D\mu^j + h^{ac}\,\epsilon_{cb}\,\mu_aD\mu^b \big) \nonumber 
					 \\[4pt]
		& &\qquad\qquad \quad \;  + e^{2\phi} X  \left(v_2\,  \textrm{Re} \,  \Omega^\6_{ijk} -v_1\,  \textrm{Im} \,  \Omega^\6_{ijk} \right) \mu^i D\mu^j\wedge D\mu^k \Big]\wedge H_\2^1 + d A_\textrm{scalars}  \; . \nonumber 
\end{eqnarray}
}In this expression, $H_{\4}^{1}$, $H_{\4}^{ab}$, etc., turn out to reproduce the $D=4$ four-, three- and magnetic two-form field strengths (\ref{magnH2})--(\ref{H3def}) of the restricted tensor hierarchy (\ref{fieldContentHierarchy}). This provides a $D=11$ crosscheck of the $D=4$ calculation of section \ref{sec:RestTDH}. The terms that contain the electric two-form field strengths $H^0_\2$, $H^1_\2$, come from the vector contributions in the covariant derivatives $D\mu^i$ and $D\mu^a$ in (\ref{11D3formEmbcoordsInternal}). Finally, $d A_\textrm{scalars}$ contains two types of terms. The first type includes contributions of covariant derivatives of $D=4$ scalars, wedged with three-forms on the internal $S^7$. The second type includes internal four-forms with coefficients that depend on the $D=4$ scalars algebraically only. The presence in $\hat{A}_{\3}$ of $J^\6_{ij}$, $\Omega^\6_{ijk}$ and $h_{ab}$ breaks the symmetry of the full $D=11$ configuration to SU(3), in agreement with the symmetry of the $D=4$ model. 

The above expressions give the complete embedding of the SU(3)--invariant, restricted tensor hierarchy (\ref{fieldContentHierarchy}) into $D=11$ supergravity. As such, these expressions contain redundant $D=4$ degrees of freedom. As argued in \cite{Varela:2015ywx}, these redundancies can be eliminated at the level of the $D=11$ four-form field strength by making use of the $D=4$ duality relations. Indeed, regarding the tensor field strengths in (\ref{11D4formEmbcoords}) as shorthand for the dualisation conditions (\ref{duality:vector/vector})--(\ref{duality:4forms}), equations (\ref{11DmetricEmbcoords}), (\ref{11D4formEmbcoords}) then express the embedding into $D=11$ supergravity exclusively in terms of the dynamically independent (metric, electric-vector and scalar) degrees of freedom that enter the $D=4$ Lagrangian (\ref{eq:Lagrangian}).

In particular, the Freund-Rubin term (the first two contributions on the r.h.s.~of (\ref{11D4formEmbcoords})), can be simplified by using the identities (\ref{eq:Potfrom4forms}), (\ref{idsEomsFromBianchisFourForms}) that relate the dualised four-form field strengths (\ref{duality:4forms}) to the scalar potential (\ref{eq:Lagrangian}) and its derivatives:
{\setlength\arraycolsep{0pt}
\begin{eqnarray} \label{FRterm}
H_{\4}^{1} \,  \mu_i \mu^{i} +  H_{\4}^{ab} \, \mu_{a} \mu_{b} &=&-\tfrac{1}{4g}  \Big[ V +\tfrac{1}{6} (\mu_i \mu^i -3 \mu_a \mu^a) \,  k^\alpha [H_0] \, \partial_\alpha V  + \big((\mu^{7})^{2}-(\mu^{8})^{2}\big) \, k^u[H_{1}] \, \partial_u V  \nonumber   \\[4pt]
&& \qquad + \sqrt{2} \, \mu^7\mu^8 \big( k^u [E_2] + k^u [F_2] \big) \, \partial_u V \Big] \textrm{vol}_4  \; .
\end{eqnarray}
}At a critical point, the terms in derivatives of the potential drop out and the Freund-Rubin term becomes proportional to the AdS$_4$ cosmological constant, in agreement with the general $\cN=8$ discussion of \cite{Varela:2015ywx}. See also \cite{Godazgar:2015qia} for a related discussion. All the Freund-Rubin terms that we write for the truncations to specific subsectors in section \ref{sec:SU3subsectorsinD=11} and for the concrete AdS$_4$ solutions in section \ref{sec:11Dsolutions} agree with the generic expression (\ref{FRterm}).

%%%%%%%%%%%%%%%
\subsection{Uplift of some further subsectors} \label{sec:SU3subsectorsinD=11} 
%%%%%%%%%%%%%%%

The uplifting formulae of section \ref{sec:SU3sectorinD=11} simplify by imposing a symmetry  enlargement, carried over to $D=11$ by restricting the $D=4$ fields as in section \ref{sec:D=4subsectors}. Introducing intrinsic $S^7$ angles by solving the constraint (\ref{eq:S7inR8}) is also facilitated in further subsectors, as some intrinsic angles are better suited than others to make the relevant symmetry apparent in $D=11$. See appendix \ref{sec: geometryS7} for some relevant geometric structures on $S^7$.

\subsubsection{$\textrm{SU}(3) \times \textrm{U}(1)^2$--invariant sector} \label{sec:SU3XU1SqinD=11} 

For the $\textrm{SU}(3) \times \textrm{U}(1)^2$--invariant sector (\ref{VMTruncation}), the embedding formulae for the $D=11$ metric, (\ref{11DmetricEmbcoords}), and three-form, (\ref{11D3formEmbcoords}), (\ref{11D3formEmbcoordsInternal}), become
{\setlength\arraycolsep{1pt}
\begin{eqnarray}	\label{eq: metricSTU}
	d\hat{s}^2_{11}	&=& e^{-\varphi}X^{1/3}\Delta_1^{2/3}ds^2_4+g^{-2} \Big[ X^{-2/3}\Delta_1^{2/3}d\alpha^2 +X^{1/3}\Delta_1^{-1/3}\cos^2\alpha\, ds^2(\mathbb{CP}^2) 
		\nonumber \\
					&& \qquad \qquad \qquad \qquad \qquad \;  +e^{2\varphi}X^{-2/3}\Delta_1^{2/3}\Delta_2^{-1}\sin^2\alpha\cos^2\alpha\left(D\tau_-+\sigma\right)^2  \\
					&& \qquad \qquad \qquad \qquad \qquad \; +X^{-2/3} \Delta_2  \, \Delta_1^{-4/3} \Big( D\psi_-+  \Delta_3 \Delta_2^{-1} \, \cos^2 \alpha \, \big (D\tau_-+\sigma\big ) \Big)^2	 \Big] \,, \nonumber	\\[16pt]
\label{eq:3formSTU}
	\hat{A}_\3 &=& C_1\cos^2\alpha+C_{77}\sin^2\alpha
				+\tfrac{1}{12} \, g^{-1} \, \sin 2\alpha  \,  \big( 4B_{77}+A^1\wedge\tilde{A}_1-3A^0\wedge\tilde{A}_0 \big) \wedge d\alpha \nonumber	\\[4pt]
				&& -\tfrac{1}{6} \, g^{-2}  \sin2\alpha \, (\tilde{A}_1 + 3\tilde{A}_0)\wedge d\alpha\wedge D\psi_- \nonumber \\[4pt]
				&& +\tfrac{1}{3} \, g^{-2}  \cos\alpha \, \tilde{A}_1\wedge\left[ \cos \alpha \, \bm{J}^\4- \sin \alpha\,d\alpha\wedge(D\tau_-+\sigma)\right] \nonumber	\\[4pt]
				&& +\tfrac12 \, g^{-3}\chi e^{2\varphi} X^{-1}\sin 2\alpha \, d\alpha\wedge D\psi_-\wedge \left(D\tau_-+\sigma\right) \nonumber \\[4pt]
				&& -g^{-3}\chi e^{2\varphi} \Delta_1^{-1} \cos^4\alpha\, (D\tau_-+\sigma) \wedge\bm{J}^\4 \nonumber \\[4pt]
				&& - g^{-3}\chi e^{2\varphi} \Delta_1^{-1}   \cos^2\alpha  \cos2\alpha\, D\psi_- \wedge\bm{J}^\4  \, .  
\end{eqnarray}
}In these expressions, $\alpha$, $\tau_-$, $\psi_-$ are angles on $S^7$ whose relation to the constrained coordinates $\mu^A$ of $\mathbb{R}^8$ is given in appendix~\ref{sec: geometryS7}. The covariant derivatives for the last two are
\begin{equation} \label{coverDerAngles1}
D\psi_-=d\psi_- - g A^0	\; , \qquad
D\tau_-=d\tau_-+g(A^0 +  A^1) \; .
\end{equation}
The line element $ds^2(\mathbb{CP}^2)$ and the two-form $\bm{J}^\4$ respectively correspond to the Fubini-Study metric, normalised so that its Ricci tensor is six times the metric, and the K\"ahler form, with potential one-form $\sigma$ such that $d \sigma = 2 \bm{J}^\4$, on the the complex projective plane. Finally, $\Delta_1$, $\Delta_2$ and $\Delta_3$ are the following functions of the $S^7$ angle $\alpha$ and the $\textrm{SU}(3) \times \textrm{U}(1)^2$--invariant, $D=4$ vector multiplet scalars
\begin{equation}
	\begin{aligned}
	\Delta_1 & =X\sin^2\alpha+e^{2\varphi}\cos^2\alpha \; ,  \\[2pt]
		\Delta_2&=e^{2\varphi}\left[\sin^4\alpha+\left(e^{2\varphi}+2\chi^2e^{2\varphi}+e^{-2\varphi}X^2\right)\sin^2\alpha\cos^2\alpha+\cos^4\alpha\right]		\; ,  \\[2pt]
		\Delta_3&= \left[X^2+\chi^2e^{4\varphi}\right]\sin^2\alpha + e^{2\varphi}\cos^2\alpha \; .
	\end{aligned}
\end{equation}
The function $\Delta_1$ here is simply the particularisation of (\ref{Deltatilde}) to the present case. 

The four-form field strength corresponding to (\ref{eq:3formSTU}) can be computed to be 
{\setlength\arraycolsep{1pt}
\begin{eqnarray}	\label{eq: 4formSTU}
		\hat{F}_{\4}&=& 2g \Big[2\left(e^{\varphi}\cos^2\alpha+e^{-\varphi}X\sin^2\alpha\right)\vphantom{2^{2^2}}+Xe^{-\varphi} \Big]\vol_4+g^{-1} \,  \sin2\alpha \, (*d\varphi - e^{2\varphi}\chi *d\chi  \big)\wedge d\alpha \nonumber 	\\[4pt]
				&& -\tfrac16 g^{-2}\Big[\sin 2\alpha \, (\tilde{H}_1 + 3\tilde{H}_0) \wedge  d\alpha\wedge D\psi_- \nonumber \\%[2pt] 
				&& \qquad \quad \; \;  - 2 \tilde{H}_1 \wedge \left(\cos^2 \alpha\, \bm{J}^\4-\sin\alpha\cos\alpha\, d\alpha\wedge\left(D\tau_-+\sigma\right)\right)\Big] \nonumber \\[4pt]
				&&+\tfrac12  \, g^{-2}\chi e^{2\varphi} \Big[ X^{-1}\sin 2\alpha\,d\alpha\wedge\left(H^0\wedge(D\tau_-+\sigma)+(H^0+H^1)\wedge D\psi_- \right) \nonumber \\[4pt]
				&& -2 \Delta_1^{-1}\cos^4\alpha \, (H^0+H^1)\wedge\bm{J}^\4+2\Delta_1^{-1}\cos^2\alpha\cos2\alpha \, H^0\wedge\bm{J}^\4 \Big] \nonumber \\[4pt]
&& +g^{-3}\Big\{\frac1{2}e^{2\varphi}\,X^{-2}\sin 2\alpha\Big[2\chi\,d\varphi-(X-2)d\chi\Big] \wedge d\alpha\wedge D\psi_-\wedge \left(D\tau_-+\sigma\right) \nonumber\\
					& & - e^{2\varphi}\,\Delta_1^{-2} \cos^4\!\alpha\Big[2\chi\,\sin^2\!\alpha\,d\varphi+\big(e^{2\varphi}\cos^2\!\alpha-(X-2)\sin^2\!\alpha\big)d\chi\Big] \wedge (D\tau_-+\sigma) \wedge\bm{J}^\4 \nonumber \\[4pt]
					& & - e^{2\varphi}\,\Delta_1^{-2} \cos^2\alpha  \cos2\alpha \Big[2\chi\,\sin^2\!\alpha\,d\varphi+\big(e^{2\varphi}\cos^2\!\alpha-(X-2)\sin^2\!\alpha\big)d\chi\Big] \wedge D\psi_- \wedge\bm{J}^\4 \nonumber\\[4pt]
					&& + \chi e^{2\varphi} X^{-1}\sin 2\alpha \, d\alpha\wedge D\psi_-\wedge \bm{J}^\4 -2 \chi e^{2\varphi} \Delta_1^{-1} \cos^4\alpha\, \bm{J}^\4 \wedge\bm{J}^\4 \nonumber \\[4pt]
				&& + 2 e^{2 \varphi } \chi(\Delta_1 +X)\,\Delta_1^{-2} \sin\!\alpha\cos^3\!\alpha\,d\alpha\wedge(D\tau_-+\sigma) \wedge\bm{J}^\4\nonumber\\[4pt]
				&&  +\tfrac{1}{2}e^{2 \varphi } \chi \, \Delta_1 ^{-2}
				\sin\!2\alpha\Big[4 e^{2 \varphi } \cos ^4 \!\alpha +X \left((\sin 2 \alpha)^2 +2 \cos 2 \alpha \right)\Big]\, d\alpha\wedge D\psi_- \wedge\bm{J}^\4 \Big\}\; .
\end{eqnarray}
}Here, we have explicitly made use of the dualisation conditions (\ref{duality:3forms}), (\ref{duality:4forms}) for the three- and four-form field strengths, particularised to $\textrm{SU}(3) \times \textrm{U}(1)^2$--scalars via (\ref{VMTruncation}). The magnetic two-form field strengths $\tilde{H}_\Lambda$, $\Lambda = 0 ,1$, stand for the dualised expressions (\ref{duality:vector/vector}). 

As noted in section \ref{sec:D=4subsectors}, the $\textrm{SU}(3) \times \textrm{U}(1)^2$--invariant sector coincides with the gauged STU model with all three vector multiplets identified. This was embedded in $D=11$ supergravity in \cite{Azizi:2016noi} (see also \cite{Cvetic:1999xp}), along with the entire STU model. Our uplifting formulae \eqref{eq: metricSTU}, \eqref{eq: 4formSTU}, obtained instead from the $D=11$ embedding of the SU(3) sector, are in perfect agreement with (6.22)-(6.24) of \cite{Azizi:2016noi}. This can be seen by using the $D=4$ redefinitions \eqref{TranslationVMTrunc}, which also imply $\tilde{H}_{0 \, \textrm{here}}=\tilde{R}_\textrm{there}$ and $\tilde{H}_{1 \, \textrm{here}}=-R_\textrm{there}$, along with the $S^7$ angle and one-form identifications 
\begin{equation}
	\xi_{\text{there}}=\alpha_{\text{here}}+\tfrac\pi2\, \quad 
	{\phi_1}_{\text{there}}={\psi_-}_{\text{here}}\,,	\quad
	\psi_{\text{there}}={\psi_-}_{\text{here}}+{\tau_-}_{\text{here}}\,,	\quad
	B_{\text{there}}=\sigma_{\text{here}}\,,	
\end{equation}
or, in terms of the $\psi$, $\tau$ defined in equation (\ref{IntrinsicCoordsS71}) of appendix \ref{sec: geometryS7}, ${\phi_1}_{\text{there}}=-\psi$, $\psi_{\text{there}}=\tau$.

\subsubsection{$\textrm{SU}(4)$--invariant sectors} \label{SU4uplift}

While the deformations inflicted on the internal $S^7$ by the SU(3)--invariant $D=4$ fields are inhomogeneous, enlarging the symmetry to $\textrm{SU}(4)_c$ and $\textrm{SU}(4)_s$ results in the deformations becoming homogeneous.

For the $\textrm{SU}(4)_c$--invariant $D=4$ fields {(\ref{SU4csector}), the $D=11$ embedding formulae (\ref{11DmetricEmbcoords}), (\ref{11D3formEmbcoords}), (\ref{11D3formEmbcoordsInternal}) simplify to
{\setlength\arraycolsep{1pt}
\begin{eqnarray} \label{11DEmbcoordsSU4c}
d\hat{s}^2_{11} &=& e^{\frac43\phi+\varphi}\, ds^2_4+g^{-2}\left[e^{-\frac23\phi}ds^2(\mathbb{CP}_+^3)+e^{\frac43\phi-2\varphi}(\bm{\eta}_+^\7+gA)^2\right]\,, \\[4pt]
\label{11DEmbcoordsSU4cA3}
\hat{A}_\3 & =&C^1 +\tfrac{1}{2} \, g^{-1}  \, B^0\wedge(\bm{\eta}_+^\7+gA)+g^{-2}\tilde{A}_0\wedge\bm{J}_+^\7	 \nonumber \\
					& &-g^{-3}\Big[ \chi\,\bm{J}_+^\7\wedge(\bm{\eta}_+^\7+gA)- \tfrac{1}2\,\zeta\,\text{Re}\,\bm{\Omega}_+^\7- \tfrac{1}2\,\tilde{\zeta}\,\text{Im}\,\bm{\Omega}_+^\7\Big]\, ,
\end{eqnarray}
}where $\phi,\ \varphi$ stand for the expressions in terms of $\chi,\ \zeta,\ \tilde{\zeta}$ given in (\ref{SU4csector}). Here, $ds^2(\mathbb{CP}_+^3)$ is the Fubini-Study metric on $\mathbb{CP}^3$ normalised so that the Ricci tensor is eight times the metric, and $\bm{\eta}^\7_+$, $\bm{J}^\7_+$, $\bm{\Omega}^\7_+$ are the homogeneous Sasaki-Einstein forms on $S^7$ defined in appendix \ref{sec: geometryS7}. The four-form field strength corresponding to (\ref{11DEmbcoordsSU4cA3}) reads 
{\setlength\arraycolsep{2pt}
\begin{eqnarray} \label{11DEmbcoordsSU4cF4}
		\hat{F}_\4&=& -6g\,e^{4\phi+3\varphi}\Big[-1+ \chi^2+\tfrac13 \big(\zeta^2+\tilde{\zeta}^2\big)\Big]\vol_4 +\tfrac{1}{2} \, g^{-1} \, e^{4\phi}*\left(\tilde{\zeta}D\zeta-\zeta D\tilde{\zeta}\right)\wedge(\bm{\eta}_+^\7+gA) 		\nonumber \\[4pt]
		&& 
		+\frac{g^{-2}(1-\chi^2)}{1+3\chi^2}\left[2\chi \, F-\sqrt{1-\chi^2}\, *F\right]\wedge\bm{J}_+^\7 	\nonumber \\[4pt]
		&&
-g^{-3}\left[ d\chi\wedge\bm{J}_+^\7\wedge(\bm{\eta}_+^\7+gA)  - \tfrac12D\zeta\wedge\text{Re} \, \bm{\Omega}_+^\7 - \tfrac12D\tilde{\zeta}\wedge\text{Im}\,\bm{\Omega}_+^\7 \right]	\nonumber \\[4pt]
		&& -  2 g^{-3}   \chi\bm{J}_+^\7\wedge\bm{J}_+^\7 -  2 g^{-3}  \big( \tilde{\zeta}\, \text{Re}\,\bm{\Omega}_+^\7 -  \zeta\, \text{Im}\,\bm{\Omega}_+^\7 \big) \wedge(\bm{\eta}_+^\7+gA)   \, , 
\end{eqnarray}
}with, again, $\phi,\ \varphi$ written in terms of $\chi,\ \zeta,\ \tilde{\zeta}$ as in (\ref{SU4csector}). As noted in section \ref{sec:D=4subsectors} following \cite{Bobev:2010ib}, the $\textrm{SU}(4)_c$--invariant sector of SO(8) supergravity coincides with the model considered in  \cite{Gauntlett:2009bh}. Using the redefinitions \eqref{TranslationSU4sTrunc} and straightforwardly identifying our Sasaki-Einstein structure with theirs, our uplifting formulae (\ref{11DEmbcoordsSU4c}), (\ref{11DEmbcoordsSU4cF4}) do indeed match (2.2), (2.3) of \cite{Gauntlett:2009bh} when the identifications of their equation (4.1) are taken into account.

The $\textrm{SU}(4)_s$--sector coincides with minimal $\cN=2$ gauged supergravity, (\ref{minimalN=2}). The $D=11$ uplift of this sector can be achieved by bringing the restrictions (\ref{SU4ssector}) to the general formulae of section \ref{sec:SU3sectorinD=11} or, equivalently, by further setting $\varphi = \chi = 0$, $A^1 = - A^0 \equiv \frac14 \bar{A}$, and $\tilde{A}_1 = - 3 \tilde{A}_0$ in the uplifting formulae of section \ref{sec:SU3XU1SqinD=11}. Using the rescaled fields (\ref{IdsMinSugraSO8}) and the $D=4$ field strengths (\ref{eq:SU4sFS}), and combining the resulting expressions in terms of the Sasaki-Einstein forms $\bm{J}^\7_-$, $\bm{\eta}^\7_-$ specified in appendix \ref{sec: geometryS7}, the $D=11$ uplift of the $\textrm{SU}(4)_s$--sector can be written as
{\setlength\arraycolsep{2pt}
\begin{eqnarray}
\metriceleven & = & \tfrac{1}{4} \, d \bar{s}_{4}^{2} +g^{-2} \,  \big(ds^{2} (\cpthree_- ) +( \bm{\eta}_-^\7 + \tfrac14 g \bar{A} )^{2} \big)\,, \nonumber \\[5pt]
\hat{F}_{(4)} & = & \tfrac{3}{8} \,  g \, \overline{\text{vol}}_{4}\,-\tfrac{1}{4}\, g^{-2} \, \bar{\ast}\bar{F} \wedge\bm{J}^\7_- \; .
\end{eqnarray}
}This coincides with the consistent truncation of $D=11$ supergravity down to minimal $\cN=2$ gauged supergravity obtained in \cite{Gauntlett:2007ma}, with straightforward identifications. An alternate $D=11$ embedding of minimal $\cN=2$ supergravity will be given in section \ref{sec:MinN=2fromD=11}.

\subsubsection{G$_2$-invariant sector} \label{sec:G2sectorin11D}

The $D=11$ embedding formulae of section \ref{sec:SU3sectorinD=11} particularised to the G$_2$--invariant sector (\ref{eq:G2sector}) become, in the relevant set of intrinsic coordinates described in appendix \ref{sec: geometryS7},
{\setlength\arraycolsep{2pt} 
\begin{eqnarray} \label{eq:D=11G2}
	\metriceleven & = & e^{-\varphi}X^{1/3}\Delta_1^{2/3}ds_{4}^{2}+g^{-2}X^{1/3}\Delta_1^{-1/3}\Big( e^{2\varphi}X^{-3}\Delta_1d\beta^{2} + \sin^{2}\!\beta\,ds^{2}(S^6) \Big) \; , \nonumber \\[5pt]
\hat{A}_\3 & = &C_1\sin^2\beta+C_{88}\cos^2\beta+4g^{-1}\sin\beta\cos\beta\,B_{77}\wedge d\beta \\[5pt]
	&+& g^{-3} \, \chi \, \Delta_1^{-1} \sin^{2}\!\beta \Big[e^{2\varphi} X^{-1}\Delta_1\cj\wedge d\beta+X^{2}\sin\beta\cos\beta\,\text{Re}\,\Upomega+e^{2\varphi}X\,\sin^{2}\!\beta\,\text{Im}\,\Upomega \Big]\,, \nonumber
\end{eqnarray}
}where $\beta$ is an angle on $S^7$, $ds^{2}(S^6)$ is the round metric on $S^6$ normalised so that the Ricci tensor equals five times the metric, $\cj$ and $\Upomega$ are the homogeneous nearly-K\"ahler forms on $S^6$ and the function $\Delta_1$ is, from (\ref{Deltatilde}) with (\ref{IntrinsicCoordsS72}),
\begin{eqnarray} \label{eq:DeltaG2}
\Delta_1 = X \big(e^{-2\varphi}X^{2}\cos^{2}\!\beta+e^{2\varphi}\sin^{2}\!\beta \big) \; .
\end{eqnarray}
The associated four-form field strength reads
{\setlength\arraycolsep{1pt}
\begin{eqnarray} \label{eq:D=11G2F4}
\hat{F}_\4 & = & -g \, e^{-3\varphi}X^{2} \Big[ \big[ (X-2)X^{2}+e^{4\varphi}(7X-12) \big]\sin^{2}\!\beta \nonumber \\
&& \qquad  \qquad \quad \; +e^{-4\varphi}X^{2}\big[X^{3}+7e^{4\varphi}(X-2)\big]\cos^{2}\!\beta\Big] \text{vol}_{4}\nonumber \\
 && -4g^{-1} \, \sin\!\beta\,\cos\beta \,   \big( * d\varphi - e^{2\varphi}\chi *  d\chi \big) \wedge d\beta\nonumber \\[4pt]
&&+  g^{-3} e^{2\varphi}  X^{-2}  \sin^2\beta \, \big(2\chi d\varphi-(X-2)d\chi\big)\wedge\cj\wedge d\beta	\nonumber\\[4pt]
	&&+ 2g^{-3} \chi X \Delta_1^{-2} \sin^3 \beta \cos\beta \, \big( \Delta_1 - 2 e^{2\varphi} X \sin^2 \beta \big) \, d\varphi \wedge \text{Re}\,\Upomega  \nonumber\\[4pt]
	&& + 4g^{-3} \chi X^3 \Delta_1^{-2} \sin^4 \beta \cos^2\beta d\varphi \wedge \text{Im}\,\Upomega  \nonumber\\[4pt]
					&& + g^{-3}  X^2 \, \Delta_1^{-2}  \sin^3\beta \cos\beta  \left[e^{2 \varphi } (3 X-2) \sin ^2\beta-e^{-2\varphi} X^2 (X-2) \cos ^2\beta\right]d\chi \wedge\,\text{Re}\,\Upomega	\nonumber\\[4pt]
					&& +  g^{-3} X^2 \, \Delta_1^{-2} \sin ^4\beta \big[ e^{4 \varphi } \sin ^2\beta-X (3 X-4) \cos ^2\beta \big]d\chi \wedge\,\text{Im}\,\Upomega \nonumber \\[4pt]
 & &+  g^{-3} e^{-2\varphi} \chi X  \Delta_1^{-2} \sin^4\beta \big[ e^{4\varphi}\left(3e^{4\varphi }+X^2\right)\sin^2\beta +X^2\left(5e^{4\varphi}-X^2\right)\cos^2\beta \big]\,\text{Re}\,\Upomega\wedge d\beta\nonumber \\[4pt]
 & & -2g^{-3} \chi X^2 \Delta_1^{-2}  \sin^3\beta \cos \beta \big[ ( e^{4\varphi} +X^2 ) \sin^2\beta +2X^2\cos^2\beta \big]\,\text{Im}\,\Upomega\wedge d\beta  \nonumber \\[4pt]
 &  & -2g^{-3}e^{2\varphi}  \chi X \Delta_1^{-1}  \sin^4\beta \,\mathcal{J}\wedge\mathcal{J} \; .
\end{eqnarray}
}In order to obtain this expression, we have again made explicit use of the dualisation conditions (\ref{duality:3forms}), (\ref{duality:4forms}) for the three- and four-form field strengths, particularised to the G$_2$--invariant sector (\ref{eq:G2sector}). The $D=11$ uplift of the various SO(7)--invariant sectors can be straightforwardly obtained by bringing (\ref{SO7vsector})--(\ref{SO7ssector}) to (\ref{eq:D=11G2})--(\ref{eq:D=11G2F4}). See \cite{Godazgar:2015qia} for a previous $D=11$ uplift of the G$_2$--invariant sector.

%%%%%%%%%%%%%%%
\subsection{Minimal $\cN=2$ gauged supergravity from $D=11$} \label{sec:MinN=2fromD=11}
%%%%%%%%%%%%%%%

It was noted in section~\ref{sec:D=4subsectors} that the SU$(4)_s$ sector coincides with minimal $\cN=2$ gauged supergravity. In section \ref{SU4uplift}, the corresponding $D=11$ uplift was obtained  and shown to coincide with the consistent embedding of \cite{Gauntlett:2007ma}. It was also discussed at the end of section~\ref{sec:D=4subsectors} that the SU(3)--sector admits an alternative truncation to minimal $\cN=2$ supergravity, by fixing the scalars to their vevs (\ref{eq:N=2vevs}) at the $\cN=2$, $\textrm{SU}(3) \times \textrm{U}(1)_c$--invariant point and selecting the $\cN=2$ graviphoton as in (\ref{eq:N=2CPWVectors}). Bringing these $D=4$ identifications to the general SU(3)--invariant consistent uplifting formulae of section \ref{sec:SU3sectorinD=11}, we obtain a new embedding of pure $\cN=2$ gauged supergravity into $D=11$. 

We find it convenient to present the result in local intrinsic $S^7$ coordinates $\psi^\prime$, $\tau^\prime$, $\alpha$, and in terms of a local five-dimensional Sasaki-Einstein structure $\bm{\eta}'$, $\bm{J}'$ and $\bm{\Omega}'$. The former are locally related to the global coordinates $\psi$, $\tau$, $\alpha$, defined in (\ref{IntrinsicCoordsS71}), that are adapted to the topological description of $S^7$ as the join of $S^5$ and $S^1$, with $\alpha$ here identified with that in (\ref{IntrinsicCoordsS71}) and
\begin{equation} \label{eq:PrimedCoords}
	\psi= \psi' \; ,	\qquad 
	\tau=\tau'-\tfrac13 \, \psi' \; .
\end{equation}
The local five-dimensional Sasaki-Einstein structure forms $\bm{\eta}'$, $\bm{J}'$ and $\bm{\Omega}'$ are related to their globally defined counterparts $\bm{\eta}^\5$, $\bm{J}^\5$ and $\bm{\Omega}^\5$ discussed in appendix \ref{sec: geometryS7} and the global coordinate $\psi$ via
\begin{equation} \label{PrimedSE5}
\bm{\eta}' \equiv d\tau'+\sigma \equiv \bm{\eta}^\5 +\tfrac13 d\psi \; , \qquad
\bm{J}' \equiv \bm{J}^\5 \; , \qquad
\bm{\Omega}' \equiv e^{i (\psi+\frac\pi4 )}\,\bm{\Omega}^\5 \; .
\end{equation}
The real two-form $\bm{J}'$ coincides with the K\"ahler form on $\mathbb{CP}^2$, $\sigma$ is a one-form on the latter such that $d\sigma  = 2 \bm{J}'$ (given {\it e.g.} by (\ref{sigmaSE5})) and the constant phase $e^{i \frac\pi4 }$ in the complex two-form $\bm{\Omega}'$ has been chosen for convenience, in order to simplify the resulting expressions. The primed forms defined in (\ref{PrimedSE5}) satisfy the Sasaki-Einstein conditions (\ref{eq: SE5algebraic}) and (\ref{eq: SE5differential}). 

Bringing all these definitions, along with the $D=4$ restrictions (\ref{eq:N=2vevs})--(\ref{eq:N=2CPWMetric}), to the uplifting formulae (\ref{11DmetricEmbcoords}), (\ref{11D3formEmbcoords}), (\ref{11D3formEmbcoordsInternal}), we find a new consistent embedding of minimal $D=4$ $\cN=2$ gauged supergravity (\ref{minimalN=2}) into the $D=11$ metric and three-form: 
{\setlength\arraycolsep{1pt}
\begin{eqnarray}	\label{eq:metricminsugraCPW}
		d\hat{s}^2_{11}	&=& \frac13 \cdot 2^{-2/3} \, (1+2\sin^2\alpha )^{2/3} \left[d\bar{s}_4^2+g^{-2}\Big[ \,  2\,d\alpha^2 + \frac{6\cos^2\alpha}{1+2\sin^2\alpha}ds^2(\mathbb{CP}^2) \right.  \\
					&& \left.
					+\frac{18\sin^2\alpha\cos^2\alpha}{1+8\sin^4\alpha}
					\left.\bm{\eta}'\right.^2+\frac{1+8\sin^4\alpha}{\left(1+2\sin^2\alpha\right)^2} \Big( D\psi'-\frac{3\cos^2\alpha}{1+8\sin^4\alpha}\bm{\eta}'\Big)^2\Big] \right] \; , \nonumber \\[10pt]
\label{eq:A3minsugraCPW}
\hat{A}_{\3} & =&  C^1-\frac{1}{2\sqrt3} \, g^{-2} \, \cos \alpha \,  \tilde{\bar{A}} \wedge \Big[\cos \alpha\ \bm{J}'-\sin\alpha \,d\alpha\wedge \bm{\eta}' \Big] \nonumber \\
				& &\quad+\frac{1}{\sqrt3} \, g^{-3} \, \cos^2\alpha \, \Big[d\alpha\wedge\text{Im}\, \bm{\Omega}'
				+\frac{\sin\alpha\cos\alpha}{1+2\sin^2\alpha}\Big(2D\psi'-3\bm{\eta}'\Big)\wedge\text{Re}\, \bm{\Omega}'\Big] \; .
\end{eqnarray}
}These expressions depend explicitly on the dynamical $D=4$ metric $d\bar{s}_4^2$ and graviphoton $\bar{A}$. The former only features in $d\hat{s}^2_{11}$ but not in $\hat{A}_{\3}$. The latter appears both in $d\hat{s}^2_{11}$ and in $\hat{A}_{\3}$, but only through the gauge covariant derivative
\begin{equation}
	D\psi'=d\psi'+\tfrac12   g \bar{A} \; .
\end{equation}
This singles out $\psi^\prime$ as the angle on the local $\cN=2$ ``Reeb" direction and thus justifies the primed coordinates (\ref{eq:PrimedCoords}) that we chose to present the result. Two other $D=4$ fields enter the consistent embedding through the three-form (\ref{eq:A3minsugraCPW}):  the magnetic dual, $\tilde{\bar{A}}$, of the $D=4$ graviphoton, and the auxiliary three-form potential $C^1$. 

The four-form field strength corresponding to $\hat{A}_{\3}$ in (\ref{eq:metricminsugraCPW}) can be computed with the help of (the primed version of) the Sasaki-Einstein conditions (\ref{eq: SE5algebraic}), (\ref{eq: SE5differential}). We find
{\setlength\arraycolsep{1pt}
\begin{eqnarray}	\label{eq:4formminsugraCPW}
		\hat{F}_\4&=& \frac{g}{2\sqrt{3} }\overline{\vol}_4+\frac{g^{-3}}{\sqrt3}\left[	-\frac{ \cos^2\alpha \, (7-10 \cos2\alpha+\cos4 \alpha)}{ \left(1+2 \sin^2\alpha\right)^2}d\alpha\wedge D\psi'\wedge\text{Re}\, \bm{\Omega}' \right.  \nonumber \\
				&&  \left. -\frac{6\cos^4\alpha}{(1+2\sin^2\alpha)^2}d\alpha\wedge\bm{\eta}'\wedge\text{Re}\, \bm{\Omega}'
				+\frac{6 \sin\alpha\cos^3\alpha}{1+2\sin^2\alpha}D\psi'\wedge\bm{\eta}'\wedge\text{Im}\, \bm{\Omega}' \right]
				  \\
				&&  +\frac{g^{-2}}{2\sqrt3}\left[ \,  \frac{2\sin\alpha\cos^3\alpha}{1+2\sin^2\alpha}\, \bar{F}\wedge \text{Re}\, \bm{\Omega}' + \cos\alpha \, \bar{*}\,\bar{F}\wedge\Big(\cos \alpha\ \bm{J}'-\sin\alpha\,d\alpha\wedge\bm{\eta}'\Big)
				\right] \; . \nonumber
\end{eqnarray}
}Again, we have made use of appropriate dualisation conditions, (\ref{eq:N=2CPWVectorsFS}), (\ref{eq:N=2CPWFourFormFS}) in this case, to express the result for the embedding (\ref{eq:4formminsugraCPW}) into the four-form only in terms of the independent $D=4$ degrees of freedom (the metric $d\bar{s}_4^2 $, the graviphoton field strength $\bar{F} = d\bar{A}$ and its Hodge dual), that appear in the Lagrangian (\ref{minimalN=2}).

The truncation (\ref{eq:metricminsugraCPW}), (\ref{eq:4formminsugraCPW}) of $D=11$ supergravity down to pure $D=4$ $\cN=2$ gauged supergravity (\ref{minimalN=2}) is consistent by construction. As a check on our formalism, we have explicitly verified consistency at the level of the Bianchi identities and equations of motion for the $D=11$ four-form: its field equations are indeed satisfied, provided the $D=4$ Bianchi, $d\bar{F} = 0$, and equation of motion, $d \bar{*} \bar{F} = 0$, of the $D=4$ graviphoton are imposed. Some details can be found in appendix \ref{app:eomMinSugra}. Moreover, these local uplifting formulae are still valid if, more generally, $\bm{\eta}'$, $\bm{J}'$, $\bm{\Omega}'$ are taken to be the defining forms of {\it any} Sasaki-Einstein five-manifold, and $ds^2(\mathbb{CP}^2)$ is replaced with the metric on the corresponding local K\"ahler-Einstein base.

%%%%%%%%%%%%%%%
%%%%%%%%%%%%%%%

\section{Recovering $D=11$ AdS$_4$ solutions} \label{sec:11Dsolutions} 

%%%%%%%%%%%%%%%
%%%%%%%%%%%%%%%

Setting the scalars to the vevs at each critical point with at least SU(3) invariance that were recorded in table \ref{table: critical loci}, and turning off the relevant tensor hierarchy fields, the consistent embedding formulae of section \ref{sec:SU3ofSO8in11D} produce AdS$_4$ solutions of $D=11$ supergravity. All these $D=11$ solutions are known, so our presentation must necessarily be brief. Our main motivation to work out these solutions is rather to test the consistency of the uplifting formulae of \cite{Varela:2015ywx} (and their particularisation to an explicit, SU(3)--invariant, subsector). Except for the more involved $D=11$ Einstein equation, we have indeed verified that the metrics and four-forms that we write below do indeed solve the eleven-dimensional field equations. Please refer to appendix \ref{EomsAdS11D} for details.

We present the solutions in the appropriate intrinsic $S^7$ angles defined in appendix \ref{sec: geometryS7}. These have already been employed in section \ref{sec:SU3subsectorsinD=11} to write the consistent $D=11$ embedding of various further subsectors. Also, AdS$_4$ is always taken to be unit radius (so that the Ricci tensor equals $-3$ times the metric). As a consequence, the metric $ds^2 (\textrm{AdS}_4 )$ that appears in the expressions below is related to the metric $ds_4^2$ that appears in the $D=4$ Lagrangian (\ref{eq:Lagrangian}) and $D=11$ embedding (\ref{11DmetricEmbcoords}) by a rescaling
\begin{equation}
ds_4^2 = -6 \, V_0^{-1} \, ds^2 (\textrm{AdS}_4 ) \; , 
\end{equation}
where $V_0$ is the cosmological constant at each critical point given in table \ref{table: critical loci}. The Freund-Rubin term is rescaled accordingly with respect to (\ref{FRterm}).

Let us first discuss the supersymmetric solutions. The $\cN=8$, SO(8) point uplifts to the Freund-Rubin solution \cite{Freund:1980xh} for which the internal four-form vanishes and the internal metric is the round, Einstein metric $ds^2 (S^7)$, given in {\it e.g.} (\ref{S7asJoinS5S1}) or (\ref{metricHopfS7}). The $\cN=2$, $\textrm{SU} (3) \times \textrm{U}(1)_c$ critical point uplifts to the $D=11$ CPW solution \cite{Corrado:2001nv}. A local form of this solution can be obtained from the expressions in section \ref{sec:MinN=2fromD=11} by turning off the $D=4$ graviphoton, $\bar{A} = 0$, $\bar{F} = 0$, and fixing the metric to $d\bar{s}^2_4 = g^{-2} ds^2 (\textrm{AdS}_4)$. As a check, we have verified that the solution in $\mathbb{R}^8$ embedding coordinates $\mu^A$, directly obtained from the formulae in section \ref{sec:SU3sectorinD=11}, perfectly agrees with the CPW solution as given in \cite{Klebanov:2009kp}. Finally, the $\cN=1$ G$_2$--invariant solution can be written, using the results and the notation of section \ref{sec:G2sectorin11D}, in terms of the homogeneous nearly-K\"ahler structure of the $S^6$ inside $S^7$ as
{\setlength\arraycolsep{2pt}
\begin{eqnarray} \label{eq:G2AdS4Sol}
	\metriceleven & = & g^{-2}\left(\frac{25}{12}\right)^{1/6}(2+\cos2\beta)^{2/3}\left[\frac5{24}\, ds^2(\textrm{AdS}_4) +\frac13d\beta^{2} +\frac{\sin^2\beta}{2+\cos2\beta}\,ds^{2}(S^6)\right] \; , 
	\nonumber \\[4pt]
\hat{F}_\4 & = & \frac1{8}\left(\frac{25}{12}\right)^{5/4}\, g^{-3} \, \vol(\textrm{AdS}_4)+\frac{\sqrt2 g^{-3}\sin^2\beta}{3^{1/4}(2+\cos2\beta)^2}\Big[\sqrt3\sin^2\beta \,\text{Re}\,\Upomega\wedge d\beta	\nonumber \\
 & &-\sin\beta\cos\beta\,(5+\cos2\beta)\,\text{Im}\,\Upomega\wedge d\beta-\sin^2\beta(2+\cos2\beta)\,\mathcal{J}\wedge\mathcal{J}\Big] \; , 
\end{eqnarray}
}with internal three-form potential
\begin{equation}
	A = \frac{\sin^2\beta}{3^{3/4}\sqrt{2}\, g^3 (2+\cos2\beta)}\Big[3\sin\beta\cos\beta\,\text{Re}\,\Upomega+ \sqrt{3} \sin^2\beta\,\text{Im}\,\Upomega+(2+\cos2 \beta)\mathcal{J}\wedge d\beta\Big] \; .
\end{equation}
This solution was first obtained by de Wit, Nicolai and Warner \cite{deWit:1984nz}.

Turning to the non-supersymmetric solutions, the SO$(7)$ critical points can again be uplifted using the results and conventions of section \ref{sec:G2sectorin11D}. The SO$(7)_v$ solution uplifts to a solution first written by de Wit and Nicolai \cite{deWit:1984va}. In our conventions, we get 
{\setlength\arraycolsep{2pt}
\begin{eqnarray}
	\metriceleven & = & 5^{-5/6} \, g^{-2} \,  (3+2\cos2\beta)^{2/3} \left[\frac34ds^2(\textrm{AdS}_4) +d\beta^{2} +\frac{5\sin^2\beta}{3+2\cos2\beta} \, ds^{2}(S^6)\right] \; , \nonumber \\
	\hat{F}_\4 &=& \tfrac{9}{8} \cdot 5^{-3/4}\, g^{-3} \, \vol(\textrm{AdS}_4) \; , 
\end{eqnarray}
}while the SO$(7)_c$ point uplifts to Englert's solution \cite{Englert:1982vs}
{\setlength\arraycolsep{2pt}
\begin{eqnarray} \label{eq:SO7cD=11Sol}
	\metriceleven & = & g^{-2}\left(\tfrac{4}{5}\right)^{1/3}\left[\tfrac3{10}ds^2(\textrm{AdS}_4)+ds^{2}(S^7)\right] \; ,  \nonumber \\
	\hat{F}_\4 & = & \frac{18}{25 \sqrt{5}\, g^3}\vol(\textrm{AdS}_4)+\frac{4\sin^4\beta}{\sqrt5\,g^3}\Big[\,\text{Re}\,\Upomega\wedge d\beta-\cot\beta\,\text{Im}\,\Upomega\wedge d\beta-\frac12\,\mathcal{J}\wedge\mathcal{J}\Big] \; , 
\end{eqnarray}
}with internal three-form
\begin{eqnarray} \label{eq:SO7cD=11SolA3}
	A=\frac{\sin^2\beta}{2 \sqrt{5}\, g^3}\Big[2\sin^2\beta\, \text{Im}\,\Upomega +2\mathcal{J}\wedge d\beta+\sin2 \beta\,\text{Re}\,\Upomega \Big] \; .
\end{eqnarray}
In the SO$(7)_c$ solution, $ds^{2}(S^7)$ is, as always, the round, SO(8)--invariant metric. It should be understood in this context as the sine-cone form (\ref{eq:S7sineCone}). Since $\textrm{SO}(7)_c \supset \textrm{SU}(4)_c$, this solution can also be re-obtained from the $\textrm{SU}(4)_c$--invariant truncation of section \ref{SU4uplift} and written in terms of the homogeneous Sasaki-Einstein structure on $S^7$. The $D=11$ metric is the same appearing in (\ref{eq:SO7cD=11Sol}) with $ds^{2}(S^7)$ now understood as the Hopf fibration (\ref{metricHopfS7}), and the four-form is given by
\begin{eqnarray} \label{SO7c4F11D}
	\hat{F}_\4 & = & \frac{18}{25 \sqrt{5}\, g^3}\vol(\textrm{AdS}_4)+\frac{2}{\sqrt{5} g^3}\Big[2\,\text{Re}\,\bm{\Omega}_+^\7\wedge \bm{\eta}_+^\7-\bm{J}_+^\7\wedge\bm{J}_+^\7\Big] \; ,
\end{eqnarray}
with internal three-form
\begin{eqnarray}
	A = -\frac{1}{\sqrt5\,g^3}\Big[\bm{J}_+^\7\wedge\bm{\eta}_+^\7+\text{Im}\,\bm{\Omega}_+^\7\Big] \; .
\end{eqnarray}
The metric in (\ref{eq:SO7cD=11Sol}) and four-form (\ref{SO7c4F11D}) for the SO$(7)_c$ solution coincide with (3.11) of \cite{Gauntlett:2009bh} upon using the redefinitions (\ref{TranslationSU4sTrunc}), and making an appropriate choice for the phase of the complex scalar $\chi_\textrm{there} \equiv -\frac{1}{\sqrt{3}} ( \tilde{\zeta}_\textrm{here} + i \zeta_\textrm{here} )$, which is unfixed at the critical point. We obtain perfect agreement with \cite{Gauntlett:2009bh} upon shifting that phase by $\pi$.

Finally, the SU$(4)_c$--invariant point gives rise to the Pope-Warner solution \cite{Pope:1984bd} in eleven dimensions. Using the results of section \ref{SU4uplift}, this solution can also be written in terms of the homogeneous Sasaki-Einstein structure on $S^7$ as
{\setlength\arraycolsep{2pt}
\begin{eqnarray} \label{SU4cSolinD=11}
	\metriceleven & = & \frac{1}{2^{1/3}\,g^2}\left[\frac3{8}ds^2(\textrm{AdS}_4)+ds^2(\mathbb{CP}_+^3)+2\bm{\eta}_+^\7\otimes\bm{\eta}_+^\7\right] \; , \nonumber \\
	\hat{F}_\4 & = & \frac{9}{32 g^3}\vol(\textrm{AdS}_4)-\frac{2}{g^3}\Big[\,\text{Re}\,\bm{\Omega}_+^\7\wedge \bm{\eta}_+^\7-\,\text{Im}\,\bm{\Omega}_+^\7\wedge \bm{\eta}_+^\7\Big] \; ,
\end{eqnarray}
}where the internal three-form potential is now
\begin{eqnarray}
	A = \tfrac12 \, g^{-3} \, \big[ \text{Re}\,\bm{\Omega}_+^\7+\text{Im}\,\bm{\Omega}_+^\7 \big] \; .
\end{eqnarray}
We again find agreement with \cite{Gauntlett:2009bh}: (\ref{SU4cSolinD=11}) coincides with (3.8) of that reference when the identifications \eqref{TranslationSU4sTrunc} are taken into account and the phase of  $\chi_\textrm{there} \equiv -\frac{1}{\sqrt{3}} ( \tilde{\zeta}_\textrm{here} + i \zeta_\textrm{here} )$, which is again unfixed at the critical point, is shifted by $\frac\pi4$.

%%%%%%%%%%
%%%%%%%%%%

\section{Discussion} \label{sec:Discussion}

%%%%%%%%%%
%%%%%%%%%%

The main goal of this paper was to test the formulae of \cite{Varela:2015ywx} for the consistent truncation \cite{deWit:1986iy} of $D=11$ supergravity \cite{Cremmer:1978km} on $S^7$ down to $D=4$ $\cN=8$ SO(8)--gauged supergravity \cite{deWit:1982ig}.~We have done so by particularising these formulae to the SU(3)--invariant sector of the $D=4$ supergravity, using an explicit parametrisation. When further restricted appropriately, our results correctly reproduce previously known consistent embeddings of sectors that preserve symmetries larger than SU(3). Our formalism thus extends previous literature and provides a unified $D=11$ embedding of the full SU(3)--invariant sector of SO(8) supergravity including all dynamical (bosonic) fields. It does so systematically, by using the restricted tensor hierarchy approach of \cite{Varela:2015ywx}.

As another crosscheck on the formulae of \cite{Varela:2015ywx}, we have re-derived the known AdS$_4$ solutions of $D=11$ supergravity that arise upon consistent uplift of the critical points of SO(8) supergravity with at least SU(3) symmetry \cite{Warner:1983vz}. Again, we have found perfect agreement with the existing literature. As a further test, we have checked that the $D=11$ field equations are indeed verified on these AdS$_4$ solutions. Moreover, we have done this in a unified way for all of them, please refer to appendix \ref{EomsAdS11D} for the details. This should again be regarded as a stringent test on the consistency of our formalism. Although we have not explicitly verified the $D=11$ Einstein equation due to its more involved structure, we have reproduced known solutions, like the ones presented in \cite{Gauntlett:2009bh}, for which the Einstein equation has been verified.

We have also obtained new embeddings of minimal $D=4$ $\cN=2$ gauged supergravity both into its parent $D=4$ $\cN=8$ SO(8)--gauged supergravity and into $D=11$ supergravity. A previously known embedding is obtained by fixing the scalars to their vevs at the SO(8) point and then selecting the graviphoton $\bar{A}$ as an appropriate combination of the two SU(3)--invariant vectors $A^\Lambda$, $\Lambda=0,1$. The resulting $D=11$ consistent uplift coincides with a previously known one, constructed in section 2 of \cite{Gauntlett:2007ma}, that is in fact valid for any Sasaki-Einstein seven-manifold. The consistency of this truncation, at least within $D=4$ theories, is guaranteed by symmetry principles. This is because this embedding of minimal $\cN=2$ supergravity into $\cN=8$ coincides with the SU$(4)_s$--invariant sector of the latter. 

More interestingly, we have shown $\cN=8$ SO(8)--supergravity to admit an alternative truncation to minimal $\cN=2$ supergravity by similarly fixing the scalars to their vevs at, now, Warner's $\cN=2$ $\textrm{SU}(3) \times \textrm{U}(1)_c$ point \cite{Warner:1983vz} and again selecting the graviphoton $\bar{A}$ appropriately. Although this alternative truncation is not driven by any apparent symmetry principle, it is nevertheless consistent. We have explicitly verified this at the level of the $D=4$ equations of motion that follow from the Lagrangian (\ref{eq:Lagrangian}), including Einstein. Using our formalism, we have then uplifted this minimal $\cN=2$ supergravity to $D=11$ in section \ref{sec:MinN=2fromD=11}. Again, we have explicitly verified the consistency of the $D=11$ embedding ---see appendix \ref{app:eomMinSugra}. Thus, we have constructed the consistent truncation of $D=11$ supergravity on the $\cN=2$ AdS$_4$ solution of CPW \cite{Corrado:2001nv} down to minimal $D=4$ $\cN=2$ gauged supergravity, predicted to exist by the general conjecture of \cite{Gauntlett:2007ma}.

%%%%%
%%%%%

\section*{Acknowledgements}

%%%%%
%%%%%

PN would like to thank IFT-Madrid for hospitality during the final stages of this project. GL is supported by an FPI-UAM predoctoral fellowship. PN and OV are supported by the NSF grant PHY-1720364. GL and OV are partially sup\-por\-ted by grants SEV-2016-0597, FPA2015-65480-P and PGC2018-095976-B-C21 from MCIU/AEI/FEDER, UE.

%%%%%
%%%%%
\appendix

%%%%%
%%%%%
\addtocontents{toc}{\setcounter{tocdepth}{1}}

%%%%%%%%%%%%%%%
%%%%%%%%%%%%%%%

\section{Details on the SU(3) sector} \label{sec:SU3details} \label{sec:Construction}

%%%%%%%%%%%%%%%
%%%%%%%%%%%%%%%

%%%%%%%%%%%%%%%
%\subsection{Construction from the $\cN=8$ theory} 
%%%%%%%%%%%%%%%

Let $t_{A}{}^B$, $t_{ABCD}$, with $A= 1, \ldots, 8$ indices in the fundamental of SL$(8,\mathbb{R})$,  be the E$_{7(7)}$ generators in the SL$(8,\mathbb{R})$ basis, in the conventions of appendix C of \cite{Guarino:2015qaa}. The $\textrm{SO}(8) \subset \textrm{SL}(8,\mathbb{R}) \subset \textrm{E}_{7(7)}$ subgroup is generated by $T_{AB} \, \equiv 2  \, t_{[A}{}^C \delta_{B]C} $. The generators of $\textrm{SU}(3) \subset \textrm{SO}(8)$ can then be taken to be $\tilde{\lambda}_\alpha$, $\alpha =1, \ldots, 8$, defined as
\begin{eqnarray} \label{SU3gens}
& \tilde{\lambda}_1 = T_{14} - T_{23} \; , \quad
\tilde{\lambda}_2 = -T_{13} - T_{24} \; , \quad
\tilde{\lambda}_3 = T_{12} - T_{34} \; , \quad
\tilde{\lambda}_4 = T_{16} - T_{25}  \; ,  \\ 
& \tilde{\lambda}_5 = -T_{15} - T_{26} \; , \quad
 \tilde{\lambda}_6 = T_{36} - T_{45} \; , \quad 
\tilde{\lambda}_7 = -T_{35} - T_{46} \; , \; \; 
\tilde{\lambda}_8 = \tfrac{1}{\sqrt{3}} \big( T_{12} + T_{34} -2 T_{56}  \big) \; . \nonumber
\end{eqnarray}
These generators indeed close into the SU(3) commutation relations
\begin{eqnarray} \label{SU3comm}
[ \tilde{\lambda}_\alpha ,  \tilde{\lambda}_\beta ] = 2 f_{\alpha \beta \gamma} \, \tilde{\lambda}_\gamma \; ,
\end{eqnarray}
with $f_{\alpha \beta \gamma} = f_{[\alpha \beta \gamma]}$ Gell-Mann's structure constants,
\begin{equation} \label{SU3strConst}
f_{123} = 1 \; , \qquad 
f_{147} = f_{165} = f_{246} = f_{257} = f_{345} = f_{376} = \tfrac12 \; , \qquad 
f_{458} = f_{678} = \tfrac{\sqrt{3}}{2} \; .
\end{equation}

Inside E$_{7(7)}$, the SU(3) generated by (\ref{SU3gens}) commutes with $\textrm{SL}(2 ,\mathbb{R}) \times \textrm{SU}(2,1)$, with the first factor generated by 
\begin{eqnarray} \label{SL2gens}
H_0 = -\tfrac12 \big(  t_i{}^i - 3 t_a{}^a \big) \; , \qquad 
E_0 = 3 \, J^{\6 ij} \epsilon^{ab} \,  t_{ijab} \; , \qquad 
F_0 = \tfrac32 \, J^{\6 ij} J^{\6 kh} \,  t_{ijkh} \; , 
\end{eqnarray}
and the second factor by
\begin{eqnarray} \label{SU21gens}
& H_1 = - t_7{}^7 + t_8{}^8    \; , \qquad 
H_2 =  J^\6_j{}^i \,  t_i{}^j \; , \nonumber \\
& E_{11} = -\sqrt{2} \; \textrm{Im} \, \Omega^{\6 ijk} \;  t_{ijk8} \; , \qquad 
E_{12} = -\sqrt{2} \; \textrm{Re} \, \Omega^{\6 ijk} \;  t_{ijk8} \; , \qquad 
E_{2} = -\sqrt{2} \; \,  t_8{}^7  \; ,  \nonumber \\
& F_{11} = \sqrt{2} \; \textrm{Re} \, \Omega^{\6 ijk} \;  t_{ijk7} \; , \qquad 
F_{12} = -\sqrt{2} \; \textrm{Im} \, \Omega^{\6 ijk} \;  t_{ijk7} \; , \qquad 
F_{2} = -\sqrt{2} \; \,  t_7{}^8 \; . 
\end{eqnarray}
These are the numerator groups in the scalar manifold (\ref{ScalManN=2}). In (\ref{SL2gens}) and (\ref{SU21gens}) we have split the indices as $A = (i, a)$, with $i = 1, \ldots , 6$ in the fundamental of SO$(6)_v$ and $a=7,8$, by effectively identifying the fundamental of SL$(8,\mathbb{R})$ with the $\bm{8}_v$ of SO(8). We have employed the SU(3)--invariant Calabi-Yau $(1,1)$ and $(3,0)$ forms
\begin{eqnarray} \label{eq:SU3invforms}
J^\6 = e^{12} + e^{34} + e^{56} \; , \qquad 
\Omega^\6 = (e^1 + i e^2) \wedge (e^3 + i e^4) \wedge (e^5 + i e^6) \; ,
\end{eqnarray}
on $\mathbb{R}^6 \subset \mathbb{R}^8$, with $e^{12} \equiv dx^1 \wedge dx^2$, etc, and $x^i$ the $\mathbb{R}^6$ Cartesian coordinates. We have also introduced the Levi-Civita tensor $\epsilon_{ab}$ in the $\mathbb{R}^2 \subset \mathbb{R}^8$ plane spanned by the $7, 8$ directions. Indices $i, j$ and $a,b$ are raised and lowered with $\delta_{ij}$ and $\delta_{ab}$. The generators (\ref{SL2gens}) and (\ref{SU21gens}) indeed commute with each other and respectively close into the $\textrm{SL}(2 ,\mathbb{R})$,
\begin{eqnarray}	\label{eq:commSL2}
[H_0 , E_0 ] = 2\,  E_0 \; , \qquad 
[H_0 , F_0 ] = -2\,  F_0 \; , \qquad 
[E_0 , F_0 ] =  H_0 \; , \qquad 
\end{eqnarray}
and $\textrm{SU}(2,1)$ commutation relations,
\begin{equation}	\label{eq:commSU21}
	\small
	\begin{alignedat}{4}
		&\left[H_1,\, H_2 \right]= 0 ,					\\
		&\left[H_1,\, E_{1i}\right]=E_{1i},				\hspace{1.9cm}
		&&\left[H_2,\, E_{1i}\right]=-3\epsilon_{ij}E_{1j},		\quad\quad
		&&\left[H_1,\, E_{2}\right]=2E_{2},				\qquad\quad
		&&\left[H_2,\, E_{2}\right]=0 ,						\\
		&\left[H_1,\, F_{1i}\right]=-E_{1i},				
		&&\left[H_2,\, F_{1i}\right]=-3\epsilon_{ij}F_{1j},		
		&&\left[H_1,\, F_{2}\right]=-2F_{2},				
		&&\left[H_2,\, F_{2}\right]=0 ,						\\
		&\left[E_{11},\, E_{12}\right]=-\sqrt{2}E_2,			
		&&\left[E_{1i},\, E_{2}\right]=0,					
		&&\left[F_{11},\, F_{12}\right]=\sqrt{2}E_2,			
		&&\left[F_{1i},\, F_{2}\right]=0	,				\\					
		&\left[E_{1i},\, F_{1j}\right]=\delta_{ij}H_1+\epsilon_{ij}H_2,
		&&\left[E_{1i},\, F_{2}\right]=\sqrt{2}\epsilon_{ij}F_{1j},
		&&\left[E_{2},\, F_{1i}\right]=\sqrt{2}\epsilon_{ij}E_{1j},
		&&\left[E_{2},\, F_{2}\right]=2H_1 \; ,
	\end{alignedat}
\end{equation}
with, here and only here, $i=1,2$. The generators of the maximal compact subgroup of $\textrm{SU}(2,1)$ are
\begin{eqnarray} \label{eq:commSU2xU1}
& K_0 \equiv E_2 - F_2 -\frac{\sqrt{2}}{3} \, H_2 \; ,  \\
& K_1 \equiv \tfrac{1}{\sqrt{8}} \big( E_{11} - F_{11} \big)   \; , \qquad 
K_2 \equiv \tfrac{1}{\sqrt{8}} \big( E_{12} - F_{12} \big)  \; , \qquad 
K_3 \equiv -\tfrac{1}{4\sqrt{2}} \big( E_{2} - F_{2} \big) -\tfrac14 \, H_2 \; , \nonumber
\end{eqnarray}
and close into the $\textrm{SU}(2) \times \textrm{U}(1)$ commutation relations
\begin{equation}
[K_0 , K_x] = 0 \; , \qquad [K_x , K_y ] = \epsilon_{xyz} \, K_z \; , \qquad x=1,2,3. 
\end{equation}
It is also interesting to note that the three different U(1)'s with which SU(3) commutes inside the SO(8) subgroups $\textrm{SO}(6)_v$, $\textrm{SU}(4)_c$ and $\textrm{SU}(4)_s$ are respectively generated by
\begin{eqnarray} \label{U1vgenerator}
\textrm{U}(1)_v & : & \qquad -J^\6_j{}^i \, t_i{}^j  \; ,  \\[4pt]
\label{U1cgenerator}
 \textrm{U}(1)_c & : & \qquad - J^\6_j{}^i \, t_i{}^j +3  \, \epsilon_b{}^a \, t_a{}^b \; , 
 \\[4pt] 
\label{U1sgenerator}
\textrm{U}(1)_s & : & \qquad - \lambda \, J^\6_j{}^i \, t_i{}^j +3 \, \epsilon_b{}^a \, t_a{}^b \; , \quad \textrm{with} \,  \lambda \in \mathbb{R} \; ,  \; \lambda \neq 1 \; .
\end{eqnarray}

With these details, the SU(3)--invariant bosonic field content and its interactions described in section \ref{sec:SU3ofSO8} can be constructed from the parent $\cN=8$ supergravity. Per the analysis above, the SU(3)--invariant scalar manifold is (\ref{ScalManN=2}). A coset representative is
\begin{equation}
{\cal V} = e^{-\chi E_0} e^{-\frac12 \varphi H_0} e^{\frac{1}{\sqrt{2}} ( a E_2 - \zeta E_{11} -\tilde{\zeta} E_{12} ) } e^{-\phi H_1} \; ,
\end{equation}
and the quadratic scalar matrix that enters the bosonic Lagrangian is ${\cal M} = {\cal V} {\cal V}^{\textrm{T}}$. The metric on (\ref{ScalManN=2}) that determines the scalar kinetic terms in the Lagrangian (\ref{eq:Lagrangian}) is then reproduced through $-\frac{1}{48} D \cM \wedge *D \cM^{-1}$. For reference, the $\textrm{SL}(2 ,\mathbb{R}) \times \textrm{SU}(2,1)$ Killing vectors of this metric, normalised to obey the commutation relations (\ref{eq:commSL2}), (\ref{eq:commSU21}), are
\begin{equation}	\label{eq:KillingSL2}
		k[H_0] =2\partial_\varphi-2\chi\partial_\chi \; , \qquad 
		k[E_0] =\partial_\chi					\; , \qquad 
		k[F_0] =2 \chi\partial_\varphi+(e^{-2 \varphi}-\chi ^2)\partial_\chi \; , 
\end{equation}
and
\begin{equation}	\label{eq:KillingSU21}
	\begin{aligned}
		k&[H_1]=\partial_\phi-2a\partial_a-\zeta\partial_{\zeta}-\tilde{\zeta}\partial_{\tilde{\zeta}} \; ,	\qquad\qquad
		k[H_2]=3\tilde{\zeta}\partial_\zeta-3\zeta\partial_{\tilde{\zeta}} \; ,						\\[3mm]
		k&[E_{11}]= \frac1{\sqrt{2}}\left(\tilde{\zeta}\partial_a-2\partial_{\zeta}\right)	 \;  ,		\qquad
		k[E_{12}]= \frac1{\sqrt{2}}\left(\zeta\partial_a+2\partial_{\tilde{\zeta}}\right) \; ,			\qquad
		k[E_2]= \sqrt{2}\partial_a	\; , 											\\[3mm]
		k[F_2]&=\sqrt{2} \left(a\partial_\phi-e^{-4\phi}\left(Z^2- Y^2\right)\partial_a-\left(a\zeta-e^{-2\phi}\tilde{\zeta}Y\right)\partial_\zeta-e^{-2\phi}\left(\tilde{\zeta}Z+\zeta Y\right)\partial_{\tilde{\zeta}}\right) \; , 	\\
		k[F_{11}]&=\frac1{\sqrt{2}}\left(-\zeta\partial_\phi +\left(a\zeta-e^{-2\phi}\tilde{\zeta}Y\right)\partial_a
		-\frac{1}{2} \left(4e^{-2\phi}-\zeta ^2+3 \tilde{\zeta}^2\right)\partial_\zeta+2 \left(a+\zeta  \tilde{\zeta}\right)\partial_{\tilde{\zeta}}\right)  \; , 	\\
		k[F_{12}]&=\frac1{\sqrt{2}}\left(\tilde{\zeta}\partial_\phi-\left(a\tilde{\zeta}+e^{-2\phi}\zeta Y\right)\partial_a+2\left(a-\zeta\tilde{\zeta}\right)\partial_\zeta+\frac{1}{2} \left(4e^{-2\phi}+3 \zeta ^2-\tilde{\zeta}^2\right)\partial_{\tilde{\zeta}}\right)  \; .
	\end{aligned}
\end{equation}

Moving on, we need to specify how the SU(3)--invariant tensor fields in (\ref{fieldContentHierarchy}) are embedded into their $\cN=8$ counterparts. Recall that the restricted $\cN=8$ tensor hierarchy contains $\bm{28}^\prime$ electric vectors ${\cal A}^{AB}$, $\bm{28}$ magnetic vectors $\tilde{{\cal A}}_{AB}$, $\bm{63}$ two-forms ${\cal B}_A{}^B$ and $\bm{36}$ three-forms ${\cal C}^{AB}$, in representations of SL$(8,\mathbb{R})$ \cite{Varela:2015ywx}. In order to determine the embedding of the SU(3)--invariant vectors $A^\Lambda$, $\tilde{A}_\Lambda$, $\Lambda=0,1$, into their $\cN=8$ counterparts, we note that SU(3) commutes inside $\textrm{SO}(8) \subset \textrm{E}_{7(7)}$ with the U$(1)^2$ generated, in the notation of (\ref{SU21gens}), by $(E_2 - F_2)$ and $H_2$ or, equivalently, by $K^0$ and $K^3$ defined in (\ref{eq:commSU2xU1}). These are the Cartan generators  of the maximal compact subgroup $\textrm{SU}(2) \times \textrm{U}(1)$ of the hypermultiplet scalar manifold. Splitting again the $\cN=8$ index as below (\ref{SU21gens}), $A= (i,a)$, and fixing the normalisations for convenience we have the following embedding into the $\cN=8$ vectors:
\begin{equation} \label{eq:VecEmbedN=8}
\ca^{ij}= A^{1}J^{\6 ij} \; , \qquad 
\ca^{ab}= \epsilon^{ab}A^{0} \; , \qquad 
\tilde{\ca}_{ij}=\tfrac{1}{3}  \tilde{A}_{1}J_{\6 ij} \; , \qquad 
\tilde{\ca}_{ab}= \tilde{A}_{0} \, \epsilon_{ab} \; .
\end{equation}
Similarly, for the two-form potentials we define
\begin{equation} \label{eq:twoformsembedN=8}
\cb_{i}{}^{j} =  -\tfrac{1}{12} \, B_a{}^a \, \delta_{i}\,^{j} +\tfrac13 \, B^{2} \, J^\6_{i}{}^{j} \; , \qquad 
\cb_{a}{}^{b}=\tfrac{1}{2}  \, B_{a}{}^{b} - \tfrac{1}{2} \, B^{0} \, \epsilon_{a}{}^{b} \; ,
\end{equation}
and for the three-form potentials,
\begin{equation} \label{eq:threeformsembedN=8}
\cc^{ij} = C^{1} \, \delta^{ij} \; , \qquad
\cc^{ab} = C^{ab} \; .
\end{equation}
The field strengths and couplings brought to section \ref{sec:SU3ofSO8} can be obtained by inserting these expressions into the $\cN=8$ equations given in \cite{Varela:2015ywx}. For example, the gauge covariant derivative acting on the scalars reduce to $ D = d  + \tfrac{1}{\sqrt{2}} g (k[E_2] - k[F_2]) \, A^0 - g \, k[H_2] \, A^1 $, and this in turn reproduces (\ref{eq:covder}) upon use of the relevant Killing vectors in (\ref{eq:KillingSU21}).

%%%%%%%%%%%%%%%
%%%%%%%%%%%%%%%

\section{Intrinsic coordinates and geometric structures on $S^7$} 	\label{sec: geometryS7}

%%%%%%%%%%%%%%%
%%%%%%%%%%%%%%%

There are various sets of intrinsic coordinates that prove useful in our context, each of them adapted to different geometric structures on $S^7$. The expressions below have been used to particularise the general SU(3)--invariant consistent embedding formulae of section \ref{sec:SU3sectorinD=11} to the further subsectors of section \ref{sec:SU3subsectorsinD=11} and the AdS$_4$ solutions of section \ref{sec:11Dsolutions}.

\subsection{$S^7$ as the join of $S^1$ and a Sasaki-Einstein $S^5$ }

The first set of coordinates solves the constraint (\ref{eq:S7inR8}) by splitting $\mu^A$, $A= 1, \ldots, 8$, as
\begin{equation} \label{IntrinsicCoordsS71}
	\mu^i=\cos\alpha\,\tilde{\mu}^i\,, \quad i =1 , \ldots , 6 \; , \qquad
	\mu^7=\sin\alpha\cos\psi\,,		\qquad
	\mu^8=\sin\alpha\sin\psi\,,	
\end{equation}
with $ 0 \leq \alpha \leq \pi/2$,  $ 0 \leq \psi < 2\pi $, and $\tilde{\mu}^i$, $i=1, \ldots , 6$, defining in turn an $S^5$, {\it i.e.}~subject to the constraint $\delta_{ij}\tilde{\mu}^i\tilde{\mu}^j=1$. The intrinsic coordinates (\ref{IntrinsicCoordsS71}) are adapted to the topological description of $S^7$ as the join of $S^5$ and $S^1$, for which the round, Einstein, SO(8)--invariant metric,
\begin{equation} \label{RoundMetricS7}
ds^2 (S^7) = \delta_{AB} \, d\mu^A d\mu^B \; ,
\end{equation}
on $S^7$ displays only a manifest $\textrm{SO}(6)_v \times \textrm{SO}(2)$ symmetry,
\begin{equation} \label{S7asJoinS5S1}
ds^2 (S^7) = d\alpha^2 + \cos^2 \alpha \,  ds^2(S^5)+ \sin^2 \alpha \, d\psi^2 \; ,
\end{equation}
with $ds^2(S^5) = \delta_{ij} \, d\tilde{\mu}^i d\tilde{\mu}^j$ the round, Einstein metric on $S^5$ normalised so that the Ricci tensor equals four times the metric. This $S^5$ comes naturally equipped with the Sasaki-Einstein structure ($\bm{\eta}^\5$, $\bm{J}^\5$, $\bm{\Omega}^\5$) endowed upon it from the Calabi-Yau forms $J^\6$, $\Omega^\6$, (\ref{eq:SU3invforms}), on the $\mathbb{R}^6$ factor of $\mathbb{R}^8 = \mathbb{R}^6 \times \mathbb{R}^2 $ in which $S^5$ is embedded,
\begin{equation} \label{eq:SE5}
	\bm{\eta}^\5 =J^\6_{ij}\tilde{\mu}^id\tilde{\mu}^j\,,			\qquad
	\bm{J}^\5 =\tfrac12J^\6_{ij} \, d\tilde{\mu}^i\wedge d\tilde{\mu}^j\,,		\qquad
	\bm{\Omega}^\5 =\tfrac12 \, \Omega^\6_{ijk} \, \tilde{\mu}^id\tilde{\mu}^j\wedge d\tilde{\mu}^k \; .
\end{equation}
These satisfy
\begin{equation}		\label{eq: SE5algebraic}
\bm{J}^\5 \wedge\bm{\Omega}^\5=0 \ , \qquad
\tfrac12 \, \bm{J}^\5 \wedge\bm{J}^\5 \wedge \bm{\eta}^\5 =\tfrac14\bm{\Omega}^\5 \wedge\bm{\bar{\Omega}}^\5 \wedge \bm{\eta}^\5 = \textrm{vol} ( S^5) \; ,
\end{equation}
and
\begin{equation}		\label{eq: SE5differential}
d\bm{\eta}^\5=2\bm{J}^\5 \ , \qquad
d\bm{\Omega}^\5=3i \bm{\eta}^\5  \wedge \bm{\Omega}^\5 \; .
\end{equation}
It is also useful to relate the Calabi-Yau forms $J^\6$ and $\Omega^\6$ written in terms of constrained $\mathbb{R}^8$ coordinates $\mu^A = (\mu^i, \mu^a)$, $i=1, \ldots, 6$, $a=7,8$, to the intrinsic $S^7$ coordinate $\alpha$ in (\ref{IntrinsicCoordsS71}) and Sasaki-Einstein forms (\ref{eq:SE5}):
\begin{eqnarray}
	J_{ij}^\6\mu^id\mu^j & = & \cos^2\!\alpha\,\bm{\eta}^\5 \; ,	\nonumber \\
	\tfrac12\, J_{ij}^\6 d\mu^i\wedge d\mu^j & = & \cos^2\!\alpha\,\bm{J}^\5-\sin\alpha\cos\alpha\, d\alpha\wedge\bm{\eta}^\5 \; ,	\nonumber \\
	\tfrac12\,\Omega_{ijk}^\6 \mu^i d\mu^j\wedge d\mu^k & = & \cos^3\!\alpha\,\bm{\Omega}^\5 \; ,	\nonumber \\
	\tfrac16\,\Omega_{ijk}^\6 d\mu^i\wedge d\mu^j\wedge d\mu^k & = & i\cos^3\!\alpha\,\bm{\Omega}^\5\wedge\bm{\eta}^\5-\sin\alpha\cos^2\!\alpha\, d\alpha\wedge\bm{\Omega}^\5  \; .
\end{eqnarray}

The round metric $ds^2(S^5)$ in (\ref{S7asJoinS5S1}) naturally adapts itself to the Sasaki-Einstein structure (\ref{eq:SE5}) when written as
\begin{equation} \label{metricHopfS5}
ds^2(S^5) = ds^2(\mathbb{CP}^2)+(d \tau+\sigma)^2 \; , 
\end{equation}
with $ds^2(\mathbb{CP}^2)$ the Fubini-Study metric on the complex projective plane, normalised so that the Ricci tensor equals six times the metric, $0 \leq \tau < 2\pi$ an angle on the $S^5$ Hopf fiber, and $\sigma$ a one-form on $\mathbb{CP}^2$ such that $d\sigma = 2 \bm{J}^\4$ with $\bm{J}^\4$ the K\"ahler form on $\mathbb{CP}^2$, so that $\bm{\eta}^\5 \equiv d\tau + \sigma$ and $\bm{J}^\5 \equiv \bm{J}^\4$. For completeness, we note that $ds^2(\mathbb{CP}^2)$ can be written in terms of complex projective coordinates $\xi^i$, $i=1,2$, as
\begin{eqnarray} \label{FSCP2metric}
ds^2(\mathbb{CP}^2) = \frac{d\bar{\xi}_i \, d\xi^i }{1 + \bar{\xi}_k \xi^k}  -\frac{ (\bar{\xi}_i d\xi^i )( \xi^j d\bar{\xi}_j ) }{( 1 + \bar{\xi}_k \xi^k )^2 } \; , 
\end{eqnarray}
by introducing complex coordinates on $\mathbb{R}^6 = \mathbb{C}^3$ through
\begin{equation} \label{eq:FSmuS5}
 \tilde{\mu}^1 + i \tilde{\mu}^2 = \tfrac{1}{\sqrt{ 1 + \bar{\xi}_i \xi^i}} \, e^{i \tau} \xi^1 \; , \quad 
 \tilde{\mu}^3 + i \tilde{\mu}^4 = \tfrac{1}{\sqrt{ 1 + \bar{\xi}_i \xi^i}} \, e^{i \tau} \xi^2 \; , \quad 
 \tilde{\mu}^5 + i \tilde{\mu}^6 = \tfrac{1}{\sqrt{ 1 + \bar{\xi}_i \xi^i}} \, e^{i \tau}\; .  
\end{equation}
In these coordinates, the one-form $\sigma$ in (\ref{metricHopfS5}) reads
\begin{equation} \label{sigmaSE5}
\sigma = \frac{i}{2} \, \frac{\xi^i d\bar{\xi}_i - \bar{\xi}_i d\xi^i }{ 1 + \bar{\xi}_k \xi^k } \; .
\end{equation}

\subsection{$S^7$ with its homogeneous Sasaki-Einstein structure}

A second set of intrinsic coordinates on $S^7$ can be chosen that adapt themselves to its two natural, homogeneous seven-dimensional Sasaki-Einstein structures. These descend on $S^7$ from the Calabi-Yau forms $J^\8_\pm$, $\Omega^\8_\pm$ on $\mathbb{R}^8$, 
\begin{eqnarray} \label{eq:SU4invforms}
&& J^\8_\pm = J^\6 \pm e^{78} =  e^{12} + e^{34} + e^{56}  \pm e^{78} \; , \nonumber \\
&& \Omega^\8_\pm = \Omega^\6 \wedge  (e^7 \pm i e^8) = (e^1 + i e^2) \wedge (e^3 + i e^4) \wedge (e^5 + i e^6)\wedge (e^7 \pm i e^8) ,
\end{eqnarray}
that are invariant under $\textrm{SU}(4)_c$ for the $+$ sign and $\textrm{SU}(4)_s$ for the $-$ sign. In terms of the constrained coordinates $\mu^A$, $A=1, \ldots, 8$, that define $S^7$ as the locus (\ref{eq:S7inR8}) in $\mathbb{R}^8$, the Sasaki-Einstein structure forms induced on $S^7$ are
\begin{equation} \label{eq:SE7}
	\bm{\eta}^\7_\pm =J^\8_{\pm \, AB} \, \mu^A d \mu^B \,,			\quad
	\bm{J}^\7_\pm =\tfrac12 J^\8_{\pm \, AB } \, d\mu^A \wedge d\mu^B \,,		\quad
	\bm{\Omega}^\7_\pm =\tfrac16 \, \Omega^\8_{\pm \, ABCD} \, \mu^A d\mu^B \wedge d\mu^C \wedge d\mu^D \; .
\end{equation}
These are subject to
\begin{equation}		\label{eq: SE7algebraic}
\bm{J}_\pm^\7 \wedge\bm{\Omega}_\pm^\7=0 \ , \qquad
 \, \bm{J}_\pm^\7 \wedge\bm{J}_\pm^\7\wedge\bm{J}_\pm^\7 \wedge \bm{\eta}_\pm^\7 = \tfrac{3i}{4} \, \bm{\Omega}_\pm^\7 \wedge\bm{\bar{\Omega}}_\pm^\7 \wedge \bm{\eta}_\pm^\7 =\mp 6 \, \textrm{vol} ( S^7) \; ,
\end{equation}
and
\begin{equation}		\label{eq: SE7differential}
d\bm{\eta}_\pm^\7=2\bm{J}_\pm^\7 \ , \qquad
d\bm{\Omega}_\pm^\7 = 4i \bm{\eta}_\pm^\7 \wedge \bm{\Omega}_\pm^\7 \; .
\end{equation}
The seven-dimensional Sasaki-Einstein structure (\ref{eq:SE7}) is related to its five-dimensional counterpart (\ref{eq:SE5}) and the angles (\ref{IntrinsicCoordsS71}) through
\begin{eqnarray} \label{SE7FromSE5}
	\bm{\eta}_\pm^\7 	& = & \cos^2\!\alpha\,\bm{\eta}^\5 \pm \sin^2\!\alpha\,d\psi	\; , \nonumber \\
	\bm{J}_\pm^\7 		& = & \cos^2\!\alpha\,\bm{J}^\5 \pm \sin\alpha\cos\alpha\, d\alpha\wedge(d\psi \mp \bm{\eta}^\5)	\; ,  \nonumber \\
	\bm{\Omega}_\pm^\7& = & e^{\pm i\psi}\cos^2\!\alpha\,\left[d\alpha \pm i\cos\!\alpha\sin\!\alpha(d\psi \mp \bm{\eta}^\5)\right]\wedge\bm{\Omega}^\5 \; .	
\end{eqnarray}

The round metric on $S^7$ adapted to seven-dimensional Sasaki-Einstein structure reads, similarly to (\ref{metricHopfS5}), 
\begin{equation} \label{metricHopfS7}
ds^2(S^7) = ds^2(\mathbb{CP}_\pm^3)+ \big(d \psi_\pm + \sigma_\pm \big)^2 \; , 
\end{equation}
where $ds^2(\mathbb{CP}_\pm^3)$ is the Fubini-Study metric, normalised so that the Ricci tensor equals eight times the metric. The $\pm$ refers to two different embeddings of $\mathbb{CP}^3$ into $S^7$, with isometry group $\textrm{SU}(4)_c \subset \textrm{SO}(8)$ for the $+$ sign and $\textrm{SU}(4)_s \subset \textrm{SO}(8)$ for the $-$ sign. The angles $\psi_\pm$ have period $2\pi$ and the one-forms $\sigma_\pm$ in (\ref{metricHopfS7}) obey $d\sigma_\pm = 2 \bm{J}_\pm^\7$ so that  $\bm{\eta}_\pm^\7 \equiv d\psi_\pm + \sigma_\pm$. It is also useful to make manifest the $\mathbb{CP}^2$ that resides inside $\mathbb{CP}^3_\pm$, which is equipped with the complex projective coordinates $\xi^i$, $i=1,2$, that appear in (\ref{eq:FSmuS5}) and the metric (\ref{FSCP2metric}). This can be achieved by writing
\begin{eqnarray} \label{eq:FSmuS7pm}
& \mu^1 + i \mu^2 = \tfrac{1}{\sqrt{ 1 + \bar{\xi}_i \xi^i}} \, \cos\alpha \, e^{i (\psi_{\pm} + \tau_{\pm})} \, \xi^1 \; , \qquad &
 \mu^3 + i \mu^4 = \tfrac{1}{\sqrt{ 1 + \bar{\xi}_i \xi^i}} \, \cos\alpha \, e^{i (\psi_{\pm} + \tau_{\pm})} \, \xi^2 \; , \nonumber \\
& \mu^5 + i \mu^6 = \tfrac{1}{\sqrt{ 1 + \bar{\xi}_i \xi^i}} \, \cos\alpha \, e^{i (\psi_{\pm} + \tau_{\pm})} \; ,   \qquad &
\mu^7 + i \mu^8 =  \, \sin\alpha \, e^{{\pm}i \psi_{\pm} } \; ,  
\end{eqnarray}
where $\tau_\pm$ are angles of period $2\pi$. The metrics $ds^2(\mathbb{CP}_\pm^3)$ and one-forms $\sigma_\pm$ inside the round $S^7$ metric (\ref{metricHopfS7}) can be written in terms of the coordinates (\ref{eq:FSmuS7pm}) as 
\begin{equation} \label{FSCP3}
ds^2(\mathbb{CP}_\pm^3) = d\alpha^2 + \cos^2 \alpha \, ds^2(\mathbb{CP}^2) + \cos^2 \alpha \sin^2\alpha \, (d\tau_\pm + \sigma)^2 \; ,
\end{equation}
and
\begin{equation} \label{sigmaCP3}
\sigma_\pm = \cos^2 \alpha \, ( d\tau_\pm + \sigma) \; ,
\end{equation}
with $ds^2(\mathbb{CP}^2)$ and $\sigma$ respectively given by (\ref{FSCP2metric}) and (\ref{sigmaSE5}). The round $S^7$ metrics (\ref{S7asJoinS5S1}) with (\ref{metricHopfS5}) and (\ref{metricHopfS7}) with (\ref{FSCP3}) are of course diffeomorphic: they are brought into each other by the change of coordinates
\begin{equation} \label{eq:changepm}
 \psi = \pm \psi_{\pm}  \; , \qquad \tau = \tau_{\pm} + \psi_{\pm} \; .
\end{equation}

\subsection{$S^7$ as the sine-cone over a nearly-K\"ahler $S^6$}

A third and final set of intrinsic angles on $S^7$ is better suited to describe the solutions with at least G$_2$ symmetry. First split the $\mu^A$, $A=1, \ldots , 8$, as $\mu^A = (\mu^I, \mu^8)$, with $I=1, \ldots , 7$, and then let 
\begin{equation}\label{IntrinsicCoordsS72}
\mu^{I} = \sin\beta\, \tilde{\nu}^{I}\, \; ,  \qquad 
\mu^{8} =  \cos\beta \; , 
\end{equation}
where $0 \leq \beta \leq \pi/2$, and $\tilde{\nu}^I$, $I=1, \ldots , 7$, define an $S^6$ through the constraint $\delta_{IJ}\tilde{\nu}^I \tilde{\nu}^J=1$. In these coordinates, the round metric (\ref{RoundMetricS7}) takes on the local sine-cone form
\begin{equation} \label{eq:S7sineCone}
ds^2(S^7) = d\beta^2 + \sin^2 \beta \, ds^2(S^6) \; ,
\end{equation}
where $ds^2(S^6) = \delta_{IJ} \, d\tilde{\nu}^I d\tilde{\nu}^J$ is the round, Einstein metric on $S^6$ normalised so that the Ricci tensor equals five times the metric. This $S^6$ is naturally endowed with the homogeneous nearly-K\"ahler structure\footnote{The typography we use for the nearly-K\"ahler forms on $S^6$ differentiates them from the Calabi-Yau forms (\ref{eq:SU3invforms}) on $\mathbb{R}^6$. For that reason, we omit labels $^\6$ for the former. Similarly, we omit labels $^\7$ for the associative and co-associative forms on $\mathbb{R}^7$.} $(\cj , \Upomega )$ inherited from the closed associative and co-associative forms, 
{\setlength\arraycolsep{2pt}
\begin{eqnarray}
\psi &=& e^{127} +e^{347} +e^{567} +e^{135} -e^{146} -e^{236} -e^{245} \; , \\
\tilde{\psi} &=& e^{1234} +e^{1256} +e^{3456} +e^{1367} +e^{1457} +e^{2357} -e^{2467} \; , 
\end{eqnarray}
}on the $\mathbb{R}^7$ factor of $\mathbb{R}^8 = \mathbb{R}^7 \times \mathbb{R} $ in which $S^6$ is embedded:
\begin{eqnarray} \label{JOmegaintermsofmu}
{\cal J} = \tfrac12 \, \psi_{IJK} \,  \tilde{\nu}^I d\tilde{\nu}^J \wedge d\tilde{\nu}^K \; , \quad 
  \Upomega  = \tfrac16 \left( \psi_{JKL} -i \, \tilde{\psi}_{IJKL} \, \tilde{\nu}^I \right)  d \tilde{\nu}^J \wedge d\tilde{\nu}^K \wedge d\tilde{\nu}^L \; . 
\end{eqnarray}
The nearly-K\"ahler forms are subject to 
\begin{eqnarray} \label{SU3str}
{\cal J} \wedge \Upomega = 0 \; , \qquad \Upomega \wedge \bar \Upomega = -\tfrac{4i}{3} {\cal J} \wedge {\cal J} \wedge {\cal J}  = -8i \,  \textrm{vol} (S^6 )   \; ,  
\end{eqnarray}
and
\begin{eqnarray} \label{SU3strDif}
d {\cal J} = 3 \,  \textrm{Re} \,  \Upomega   \; , \quad d \, \textrm{Im} \, \Upomega =  -2 \,  {\cal J} \wedge {\cal J} \ . 
\end{eqnarray}
It is also useful to note the following relations between the associative and co-associative forms $\psi$, $\tilde{\psi}$ written in constrained $\mathbb{R}^8$ coordinates $\mu^A = (\mu^I, \mu^8)$, the $S^7$ coordinate $\beta$ in (\ref{IntrinsicCoordsS72}), and the nearly-K\"ahler forms (\ref{JOmegaintermsofmu}):
\begin{eqnarray}\label{eq:G2ToNK}
\tfrac{1}{2}\psi_{IJK}\mu^{I}d\mu^{J}\wedge d\mu^{K}\wedge d\mu^8 & = & -\sin^{4}\!\beta\,\mathcal{J}\wedge d\beta \; , \nonumber \\
\tfrac{1}{6}\psi_{IJK}d\mu^{I}\wedge d\mu^{J}\wedge d\mu^{K} & = & \sin^{3}\!\beta\,\text{Re}\,\Upomega+\sin^{2}\!\beta\,\cos\beta\,\mathcal{J}\wedge d\beta \; , \nonumber \\
\tfrac{1}{6}\tilde{\psi}_{IJKL}\mu^{I}d\mu^{J}\wedge d\mu^{K}\wedge d\mu^{L} & = & -\sin^{4}\!
\beta\,\text{Im}\,\Upomega  \; , \\
\tfrac{1}{24}\tilde{\psi}_{IJKL}d\mu^{I}\wedge d\mu^{J}\wedge d\mu^{K}\wedge d\mu^{L} & = & \tfrac12\sin^{4}\!\beta\,\mathcal{J}\wedge\mathcal{J}+\sin^{3}\!\beta\,\cos\beta\,\text{Im}\,\Upomega\wedge d\beta  \; . \nonumber
\end{eqnarray}
Finally, the following relations hold between the associative and co-associative forms on $\mathbb{R}^8 = \mathbb{R}^7 \times \mathbb{R} $ and the Calabi-Yau forms $\mathbb{R}^8 = \mathbb{R}^6 \times \mathbb{R}^2$: 
{\setlength\arraycolsep{2pt}
\begin{eqnarray}\label{eq:G2toCY3}
	\tfrac{1}{2}\psi_{IJK}\mu^{I}d\mu^{J}\wedge d\mu^{K}
			& = & J^\6_{ij}\mu^i\, d\mu^j\wedge d\mu^{7}+\tfrac12\big(J^\6_{jk}\,\mu^{7}+\text{Re}\,\Omega^\6_{ijk}\,\mu^i\big)d\mu^j\wedge d\mu^k \; , \nonumber	 \\
	\tfrac{1}{6}\psi_{IJK}d\mu^{I}\wedge d\mu^{J}\wedge d\mu^{K} 
			& = &  \tfrac16\text{Re}\,\Omega^\6_{ijk}\,d\mu^i\wedge d\mu^j\wedge d\mu^k+\tfrac12J^\6_{ij}d\mu^i\wedge d\mu^j\wedge d\mu^{7}	\; , \nonumber	 \\
	\tfrac16\tilde{\psi}_{IJKL}\mu^{I}d\mu^{J}\wedge d\mu^{K}\wedge d\mu^{L} 
			& = & -\tfrac16\text{Im}\,\Omega^\6_{ijk}\,\mu^7d\mu^i\wedge d\mu^j\wedge d\mu^k+\tfrac12J^\6_{ij}\, J^\6_{kl}\mu^i\, d\mu^j\wedge d\mu^k\wedge d\mu^l	\nonumber \\
			&& +\tfrac12\text{Im}\,\Omega^\6_{ijk}\, \mu^i\, d\mu^j\wedge d\mu^k\wedge d\mu^{7} \; . 
\end{eqnarray}
}These expressions come handy to derive the G$_2$--invariant consistent uplifting formulae of section \ref{sec:G2sectorin11D} from the general expressions of section \ref{sec:SU3sectorinD=11}. They are also useful to rewrite the solutions (\ref{eq:G2AdS4Sol})--(\ref{eq:SO7cD=11SolA3}) with at least G$_2$ symmetry in the form (\ref{eq:MetricPulledBack})--(\ref{eq:fINTermsOfL}), in order to verify that they satisfy the equations of motion.

%%%%%%%%%%%%%%%
%%%%%%%%%%%%%%%

\section{Consistency of the minimal $\cN=2$ truncation} \label{app:eomMinSugra}

%%%%%%%%%%%%%%%
%%%%%%%%%%%%%%%

We have explicitly verified at the level of the $D=4$ field equations that the restrictions (\ref{eq:N=2vevs})--(\ref{eq:N=2CPWFourFormFS}) define a consistent truncation of the SU(3)--invariant theory (\ref{eq:Lagrangian}) to minimal $\cN=2$ gauged supergravity (\ref{minimalN=2}). In turn, the consistency of the $D=11$ embedding of the entire SU(3) sector described in section \ref{sec:SU3sectorinD=11} guarantees the consistency of the new uplift of minimal $\cN=2$ supergravity given in section \ref{sec:MinN=2fromD=11}. We have nevertheless checked consistency explicitly at the level of the Bianchi identity and the equation of motion of the $D=11$ four-form $\hat{F}_\4 = d \hat{A}_\3$,
\begin{eqnarray} \label{F4FieldEqs}
d \hat{F}_\4 = 0 \; , \qquad
d \hat{*} \hat{F}_\4 + \tfrac12 \hat{F}_\4\wedge\hat{F}_\4=0 \; .
\end{eqnarray}
The configuration (\ref{eq:metricminsugraCPW}), (\ref{eq:4formminsugraCPW}) does solve the $D=11$ field equations (\ref{F4FieldEqs}) provided the Bianchi identity and the Maxwell equation for the $D=4$ graviphoton,
\begin{eqnarray} \label{F2FieldEqs}
d \bar{F} = 0 \; , \qquad
d \bar{*} \bar{F} =0 \; ,
\end{eqnarray}
are imposed. 

It is straightforward to see that the $D=11$ Bianchi identity is satisfied. Hitting (\ref{eq:4formminsugraCPW}) with the differential operator we obtain, after using (\ref{F4FieldEqs}) and the algebraic and differential conditions for the local five-dimensional Sasaki-Einstein structure (\ref{PrimedSE5}) (that is, (\ref{eq: SE5algebraic}), (\ref{eq: SE5differential}) written for the primed forms $\bm{\eta}'$, $\bm{J}'$ and $\bm{\Omega}'$), 
{\setlength\arraycolsep{0pt}
\begin{eqnarray}
		d\hat{F}_{(4)}&=& \frac{g^{-3}}{\sqrt3}\left[\frac{ \cos^2\alpha (7-10 \cos2\alpha+\cos4 \alpha)}{ \left(1+2 \sin^2\alpha\right)^2}
				d\alpha\wedge \Big(\frac g{2}\bar{F}\wedge\text{Re}\, \bm{\Omega}' +3D\psi'\wedge\text{Im}\, \bm{\Omega}'\wedge\bm{\eta}'\Big)  \right. \nonumber \\
				&&
				\qquad \; \left. +6\,\partial_{\alpha}\left(\frac{\sin\alpha\cos^3\alpha}{1+2\sin^2\alpha}\right)d\alpha\wedge D\psi'\wedge\bm{\eta}'\wedge\text{Im}\, \bm{\Omega}' +g \frac{3\sin\alpha\cos^3\alpha}{1+2\sin^2\alpha}  \bar{F}\wedge\bm{\eta}' \wedge\text{Im}\, \bm{\Omega}'  \right]
				\nonumber \\
				&&  +\frac{g^{-2}}{2\sqrt3}\left[2\,\partial_{\alpha}\left(\frac{\sin\alpha\cos^3\alpha}{1+2\sin^2\alpha}\right)d\alpha \wedge \bar{F}\wedge \text{Re}\, \bm{\Omega}'	
				-\frac{6\sin\alpha\cos^3\alpha}{1+2\sin^2\alpha}\bar{F}\wedge \text{Im}\, \bm{\Omega}'\wedge\bm{\eta}'	\right]  . 
\end{eqnarray}
}Terms with the same form dependence cancel each other, thus leading to $d\hat{F}_{(4)} =0$.

Moving on to the equation of motion, we find it useful for the calculation to introduce the obvious frame that can be read off from \eqref{eq:metricminsugraCPW},
\begin{equation}
	\begin{aligned}
		&\hat{e}^\alpha=\frac{\left(1+2\sin^2\alpha\right)^{1/3}}{2^{1/3}\sqrt3}\; \bar{e}^\alpha,		
				\hspace{3.2cm} \text{with } \bar{e}^\alpha \text{ a vierbein for } d\bar{s}_4^2\,,	\\[.7em]
		&\hat{e}^p=\frac{2^{1/6} \cos\alpha}{g \left(1+2 \sin ^2\alpha\right)^{1/6}}\; e^p,		
				\hspace{3cm} \text{with } e^p \text{ a vierbein for } ds^2(\mathbb{CP}^2)\,,		\\[.7em]
		&\hat{e}^8=\frac{2^{1/6} \left(1+2 \sin ^2\alpha\right)^{1/3}}{\sqrt{3}\, g}\,d\alpha\,,			\\[.7em]
		&\hat{e}^9=\frac{2^{1/6}\sqrt{3} \sin\alpha\cos \alpha \left(1+2 \sin ^2\alpha\right)^{1/3}}{g \left(1+8\sin^4\alpha\right)^{1/2}}\,\bm{\eta}'\,,		\\[.7em]
		&\hat{e}^{10}=\frac{\left(1+8\sin^4\alpha\right)^{1/2}}{2^{1/3} \sqrt{3}\, g \left(1+2 \sin ^2\alpha\right)^{2/3}}\Big( D\psi'-\frac{3\cos^2\alpha}{1+8\sin^4\alpha}\bm{\eta}'\Big)\,,
	\end{aligned}
\end{equation}
\noindent with $\alpha=0,\,1,\,2,\,3$ and $p=4,\,5,\,6,\,7$. Using this frame, the Hodge dual of $\hat{F}_\4$ reads
	\begin{align} \label{eq:StarF4D=11}
		\hat{*} \hat{F}_\4	&=-\frac{3^{3/2} \cos ^4\alpha}{g^3\, (1+2\sin^2\alpha)^2}\;\hat{e}^{8910}\wedge\bm{J}'\wedge\bm{J}'
					-\frac{\left(1+2\sin^2\alpha\right)^{2/3} \cos^2\alpha}{2^{1/6} \cdot3^{3/2}\, g}\;\overline{\text{vol}_{4}}\wedge\hat{e}^8\wedge\,\text{Im}\,\bm{\Omega}'		\notag\\[.5em]
					&+\frac{\cos ^2\alpha\, (7-10 \cos2 \alpha+\cos 4 \alpha)}{2^{7/6} \cdot3^{3/2} \, g \left(1+2 \sin ^2\alpha\right)^{1/3} \left(1+8\sin^4\alpha\right)^{1/2}}\;
					\overline{\text{vol}_{4}}\wedge\hat{e}^9\wedge\,\text{Re}\,\bm{\Omega}'																\notag\\[.5em]
					&\quad-\frac{\cos^3\alpha\, (7-10 \cos2 \alpha+\cos4 \alpha)}{3^{3/2}\cdot 2^{5/3}\, g\,\sin \alpha\, \left(1+2\sin^2\alpha\right)^{4/3} \left(1+8\sin^4\alpha\right)^{1/2}}\;
					\overline{\text{vol}_{4}}\wedge\hat{e}^{10}\wedge\,\text{Re}\,\bm{\Omega}'		\notag\\[.5em]
					&\qquad+\frac{\sin\alpha\cos^3\alpha}{\sqrt{3}\, g^2\, \left(1+2\sin^2\alpha\right)}\;\bar{*}\,\bar{F}\wedge\hat{e}^{8910}\wedge\,\text{Re}\,\bm{\Omega}'
					-\frac{\cos^2\alpha}{2 \sqrt{3}\, g^2}\;\bar{F}\wedge\hat{e}^{8910}\wedge\bm{J}'			\\[.5em]
					&\qquad\quad+\frac{\left(1+8\sin^4\alpha\right)^{1/2} \cos^4\alpha}{2^{5/3}\cdot 3^{1/2}\, g^4 \left(1+2\sin^2\alpha\right)^{4/3}}\;\bar{F}\wedge\hat{e}^{10}\wedge\bm{J}'\wedge\bm{J}'\,,	\notag
	\end{align}
where $\hat{e}^{8910}=\hat{e}^{8}\wedge \hat{e}^{9}\wedge \hat{e}^{10}$. Computing the differential of (\ref{eq:StarF4D=11}) with the help of the Sasaki-Einstein conditions satisfied by  $\bm{\eta}'$, $\bm{J}'$ and $\bm{\Omega}'$, as well as $\hat{F}_\4 \wedge \hat{F}_\4$ from (\ref{eq:4formminsugraCPW}) and putting everything together, we find that the $D=11$ equation of motion in (\ref{F4FieldEqs}) is indeed satisfied on the $D=4$ field equations (\ref{F2FieldEqs}).

%%%%%%%%%%%%%%%
%%%%%%%%%%%%%%%

\section{$D=11$ equations of motion on the AdS$_4$ solutions} \label{EomsAdS11D}

%%%%%%%%%%%%%%%
%%%%%%%%%%%%%%%

The AdS$_4$ solutions that we brought to section \ref{sec:11Dsolutions} of the main text are obtained from the consistent uplifting formulae of section \ref{sec:SU3sectorinD=11} by turning off the relevant tensor hierarchy fields, fixing the $D=4$ scalars to the vevs recorded in table \ref{table: critical loci}, and fixing the $\mathbb{R}^8$ embedding coordinates $\mu^A$, $A= 1, \ldots, 8$, in terms of various sets of intrinsic angles on $S^7$ discussed in appendix \ref{sec: geometryS7}. The particular choice of intrinsic coordinates for each solution was made on a case-by-case basis, as specific sets of coordinates are more suitable than others to highlight the specific symmetry of a solution. While this is obviously the best approach for the sake of presentation, it is definitely inconvenient to check the $D=11$ equations of motion, as one would also need to proceed on a case-by-case basis for each solution.

In order to check that the $D=11$ equations of motion hold it is more convenient to proceed differently. Firstly, leave the $D=4$ scalars as temporarily unfixed constants, and make a choice of intrinsic $S^7$ coordinates (regardless of whether they would be well adapted to specific sectors). For this purpose, we have chosen the intrinsic coordinates (\ref{IntrinsicCoordsS71}). The $D=11$ metric and four-form then get expanded in terms of the global five-dimensional Sasaki-Einstein structure $\bm{\eta}^\5$, $\bm{J}^\5$, $\bm{\Omega}^\5$ specified in appendix \ref{sec: geometryS7}, with coefficients that depend on the $D=4$ scalars along with the $S^7$ angles $\alpha$ and $\psi$. Secondly, plug these expressions into the $D=11$ field equations (\ref{F4FieldEqs}) and obtain, with the help of the Sasaki-Einstein relations (\ref{eq: SE5algebraic}), (\ref{eq: SE5differential}),  the set of equations that the coefficients must obey for the $D=11$ equations to hold. Finally, verify that these equations are satisfied when the $D=4$ scalars are fixed to the critical points recorded in table \ref{table: critical loci}.

Proceeding this way, we find that the $D=11$ metric (\ref{11DmetricEmbcoords}) can be written in terms of the intrinsic angles \eqref{IntrinsicCoordsS71} as
\begin{eqnarray} \label{eq:MetricPulledBack}
\metriceleven & = & \Delta^{-1}ds_{4}^{2}+ds_{7}^{2}\,,\nonumber \\
ds_{7}^{2}\, & = & G_{5}d\alpha^{2}+G_{7}d\psi^{2}+2G_{6}d\alpha d\psi\nonumber \\
 & + & G_{4}ds^{2}(\cptwo)+(G_{3}+G_{4}) \, ( \bm{\eta}^{\5} )^2 -2(G_{1}d\alpha+G_{2}d\psi) \, \bm{\eta}^\5 \,,
\end{eqnarray}
where both the warp factor,
\begin{eqnarray}
\Delta^{-1} & \equiv & e^{-\varphi}X^{1/3}\Delta_{1}^{2/3}\,,
\end{eqnarray}
given by $\Delta_{1}$ in (\ref{Deltatilde}) with \eqref{IntrinsicCoordsS71}, and the coefficients of the internal metric $ds_7^2$ depend on the $S^7$ angles $\alpha$, $\psi$ and the $D=4$ scalars: 
{\setlength\arraycolsep{1pt}
\begin{eqnarray}
G_{1} & = & \frac{\Delta^{2}}{g^{2}}\Big[-\frac{1}{2}e^{-2\phi}\sin\!\alpha\,\cos^{3}\!\alpha\,(X-Y)\left(2ae^{4\phi}\cos2\psi-\sin2\psi\left(-Y^{2}-Z^{2}+e^{4\phi}\right)\right)\Big]\,,\nonumber \\
G_{2} & = & \frac{\Delta^{2}}{g^{2}}\Big[e^{-2\phi}\sin^{2}\!\alpha\,\cos^{2}\!\alpha\,(X-Y)\left(ae^{4\phi}\sin2\psi+\sin^{2}\!\psi\left(Y^{2}+Z^{2}\right)+e^{4\phi}\cos^{2}\!\psi\right)\Big]\,,\nonumber \\
G_{3} & = & \frac{\Delta^{2}}{g^{2}}\Big[Y\cos^{4}\!\alpha\,(Y-X)\Big]\,,\nonumber \\
G_{4} & = & \frac{\Delta^{2}}{g^{2}}\Big[X^{2}\sin^{2}\!\alpha\,\cos^{2}\!\alpha\,e^{-2(\varphi+\phi)}\left(ae^{4\phi}\sin2\psi+\sin^{2}\!\psi\,\left(Y^{2}+Z^{2}\right)+e^{4\phi}\cos^{2}\!\psi\right)+XY\cos^{4}\!\alpha\Big]\,,\nonumber \\
G_{5} & = & \frac{\Delta^{2}}{g^{2}}\Big\{ XY\sin^{2}\!\alpha\,\cos^{2}\!\alpha\,-\frac{1}{64}\sin^{2}(2\alpha)\left(e^{2\phi}\left(\text{\ensuremath{\zeta}}^{2}+\text{\ensuremath{\tilde{\ensuremath{\zeta}}}}^{2}\right)+4\right)\left(e^{2\phi}\left(\text{\ensuremath{\zeta}}^{2}+\text{\ensuremath{\tilde{\ensuremath{\zeta}}}}^{2}\right)-4e^{2\varphi}\chi^{2}\right)\nonumber \\
 &  & \quad+X^{2}\sin^{4}\!\alpha\,e^{-2(\varphi+\phi)}\left(ae^{4\phi}\sin2\psi+\sin^{2}\!\psi\,\left(Y^{2}+Z^{2}\right)+e^{4\phi}\cos^{2}\!\psi\right)\nonumber \\
 &  & \quad+e^{-4\phi}\cos^{2}\!\alpha\,\Big[-2ae^{4\phi}\sin\!\psi\,\cos\!\psi+\cos^{2}\!\psi\,\left(Y^{2}+Z^{2}\right)+e^{4\phi}\sin^{2}\!\psi\Big]\nonumber \\
 &  & \qquad\quad\times\Big[\sin^{2}\!\alpha\,\left(ae^{4\phi}\sin2\psi+\sin^{2}\!\psi\,\left(Y^{2}+Z^{2}\right)+e^{4\phi}\cos^{2}\!\psi\right)+\cos^{2}\!\alpha\,e^{2(\varphi+\phi)}\Big]\Big\}\,,\nonumber \\
G_{6} & = & \frac{\Delta^{2}}{g^{2}}\Big[e^{-4\phi}\sin\!\alpha\,\cos\!\alpha\,\Big(-ae^{4\phi}\cos2\psi+\sin\!\psi\,\cos\!\psi\,\left(-Y^{2}-Z^{2}+e^{4\phi}\right)\Big)\Big]\nonumber \\
 &  & \,\,\times\,\,\Big[\sin^{2}\!\alpha\,\left(ae^{4\phi}\sin2\psi+\sin^{2}\!\psi\,\left(Y^{2}+Z^{2}\right)+e^{4\phi}\cos^{2}\!\psi\right)+\cos^{2}\!\alpha\,e^{2(\varphi+\phi)}\Big]\,,\nonumber \\
G_{7} & = & \frac{\Delta^{2}}{g^{2}}\Big[e^{-4\phi}\sin^{2}\!\alpha\,\Big(ae^{4\phi}\sin2\psi+\sin^{2}\!\psi\,\left(Y^{2}+Z^{2}\right)+e^{4\phi}\cos^{2}\!\psi\Big)\Big]\nonumber \\
 &  & \,\,\times\,\,\Big[\sin^{2}\!\alpha\,\left(ae^{4\phi}\sin2\psi+\sin^{2}\!\psi\,\left(Y^{2}+Z^{2}\right)+e^{4\phi}\cos^{2}\!\psi\right)+\cos^{2}\!\alpha\,e^{2(\varphi+\phi)}\Big]\, .
\end{eqnarray}
}

Turning off the $D=4$ tensor hierarchy fields (except for the local three-form $C_{\textrm{FR}} \equiv C^1 = C^{77} = C^{88}$ whose role is merely to serve as a local potential for the Freund-Rubin term) in the three form (\ref{11D3formEmbcoords}), its pull-back on $S^7$ induced by \eqref{IntrinsicCoordsS71} reads
{\setlength\arraycolsep{1pt}
\begin{eqnarray} \label{eq:A3App}
\hat{A}_\3 & = & (L_{2}d\alpha+L_{3}d\psi)\wedge\boldsymbol{J}^\5+(L_{4}d\alpha+L_{5}d\psi)\wedge\rp\boldsymbol{\Omega}^\5+(L_{6}d\alpha+L_{7}d\psi)\wedge\ip\boldsymbol{\Omega}^\5\nonumber \\
 & + & \big(L_{8}\ip\boldsymbol{\Omega}^\5+L_{9}\rp\boldsymbol{\Omega}^\5+L_{10}\boldsymbol{J}^\5\big)\wedge\boldsymbol{\eta}^\5+L_{1}\,d\alpha\wedge d\psi\wedge\boldsymbol{\eta}^\5 + C_{\textrm{FR}} \; .
\end{eqnarray}
}
The coefficients here are given by
{\setlength\arraycolsep{1pt}
\begin{eqnarray} \label{eq:A3Coefs}
L_{1} & = & \frac{\Delta^{3}}{8g}\Big(\frac{1}{2}\chi\sin\!\alpha\,\cos^{2}\!\alpha\,e^{-\varphi-4\phi}\Big)\nonumber \\
 &  & \times\Big[\sin\!\alpha\,\sin2\alpha\,(X-Y)e^{2(\varphi+\phi)}\left(e^{2\varphi}\chi^{2}-Y+1\right)\left(ae^{4\phi}\sin2\psi+\sin^{2}\!\psi\,\left(Y^{2}+Z^{2}\right)+e^{4\phi}\cos^{2}\!\psi\right)\nonumber \\
 &  & \quad-2\left(X^{2}\sin^{2}\!\alpha\,\left(ae^{4\phi}\sin2\psi+\sin^{2}\!\psi\,\left(Y^{2}+Z^{2}\right)+e^{4\phi}\cos^{2}\!\psi\right)+Y^{2}\cos^{2}\!\alpha\,e^{2(\varphi+\phi)}\right)\nonumber \\
 &  & \quad\,\,\times\,\,\Big(\cos\!\alpha\,e^{2(\varphi+\phi)}+\sin\!\alpha\,\tan\!\alpha\,\left(\sin^{2}\!\psi\,\left(Y^{2}+Z^{2}\right)+Ze^{2\phi}\sin2\psi+e^{4\phi}\cos^{2}\!\psi\right)\Big)\Big]\,,\nonumber \\
L_{2} & = & \frac{\Delta^{3}}{g^{3}}\Big[-\chi e^{-\varphi-4\phi}X\sin\!\alpha\,\cos^{3}\!\alpha\,\Big(\sin\!\psi\,\cos\!\psi\,\left(-Y^{2}-Z^{2}+e^{4\phi}\right)-Ze^{2\phi}\cos2\psi\Big)\Big]\nonumber \\
 &  & \,\,\times\,\,\Big[X\sin^{2}\!\alpha\,\left(ae^{4\phi}\sin2\psi+\sin^{2}\!\psi\,\left(Y^{2}+Z^{2}\right)+e^{4\phi}\cos^{2}\!\psi\right)+Y\cos^{2}\!\alpha\,e^{2(\varphi+\phi)}\Big]\,,\nonumber \\
L_{3} & = & -\frac{\tan\!\alpha\,\sin\!\psi\,\left(Y^{2}+Z^{2}+2Ze^{2\phi}\cot\!\psi\,+e^{4\phi}\cot^{2}\!\psi\,\right)}{Ze^{2\phi}\cos2\psi-\sin\!\psi\,\cos\!\psi\,\left(-Y^{2}-Z^{2}+e^{4\phi}\right)}\,\,L_{2}\,,\nonumber \\
L_{4} & = & \frac{\Delta^{3}}{g^{3}}\Big(\frac{1}{2}X\cos^{2}\!\alpha\,e^{-3\varphi-2\phi}\Big)\nonumber \\
 &  & \,\,\times\,\,\Big[X\sin^{2}\!\alpha\,\left(ae^{4\phi}\sin2\psi+\sin^{2}\!\psi\,\left(Y^{2}+Z^{2}\right)+e^{4\phi}\cos^{2}\!\psi\right)+Y\cos^{2}\!\alpha\,e^{2(\varphi+\phi)}\Big]\nonumber \\
 &  & \,\,\times\,\,\Big[X\sin^{2}\!\alpha\,\left(\text{\ensuremath{\zeta}}e^{2\phi}\cos\!\psi+\sin\!\psi\,(\tilzeta Y+\zeta Z)\right)+e^{2\varphi}\cos^{2}\!\alpha\,\left(\tilzeta e^{2\phi}\sin\!\psi+\cos\!\psi\,(\zeta Y-\tilzeta Z)\right)\Big]\,,\nonumber \\
L_{5} & = & -\frac{e^{2\phi}\left(\text{\ensuremath{\tilzeta}}e^{2\phi}\cos\!\psi+\sin\!\psi\,(\text{\ensuremath{\tilzeta}}Z-\zeta Y)\right)}{\chi\left(\sin2\psi\left(-Y^{2}-Z^{2}+e^{4\phi}\right)-2Ze^{2\phi}\cos2\psi\right)}\,\,L_{2}\,,\nonumber \\
L_{6} & = & \frac{2}{\sin2\alpha}\left(e^{-2\varphi}X\sin^{2}\!\alpha+\frac{\cos^{2}\!\alpha\,\left(\cos\!\psi\,(\tilzeta Y+\zeta Z)-\zeta e^{2\phi}\sin\!\psi\right)}{\tilzeta e^{2\phi}\cos\!\psi+\sin\!\psi\,(\text{\ensuremath{\tilzeta}}Z-\zeta Y)}\right)\,\,L_{5}\,,\nonumber \\
L_{7} & = & -\frac{\text{\ensuremath{\zeta}}e^{2\phi}\cos\!\psi+\tilzeta Y\sin\!\psi+\zeta Z\sin\!\psi}{\text{\ensuremath{\tilzeta}}e^{2\phi}\cos\!\psi-\zeta Y\sin\!\psi+\tilzeta Z\sin\!\psi}\,\,L_{5}\,,\nonumber \\
L_{8} & = & -e^{-2\varphi}X\;L_{7}\,,\nonumber \\
L_{9} & = & -e^{-2\varphi}X\;L_{5}\,,\nonumber \\
L_{10} & = & \frac{\Delta^{3}}{g^{3}}\Big(-e^{\varphi}\chi Y\cos^{2}\!\alpha\Big)  \\
 &  & \,\,\times\,\,\Big[X^{2}\sin^{2}\!\alpha\,\cos^{2}\!\alpha\,e^{-2(\varphi+\phi)}\left(ae^{4\phi}\sin2\psi+\sin^{2}\!\psi\,\left(Y^{2}+Z^{2}\right)+e^{4\phi}\cos^{2}\!\psi\right)+XY\cos^{4}\!\alpha\Big]\, . \nonumber
\end{eqnarray}
}

Finally, the $D=11$ four-form $\hat{F}_\4 = d\hat{A}_\3$ is
{\setlength\arraycolsep{1pt} 
\begin{eqnarray} \label{eq:F4App}
\hat{F}_{(4)} & = & U \, \text{vol}_{4}+d\alpha\wedge d\psi\wedge\big(f_{1}\boldsymbol{J}^\5+f_{2}\rp\boldsymbol{\Omega}^\5+f_{3}\ip\boldsymbol{\Omega}^\5\big)+f_{10}\boldsymbol{J}^\5\wedge\boldsymbol{J}^\5 \\
 & + & \Big[\big(f_{4}d\alpha+f_{5}\,d\psi\big)\wedge\rp\boldsymbol{\Omega}^\5+\big(f_{6}d\alpha+f_{7}\,d\psi\big)\wedge\ip\boldsymbol{\Omega}^\5+\big(f_{8}d\alpha+f_{9}\,d\psi\big)\wedge\boldsymbol{J}^\5\Big]\wedge\boldsymbol{\eta}^\5\,, \nonumber
\end{eqnarray}
}where the Freund Rubin term is given by $U\text{vol}_{4}=H_{\4}^{1}\mu_{i}\mu^{i}+H_{\4}^{ab} \mu_{a}\mu_{b}$ evaluated on \eqref{IntrinsicCoordsS71} and on the $D=4$ dualisation conditions (\ref{duality:4forms}). The functional coefficients in (\ref{eq:F4App}) can be written in terms of the coefficients (\ref{eq:A3Coefs}) of the three form (\ref{eq:A3App}) as
\begin{eqnarray} 
f_{1}=2L_{1}+\partial_{\alpha}L_{3}-\partial_{\psi}L_{2}\,, 
& \quad & 
f_{6}=3L_{4}+\partial_{\alpha}L_{8\,,}\nonumber \\
f_{2}=\partial_{\alpha}L_{5}-\partial_{\psi}L_{4}\,, 
& \quad & 
f_{7}=3L_{5}+\partial_{\psi}L_{8\,,}\nonumber \\
f_{3}=\partial_{\alpha}L_{7}-\partial_{\psi}L_{6}\,, 
& \quad & 
f_{8}=\partial_{\alpha}L_{10}\,,\nonumber \\
f_{4}=-3L_{6}+\partial_{\alpha}L_{9\,,} 
& \quad & 
f_{9}=\partial_{\psi}L_{10}\,,\nonumber \\
f_{5}=-3L_{7}+\partial_{\psi}L_{9\,,} 
& \quad & 
f_{10}=2L_{10}\,.\label{eq:fINTermsOfL}
\end{eqnarray}

The Bianchi identity $d\hat{F}_{(4)}=0$ amounts to verifying the
following relations:
\begin{align*}
3f_{3}+\partial_{\alpha}f_{5}-\partial_{\psi}f_{4}=0\,,\qquad & -3f_{2}+\partial_{\alpha}f_{7}-\partial_{\psi}f_{6}=0\,,
\end{align*}
\begin{eqnarray}
\partial_{\alpha}f_{10}-2f_{8}=0\, , \qquad & \partial_{\alpha}f_{9}-\partial_{\psi}f_{8}=0\,,\qquad & \partial_{\psi}f_{10}-2f_{9}=0\,.
\end{eqnarray}
Of course, these conditions are automatically satisfied by construction for all values of the $D=4$ scalars upon using (\ref{eq:fINTermsOfL}).

We next compute the Hodge dual of the $\hat{F}_{(4)}$ given in (\ref{eq:F4App}) with respect to the $D=11$ metric (\ref{eq:MetricPulledBack}). We obtain
{\setlength\arraycolsep{1pt}
\begin{eqnarray}
\hat{*}\hat{F}_{(4)} & = & \Delta^{2}\text{vol}_{7}\nonumber \\
 & + & \Delta^{-2}\text{vol}_{4}\wedge\Big[(p_{1}d\alpha+p_{2}d\psi)\wedge\boldsymbol{J}^\5+(p_{4}d\alpha+p_{5}d\psi)\wedge\rp\boldsymbol{\Omega}^\5+(p_{7}d\alpha+p_{8}d\psi)\wedge\ip\boldsymbol{\Omega}^\5\nonumber \\
 &  & \qquad\qquad\;\;+\big(p_{6}\rp\boldsymbol{\Omega}^\5+p_{9}\ip\boldsymbol{\Omega}^\5 +p_{3}\boldsymbol{J}^\5\big)\wedge\boldsymbol{\eta}^\5 +p_{10}\,d\alpha\wedge d\psi\wedge\boldsymbol{\eta}^\5 \Big]\,,
\end{eqnarray}
}with coefficients
{\setlength\arraycolsep{1pt}
\begin{eqnarray}
p_{1}=\frac{1}{\Delta^2G_V}
\Big[f_{1}G_{1}-f_{9}G_{5}+f_{8}G_{6}\Big]\,, 
& \quad &
p_{6}=\frac{1}{\Delta^2G_V}
\Big[f_{5}G_{1}-f_{4}G_{2}-f_{2}\left(G_{3}+G_{4}\right)\Big]\,,
\nonumber \\
p_{2}=\frac{1}{\Delta^2G_V}
\Big[f_{1}G_{2}-f_{9}G_{6}+f_{8}G_{7}\Big]\,, 
& \quad &
p_{7}=\frac{1}{\Delta^2G_V}
\Big[f_{3}G_{1}-f_{7}G_{5}+f_{6}G_{6}\Big]\,,
\nonumber \\
p_{3}=\frac{1}{\Delta^2G_V}
\Big[f_{9}G_{1}-f_{8}G_{2}-f_{1}\left(G_{3}+G_{4}\right)\Big]\,,
&\quad  & 
p_{8}=\frac{1}{\Delta^2G_V}
\Big[f_{3}G_{2}-f_{7}G_{6}+f_{6}G_{7}\Big]\,,
\nonumber \\
p_{4}=\frac{1}{\Delta^2G_V}
\Big[f_{2}G_{1}-f_{5}G_{5}+f_{4}G_{6}\Big]\,, 
& \quad & 
p_{9}=\frac{1}{\Delta^2G_V}
\Big[f_{7}G_{1}-f_{6}G_{2}-f_{3}\left(G_{3}+G_{4}\right)\Big]\,,
\nonumber \\
p_{5}=\frac{1}{\Delta^2G_V}
\Big[f_{2}G_{2}-f_{5}G_{6}+f_{4}G_{7}\Big]\,,
& \quad & p_{10}=-2\frac{G_{V}G_{4}^{2}}{\Delta^2}f_{10}\, .
\end{eqnarray}
}Here,
\begin{equation}
G_{V} = \sqrt{-G_{7}G_{1}^{2}+2G_{2}G_{6}G_{1}-G_{3}G_{6}^{2}-G_{4}G_{6}^{2}-G_{2}^{2}G_{5}+G_{3}G_{5}G_{7}+G_{4}G_{5}G_{7}}\, 
\end{equation}
is related to the volume element corresponding to the internal metric $ds_7^2$ in 
(\ref{eq:MetricPulledBack}). With these definitions, the equation of motion  in (\ref{F4FieldEqs}) for the $D=11$ four-form becomes equivalent to the following conditions:
\begin{eqnarray} \label{eq:EOMsAdS4D=11}
Uf_{1}+\partial_{\alpha}p_{2}-\partial_{\psi}p_{1}+2p_{10}=0\,, 
& \qquad & 
Uf_{6}+\partial_{\alpha}p_{9}+3p_{4}=0\,,
\nonumber \\
Uf_{2}+\partial_{\alpha}p_{5}-\partial_{\psi}p_{4}=0\,, 
& \qquad & 
Uf_{7}+\partial_{\psi}p_{9}+3p_{5}=0\,,
\nonumber \\
Uf_{3}+\partial_{\alpha}p_{8}-\partial_{\psi}p_{7}=0\,, 
& \qquad & 
Uf_{8}+\partial_{\alpha}p_{3}=0\,,
\nonumber \\
Uf_{4}+\partial_{\alpha}p_{6}-3p_{7}=0\,, 
& \qquad & 
Uf_{9}+\partial_{\psi}p_{3}=0\,,\
\nonumber \\
Uf_{5}+\partial_{\psi}p_{6}-3p_{8}=0\,,
& \qquad & 
 Uf_{10}+2p_{3}=0\,.
\end{eqnarray}

We have verified that equations (\ref{eq:EOMsAdS4D=11}) hold when the $D=4$ scalars are evaluated at any of the critical points collected in table \ref{table: critical loci}. We have also checked that all the metric and four-forms for the explicit AdS$_4$ solutions written in section \ref{sec:11Dsolutions} can be brought to the form (\ref{eq:MetricPulledBack})--(\ref{eq:fINTermsOfL}), with the help of the relations given in appendix B. Thus, the explicit AdS$_4$ configurations of section \ref{sec:11Dsolutions} do indeed solve the $D=11$ field equations (\ref{F4FieldEqs}).

%\printbibliography		% \gl{I need to compile the bibliography using this}
\bibliography{references}
\end{document}